\documentclass{article}
\usepackage{graphicx}
\usepackage{balance}  % for  \balance command ON LAST PAGE  (only there!)

\usepackage{comment}
\usepackage{times}
\usepackage{subfigure}
\usepackage{algorithm}
\usepackage{algorithmic}
\usepackage{amsmath}

\usepackage{color} % new added

\newcommand{\SF}{ {\cal F} }
\newcommand{\SI}{ {\cal I} }

\newcommand{\SG}{ {\cal G} }

\newcommand{\SV}{ {\cal V} }

\newcommand{\SD}{ {\cal D} }
\newcommand{\SR}{ {\cal R} }
\newcommand{\SP}{ {\cal P} }
\newcommand{\SA}{ {\cal A} }

\newcommand{\SN}{ {\cal N} }

\newcommand{\ST}{ {\cal T} }
\newcommand{\SH}{ {\cal H} }

\newcommand{\HV}{ \widehat{ \mbox{V}}\mbox{ar} }
\newcommand{\HP}{ \widehat { \mbox{P} }\mbox{r} }
\newcommand{\HE}{ {\widehat E} }
\newcommand{\MP}{ \mbox{Pr} }
\newcommand{\MV}{ \mbox{Var} }

\newtheorem{theorem}{{\bf Theorem}}
\newtheorem{lemma}{{\bf Lemma}}

\newtheorem{definition}{{\bf Definition}}

\def\done{\hspace*{\fill} \rule{1.8mm}{2.5mm} \\}

\begin{document}

\title{{\bf Mathematical Modeling of Competitive Group Recommendation Systems with
Application to Peer Review Systems}}

%\numberofauthors{1} %  in this sample file, there are a *total*

\author{
Hong Xie \hspace{0.1in} John C.S. Lui\\
Computer Science \& Engineering Department\\
The Chinese University of Hong Kong\\
Email: \{hxie,cslui\}@cse.cuhk.edu.hk
}

\maketitle
\begin{abstract}
In this paper, we present a mathematical model to capture various factors which
may influence the accuracy of a competitive group recommendation system.  We
apply this model to peer review systems, i.e., conference or research grants
review, which is an essential component in our scientific community. We explore
number of important questions, i.e., how will the number of reviews per paper
affect the accuracy of the overall recommendation?  Will the score aggregation
policy influence the final recommendation?  How reviewers' preference may
affect the accuracy of the final recommendation? To answer these important
questions, we formally analyze our model.  Through this analysis, we obtain the
insight on how to design a randomized algorithm which is both computationally
efficient and asymptotically accurate in evaluating the accuracy of a
competitive group recommendation system.  We obtain number of interesting
observations: i.e., for a medium tier conference, three reviews per paper is
sufficient for a high accuracy recommendation. For prestigious conferences, one
may need at least seven reviews per paper to achieve high accuracy. We also
propose a heterogeneous review strategy which requires equal or less reviewing
workload, but can improve over a homogeneous review strategy in recommendation
accuracy by as much as 30\% . We believe our models and methodology are
important building blocks to study competitive group recommendation systems.
\end{abstract}

\section{Introduction} \label{section: introduction}

In recent years, recommendation systems \cite{rec_sys} have received a lot of
attention in both commercial and academic communities. Researchers investigate
various algorithmic and complexity issues \cite{rec_sys_item_top_n_algo,
rec_sys_eval_coll_fil_algo, rec_sys, rec_sys_rec_algo,rec_sys_ec}, at the same
time, we also see successful applications of recommendation systems in
commercial products. In general, recommendation systems take into account a
user's preference and make a recommendation so as to maximize the user's
utility. {\em Group recommendation systems
}\cite{group_rec_sys_state_of_the_art}, on the other hand, take into account
the preferences of {\em all} users in a group to make a single recommendation.
In recent years, we have seen successful group recommendations in commercial
areas \cite{group_rec_sys_sem_eff,group_rec_sys_state_of_the_art,
group_rec_sys_challenge, travel_decision_forum, group_rec_sys_opt_pre,
group_modeling,PolyLen}.

In this paper, we consider a special class of recommendation system which we
call the {\em competitive group recommendation system}: there are $N$ users and
the system will make a single recommendation to $k$ users only, where $k \!
\leq\! N$, while $N\!-\!k$ users will receive the {\em complement} of the
recommendation. Competitive group recommendation systems have many important
applications. In here, we consider an application which is dearest to many
researchers' heart: {\em peer review systems} for conferences or research grant
proposals. To a certain degree, the progress of our scientific community
depends on the accuracy this type of recommendation systems. To the best of our
knowledge, this is the first paper which provides a formal mathematical
analysis to such recommendation systems.

A peer review system can be briefly described as follows: there are $N$
candidates (papers or grant proposals), a group of reviewers is asked to review
these candidates.  Each reviewer evaluates a subset of these candidates based
on her preference, and will provide a rating for each candidate. The system
will use some policies to aggregate all ratings of all candidates, and will
only recommend a subset $k$ candidates for acceptance, while all other
candidates will receive a rejection, which is the complement of the acceptance
recommendation. For such systems, there are many interesting questions to
explore, e.g., to achieve high accuracy, how many reviews each candidate should
receive? What is the probability that the best candidate will be accepted or
rejected? How reviewers' preference may influence the final recommendation? Is
one rating aggregation policy more accurate than others?

Our contribution can be summarized as follows:
\begin{itemize}
\item We propose a mathematical model to understand the accuracy of a
    competitive group recommendation system and apply it to conference
    review systems.

\item We formally analyze the model. Through this analysis, we gain the
    insight to create a randomized algorithm to evaluate the model. We show
    our algorithm is computationally efficient and also provides
    performance guarantees.

\item We apply our model to a conference review system and show many
    interesting insights, i.e., for a medium titer conference, three
    reviews per paper can guarantee a highly accurate recommendation, but
    for prestigious conferences, we need at least seven reviews per paper.

\item We propose a two round heterogeneous review strategy which
    outperforms the homogeneous review strategy  by as much as 30\% in
    recommendation accuracy with the the same or less reviewing workload.
\end{itemize}

This is the outline of the paper. In Section \ref{section: model}, we present
the mathematical model of competitive group recommendation systems. In Section
\ref{section: analysis}, we present analysis and derive theoretical results of
the model. In Section \ref{section: monte carlo}, we propose a randomized
algorithm which is computationally efficient and provides performance
guarantees in evaluating a competitive group recommendation system. In Section
\ref{section: simulations}, we evaluate the performance of a conference review
system and explore various factors that influence its accuracy. Related work is
given in Section \ref{section: related work} and Section\ref{section:
conclusion} concludes.

\section{Mathematical Model}
\label{section: model}

Let us present the mathematical model of a competitive group recommendation
system and we focus on a particular application scenario, a conference review
system which is a representative example of peer review systems. Let
$\mathcal{P}\! = \!\{P_1, \ldots, P_N\}$ be a finite set of $N$ candidate
papers. Let $Q_i \!\in\! (1, m)$ represent the {\em intrinsic quality} of paper
$P_i$. Higher value of intrinsic quality implies higher quality. Hence, if $Q_i
\!> \! Q_j$, it means paper $P_i$ is better than $P_j$. Without any loss of
generality, let us assume $Q_1  \!> \! Q_2  \!> \! \cdots \! > \! Q_N$. It is
important to emphasize that reviewers of these papers do {\em not} have any
a-prior knowledge of $Q_i, \forall i$. The conference can only accept $k$
papers, where $1 \! \leq \! k \! \leq N$. Let $\SA^I(k)$ and $\SA(k)$ denote
the set of the $k$ accepted papers according to the intrinsic quality or
according to the conference recommendation criteria respectively. It is clear
that $\SA^I(k)\! =\! \{P_1, P_2,\cdots,P_k\}$, and if a conference
recommendation system is perfect, we should have $\SA^I(k)\! =\! \SA(k)$. But
in general, many factors or reviewers' preference may influence the final
recommendation, hence $\SA^I(k) \! \not= \! \SA(k)$. To measure the accuracy of
a recommendation system, we aim to determine how many papers in $\SA(k)$ are
also in $\SA^I(k)$. Formally, we seek to derive the following probability mass
function (pmf) : \vspace{-0.025in}
\[
    \MP[ |\SA^I(k) \cap \SA(k)| = i ], \hspace{0.2in}
    \mbox{for $i=0,1,\ldots,k$.}
\]
Intuitively, if $\MP[ |\SA^I(k) \cap \SA(k)| \! = \! k ]$ occurs with a high
probability, then the conference recommendation system is {\em accurate} and at
the same time, robust against different scoring and human factors.

Let $\mathcal{R}$ be a finite set of $M$ reviewers. We assume that the
reviewers are independent. Reviewers do not have a direct knowledge of $Q_i,
\forall i$, the intrinsic quality of papers, and they evaluate papers based on
their own preference. A reviewer submits a score for each paper after
reviewing. Scores are discrete and take on value in $\{1, \ldots, m\}$. Paper
$P_i$, for $i\!=\!1,\ldots, N$, is assigned to $n_i \! \geq\!  1$ reviewers.
Hence, paper $P_i$ receives $n_i$ reviewing scores. Let $\mathcal{S}^i \!=\!
\{S_{1}^i, \ldots, S_{n_i}^i\}$ denote the set of $n_i$ scores of paper $P_i$,
where $S_{j}^i \! \in \! \{1, \ldots, m \}$. Let $\SR(S_{j}^i)$ represent the
reviewer who submits score $S^i_j$. Let $e_{j}^i  \! \in \!  \{1, \ldots, l \}$
represent the expertise level (or familiarity) that reviewer $\SR(S_{j}^i)$
selects on topics related to paper $P_i$. Reviewer $\SR(S_{j}^i)$ submits score
$S^i_j$ in conjunction with expertise level $e^i_j$. Again, we adopt the
convention that higher values represent higher quality or expertise level.
There are number of interesting questions one can explore, i.e., how $M$, the
number of reviewers (or the size of a technical program committee), as well as
$n_i, \forall i$, the number of reviews for each paper, may affect the accuracy
of the final recommendation?

Let $\mathcal{V}$ be the voting rule that is used by the conference
recommendation system to rank papers based on their reviewing scores.
Generally, a voting rule works in two steps. It first aggregates the reviewing
scores of each paper into a combined overall score. Then it ranks all papers
based on their respective combined overall scores. Let $\gamma_i \!=\! \SV (
\mathcal{S}^i )$ be the combined overall score of paper $P_i$ derived from
$\mathcal{S}^i$ under the voting rule $\SV$. There can be many voting rules. A
simple and often used voting rule is the {\em average score rule}. In this
case, we have $\gamma_i \!=\! \sum_{S \in \mathcal{S}^i } S / |\mathcal{S}^i
|$. By sorting $\gamma_1, \ldots, \gamma_N$, we obtain a ranked list of all
papers. Again, there are number of interesting questions to explore, i.e., what
are some effective voting rules?  Can one voting rule be more accurate than
others?

Specifying the voting rule is not enough. Recall that the system can only
accept $k$ papers. It may happen that the combined overall score of the
$k^{th}$ ranked paper, equals to that of $(k\!+\!1)$-th ranked paper. In this
case, we need to specify a {\em tie-breaking rule} to decide which paper should
be selected. Let $\ST$ denote the tie-breaking rule. It is interesting to
explore whether the recommendation results are sensitive to a particular
tie-breaking rule.

To answer the above questions, let us now present probabilistic models in
describing the intrinsic quality (or the {\em self-selection effect}), the
reviewing behavior, as well as critical degree of reviewers.

\subsection{Model Intrinsic Quality via Self-selection}
\label{section_model_intrinsic_quality}

It is well-known that paper submission has the {\em self-selection effect}. In
other words, authors tend to submit their high quality papers to some highly
prestigious and selective conferences, while lower tier conferences may receive
papers with lower quality, or candidate papers have high variance in quality.
The intrinsic quality of a submitted paper can be described as a random
variable, and one can vary its mean or variance to reflect the {\em
self-selection effect}. Specifically, a high value of mean and a small value of
variance imply that the submitted papers are of high intrinsic qualities and
these qualities have small variation only. On the other hand, a low value of
mean and a large value of variance imply that the submitted papers have low
intrinsic qualities and these qualities have high variability.

We use $Q_i \!\in \! (1,m)$ to denote the intrinsic quality of paper $P_i$.
Assume $Q_1,..., Q_N$ are independent random variables. Let $\SD(Q_i)$ denote
the probability distribution of $Q_i$. The probability distribution $\SD(Q_i)$
is described by a truncated normal distribution $\mathcal{N}(q_i,
\sigma^{2}_i)$, where $q_i \! \in\! (1, m)$ is the mean and $\sigma^2_i$ is the
variance. Since the value of the intrinsic quality $Q_i$ is in $(1, m)$, thus
$\SD(Q_i)$ is obtained by truncating $\mathcal{N}(q_i, \sigma^2_i)$ to keep
those values in $(1, m)$ and scaling up the kept values by $1 / \MP[1 \! < \! X
\! < \! m]$, where $X$ is a random variable with probability distribution
$\mathcal{N}(q_i, \sigma^{2}_i)$. It should be clear that after truncation, the
mean $q_i$ and the variance $\sigma^2_i$ can still reflect the {\em
self-selection effect}. Here, we use the following parameters to reflect four
representative types of {\em self-selectivity}:

\noindent {\bf High {\em self-selectivity}}: the mean $q_i$ and variance
$\sigma_i^2$ are specified by
\begin{equation}
\label{equation:high_self_select}
    q_i = m, \: \sigma_i^2 = 1,  \hspace{0.2in}
    \mbox{for $i = 1, \ldots, N$}.
\end{equation}
This indicates that papers tend to have high intrinsic quality (or high mean),
and most of the probability mass concentrates around high intrinsic quality.
Top tier conferences fall into this category.

\noindent {\bf Medium {\em self-selectivity}}: the mean $q_i$ and variance
$\sigma_i^2$ are
\begin{equation}
\label{equation:medium_self_select}
    q_i = (m+1)/2, \: \sigma_i^2 = 1,  \hspace{0.2in}
    \mbox{for $i = 1, \ldots, N$}.
\end{equation}
This reflects that papers tend to have an average intrinsic quality and most of
the probability mass concentrates around the average intrinsic quality. Medium
tier conferences fall into this category.

\noindent {\bf Low {\em self-selectivity}}: the mean $q_i$ and variance
$\sigma_i^2$ are
\begin{equation}
\label{equation:low_self_select}
    q_i = 1, \: \sigma_i^2 = 1,  \hspace{0.2in}
    \mbox{for $i = 1, \ldots, N$}.
\end{equation}
This indicates that papers tend to have low intrinsic quality (or low mean),
and most of the probability mass concentrates around low quality. Low tier
conferences fall into this category.

\noindent {\bf Random {\em self-selectivity}}: the variance $\sigma_i^2$ is
\begin{equation}
\label{equation:random_self_select}
    \sigma_i^2 = \infty,  \hspace{0.2in}
    \mbox{for $i = 1, \ldots, N$}.
\end{equation}
So $\SD(Q_i)$ converges to a uniform distribution on $(1,m)$. This means that
the intrinsic qualities of submitted papers are uniformly distributed. Newly
started conferences fall into this category. This is because a newly started
conference has not built up a reputation yet, thus researchers are not sure if
it is a good conference, which results in random quality in submission.

\subsection{Model for Reviewing Behavior}
When reviewing a paper, a reviewer needs to evaluate its quality. Here we
assume that each reviewer is fair, unbias and critical. We consider two most
important factors that affect the evaluation. The first one is $Q_i$, the
intrinsic quality of paper $P_i$, and the second one is the critical degree of
the reviewer. Specifically, when $Q_i$ is high, the evaluated quality of $P_i$
is more likely to be high. And the higher the critical degree of the reviewer,
the more likely that the evaluated quality tends to be close to the intrinsic
quality of that paper. The reviewing behavior can be described by a random
variable and one can vary its mean and variance to reflect the intrinsic
quality and critical degree.

To illustrate, consider a paper $P_i$ and one of its score $S^i_j$. Recall that
the reviewer who submits score $S^i_j$ is denoted by $\SR(S^i_j)$. Let $c^i_j
\! \in \! [0, 1]$ denote the critical degree of reviewer $\SR(S^i_j)$. With the
usual convention, higher value represents higher critical degree. The score
$S^i_j$ is a random variable with probability distribution $\SD ( S^i_j )$,
which should have the following two properties:

\noindent {\bf Property 1:} The mean should be equal to $Q_i$. The physical
meaning is that a reviewer is unbias.

\noindent {\bf Property 2:} The variance should reflect the critical degree of
a reviewer. Specifically, the higher the critical degree, the lower the
variance for the probability distribution $\mathcal{D}(S^i_j)$.

In our study, the probability distribution $\mathcal{D}(S^i_j)$ is obtained by
mapping a  normal distribution $\SN \left(Q_i, \sigma^2(c^i_j) \right)$ to a
discrete distribution. Note that the standard variance $\sigma(c^i_j)$ is a
monotonic decreasing function of $c^i_j$ and we will specify it in later
section. The probability distribution mapping can be described by the following
two steps:

\noindent {\bf {\em Discretization}} : Transform a normal distribution into a
discrete distribution. We transform the normal distribution $\SN \left(Q_i,
\sigma^2(c^i_j) \right)$ into a discrete random variable $L$ with probability distribution \\
$\widetilde{\SN} \left(Q_i, \sigma^2(c^i_j) \right)$ with values in $\{1,...,m
\}$. The pmf of  $L$ is:
\begin{align}
\MP[ L = \ell ]
& = \frac{ \MP [ \ell - 0.5 \leq X \leq \ell + 0.5 ] }
         { \MP [ 0.5 \leq X \leq m + 0.5 ] }
  = \frac{ \Phi \left( ( \ell + 0.5 - Q_i ) / \sigma( c^i_j ) \right) -
           \Phi \left( ( \ell - 0.5 - Q_i ) / \sigma( c^i_j ) \right)
         }
         { \Phi \left( ( m + 0.5 - Q_i ) / \sigma( c^i_j ) \right) -
           \Phi \left( ( 0.5 - Q_i ) / \sigma( c^i_j ) \right)
         },  \nonumber \\
& \hspace{0.17 in} \mbox{ for $ \ell = 1 , \ldots, m $ }
\label{equation_Prob(s)}
\end{align}
where $\Phi(x) \!=\! \int^x_{-\infty} \exp( - t^2 / 2 ) / \sqrt{2\pi} dt$ and
the probability distribution of $X$ is $\SN \left( Q_i, \sigma^2(c^i_j)
\right)$. Note that this discrete distribution satisfies Property 2 but not
Property 1. In the following step, we adjust the distribution so that it
satisfies Property 1 also.

\noindent {\bf {\em Adjustment}}: Adjust the distribution $\widetilde{\SN}
\left(Q_i, \sigma^2(c^i_j) \right)$ such that its mean equals to $Q_i$. The
idea is that if $E[ L ] \! < \! Q_i$, then we increase the mean of
$\widetilde{\SN} \left(Q_i, \sigma^2(c^i_j)\right)$ by scaling up the
probability:
\[
    \MP[ L = \ell], \hspace{0.2in}
    \mbox{ for all $ \ell = \lfloor Q_i \rfloor + 1, \ldots, m $.}
\]
Else, we decrease the mean by scaling them down. Applying this idea to adjust
the mean of $\widetilde{ \SN } \left( Q_i, \sigma^2( c^i_j ) \right)$, we
obtain the probability distribution $\SD ( S^i_j )$. The pmf of $ \SD ( S^i_j )
$ is:
\begin{align}
\MP[S^i_j = \ell] & = \begin{cases}
    [ 1 - \beta ( Q_i, c^i_j ) ] \MP [ L = \ell ],
    & \forall \ell = 1, \ldots, \lfloor Q_i \rfloor  \\
    [ 1 + \alpha ( Q_i, c^i_j ) ] \MP [ L = \ell ],
    & \forall \ell = \lfloor Q_i \rfloor + 1, \ldots, m
\end{cases},
\label{equation_rating_distribution_general}
\end{align}
where $\alpha(Q_i, c^i_j)$ and $\beta(Q_i, c^i_j)$ are:
\begin{equation}
\alpha ( Q_i, c^i_j )
= \frac{ \sum_{ \ell = 1 }^{ \lfloor Q_i \rfloor }
         \MP [ L = \ell ] ( Q_i - E [ L ] )
       }
       { \sum_{ \ell = 1 }^{ \lfloor Q_i \rfloor }
         \MP [ L = \ell ] ( E [ L ] - \ell )
       },
\label{equation_alfa}
\end{equation}
\begin{equation}
\beta(Q_i, c^i_j)
= \frac{ \sum_{ \ell = \lfloor Q_i \rfloor + 1}^m
         \MP [ L = \ell ] ( Q_i - E[ L ] )
       }
       { \sum_{ \ell = 1 }^{ \lfloor Q_i \rfloor }
         \MP [ L = \ell ] ( E[ L ] - \ell )
       },
\label{equation_beta}
\end{equation}
and $L$ is a discrete random variable with probability distribution
$\widetilde{ \SN } \left( Q_i, \sigma^2 ( c^i_j ) \right)$. Note that
$\mathcal{D}(S^i_j)$ satisfies both Property 1 and 2.

\subsection{Model for Critical Degree}
To model the critical degree of a reviewer, we classify papers and reviewers
into ``{\em types}''.  Specifically, a paper can be of many types (e.g., system
paper, theory paper, etc), and reviewers can be of many types also (e.g.,
prefer system paper, or theory paper, etc). If a paper-reviewer pairing is of
the same type, then the expertise level and the critical degree of the reviewer
will be of high values, else they will be of low values.

To illustrate, consider a paper $P \! \in \! \SP$ and a reviewer $R \! \in \!
\SR$. Assume reviewer $R$ reviews paper $P$. Let $u \! \in \! [0, 1]$ denote
the matching degree between reviewer $R$ and paper $P$ and let $e, c$ denote
the corresponding expertise level and critical degree respectively. Using our
usual convention, higher value represents higher matching degree. The matching
degree couples the expertise level and the critical degree in the following
manner:
\begin{eqnarray}
  e &=& \kappa , \hspace{0.2in}
    \mbox{if $\mu \in [(\kappa - 1 ) / l , \kappa / l )$,
    for $ \kappa \!=\! 1, \ldots, l$},  \nonumber\\
  c &=& f(\mu) \in [0,1], \hspace{0.2in}
    \mbox{where $\mu \in [0,1]$.}
\label{equation_critical_function}
\end{eqnarray}
Note that $e \!=\! l$ when $\mu \!=\! 1$ and $f(\mu)$ is a monotonic increasing
function of $\mu$. There are number of choices for function $f(\mu)$, e.g.,
$f(\mu) \!=\! \mu$, or $f(\mu) \!=\! \mu^2$, etc.  We will specify it in later
section.

Again there are number of interesting questions to explore, i.e., is the
conference recommendation system sensitive to the paper-reviewer matching? Will
a small percentage of reviewers who prefer theory create a large inaccuracy in
the final recommendation of a system-oriented conference (or vice versa)?

\section{Theoretical Analysis}
\label{section: analysis}

Recall that $\SA^I(k)$ and $\SA(k)$ denote the set of $k$ accepted papers
according to the intrinsic quality or according to the conference
recommendation criteria respectively. In this section, we first derive the
following probability mass function (pmf) :
\[
    \MP[|\SA^I(k) \cap \SA(k)| = i], \hspace{0.2in}
    \mbox{for $i=0,1,\ldots,k$.}
\]
With this pmf, we can then derive the expectation $E[|\SA^I(k) \cap \SA(k)|]$
and variance $\MV[|\SA^I(k) \cap \SA(k)|]$. The above probability measures can
provide us with a lot of insights, e.g., if $\MP [|\SA^I(k) \cap \SA(k)| \! =
\! k ]$ occurs with a high probability, or $E[|\SA^I(k) \cap \SA(k)|] \!
\approx \! k$, then the conference recommendation system is very accurate and
robust against different human factors, or if $\MV [ | \SA^I ( k ) \cap \SA ( k
) | ]$ is of small value, then the conference recommendation system is very
stable likely to be close to the expectation. To derive this pmf, let us first
consider the following special case. The purpose is to show the general idea of
derivation and to illustrate the underlying computational complexity. We will
consider the derivation of the general case later.

\subsection{Derivation of the Special Case}
Let us consider a conference recommendation system which has only one type of
papers and one type of reviewers (e.g., all papers are theoretical and all
reviewers prefer theoretical topics). Hence, the critical degree of all
reviewers are the same, say $c$. The intrinsic quality of each paper is
specified as follows:
\begin{equation}
\label{equation:intrinsic_quality}
    Q_i = m - i(m-1)/(N+1), \hspace{0.2in}
    \mbox{for $i=1,\ldots,N$.}
\end{equation}
Each paper will have the same number of reviews, or $n_i \!=\! n, \, \forall
i$. The voting rule $\SV$ is the {\em average score rule} and we use a random
rule to tie-break any papers whose scores are the same.

\subsubsection{Theoretical derivation}
The score set for paper $P_i$ is $\{S^i_1, \ldots, S^i_n\}$. Recall that a
score is described by a random variable, and its probability distribution is
uniquely determined by the intrinsic quality of the corresponding paper and
critical degree of the corresponding reviewer. Since the critical degree of
each reviewer is the same $c$, thus $S^i_1, \ldots, S^i_n$ are i.i.d. random
variables. By specifying $Q_i$ and $c^i_j$ in Eq.
(\ref{equation_rating_distribution_general}) with Eq.
(\ref{equation:intrinsic_quality}) and $c^i_j = c$ respectively, we obtain the
pmf of score $S^i_j$. This is stated in the following lemma.
\begin{lemma}
\label{lemma_rating_distribution} The pmf of score $S^i_j$, for
$i\!=\!1,\ldots,N$, $j \!=\!1, \ldots,n$, is
\begin{align}
&\MP [ S^i_j = \ell ] =  \begin{cases}
    [ 1 - \beta ( m - \frac{ i ( m - 1 ) } { N + 1 }, c ) ] \MP [ L = \ell ],
    & \forall \ell = 1, \ldots, \lfloor Q_i \rfloor  \\
    [ 1 + \alpha ( m - \frac{ i ( m - 1 ) } { N + 1 }, c ) ] \MP [ L = \ell ],
    & \forall \ell = \lfloor Q_i \rfloor + 1, \ldots, m \\
\! 0, & \text{otherwise}
\end{cases},
\label{equation_rating_distribution}
\end{align}
where $L$ is a discrete random variable with probability distribution
$\widetilde{\mathcal{N}} \left(m -i( m - 1)/(N+1), \sigma^2(c) \right)$ whose
pmf is derived by Eq. (\ref{equation_Prob(s)}), and $\alpha(m -i( m - 1)/
(N+1), c)$, $\beta(m -i( m - 1)/ (N+1), c)$ are derived by Eq.
(\ref{equation_alfa}) and (\ref{equation_beta}) respectively.
\end{lemma}

The average score of each paper is $\gamma_i \! = \! \sum_{j=1}^n S^i_j/n,
\forall i$. The probability mass function (pmf) of the average score of each
paper is specified in the following lemma.

\begin{lemma}
The pmf of the averages score $\gamma_i, \forall i$, is
\begin{align}
\MP \left[ \gamma_i = \frac{ \ell }{ n } \right]
& = \sum\nolimits_{ \sum_{ j = 1 }^n s_j = \ell }
    \prod\nolimits_{ j = 1 }^{ n } \MP [ S^i_j = s_j ],
    \hspace{0.1 in} \mbox{ for all $ \ell = n, \ldots, nm$},
\label{equation_mass_prob_dis_avr_rating}
\end{align}
and its cumulative distribution function (CDF) is
\begin{align}
\MP \left[\gamma_i \leq \frac{ \ell }{ n }\right]
& = \sum\nolimits_{ \sum_{ j = 1 }^n s_j \leq \ell }
    \prod\nolimits_{ j= 1 }^n  \MP [ S^i_j = s_j ],
    \hspace{0.1 in} \mbox{ for all $ \ell = n, \ldots, nm$},
\label{equation_cumulative_prob_dis_avr_score}
\end{align}
where $\MP[ S^i_j =s_{j} ]$ is specified in Eq.
(\ref{equation_rating_distribution}). \label{lemma_prob_dis_avr_rating}
\end{lemma}

\noindent {\bf Proof:} Note that, the ratings of each paper are independent
random variables. The distribution of each rating has been derived in Lemma
\ref{lemma_rating_distribution}. Since $\gamma_i \! = \! \sum_{j=1}^n S^i_j
/n$, thus by enumerating all the cases satisfying the condition $\sum_{j=1}^n
S^i_j = \ell$, we could obtain the pmd of $\gamma_i$, or
\begin{align*}
\MP \left[ \gamma_i  = \frac{ \ell }{ n } \right]
& = \MP \left[\sum\nolimits_{j=1}^n S_j^i = \ell \right]
  = \sum\nolimits_{ \sum_{ j = 1 }^n s_j = \ell }
    \prod\nolimits_{ j = 1 }^n \MP [ S^i_j = s_j ].
\end{align*}
Similarly, by enumerating all the cases satisfying the condition $\sum_{j=1}^n
S^i_j \leq \ell$, we could obtain the CDF of $\gamma_i$, or
\begin{align*}
\MP \left[\gamma_i  \leq \frac{ \ell }{ n } \right]
& = \MP \left[ \sum\nolimits_{ j = 1 }^n S_j^i \leq \ell \right]
  = \sum\nolimits_{ \sum_{ j = 1 }^n s_j \leq \ell }
    \prod\nolimits_{ j = 1 }^n  \MP [ S^i_j = s_j ],
\end{align*}
which completes the proof. \done

Based on the results derived in Lemma \ref{lemma_rating_distribution} and Lemma
\ref{lemma_prob_dis_avr_rating}, the probability that a specific set of papers
is accepted is stated as follows.

\begin{theorem}
\label{theorem_prob_specific_set_accpt} Let $\{P_{i_1},\ldots, P_{i_k}\}$ be a
set of $k$ papers.  The probability that the accepted paper set equals to
$\{P_{i_1},\ldots, P_{i_k}\}$ is:
\begin{align}
\label{equation_prob_specific_set_accpt}
\MP[\SA(k) = \{P_{i_1},\ldots, P_{i_k}\} ]
& = \sum_{ \ell = n }^{ nm } \left(
    \prod_{ i \in \SI } \MP \left[ \gamma_i \leq \ell / n \right] -
    \prod_{ i \in \SI } \MP \left[ \gamma_i \leq (\ell - 1) / n \right] \right)
    \prod_{ j \in \overline{ \SI } }
    \MP \left[ \gamma_j \leq ( \ell - 1 ) / n \right] +   \nonumber \\
& \hspace{0.17 in} \sum_{ \SF \subseteq \SI , \: \SG \subseteq \overline{ \SI },
    \: \SF , \: \SG \neq \emptyset } { | \SF \cup \SG | \choose | \SF | }^{ -1 }
    \sum_{ \ell = n }^{ nm } \prod_{ i \in \SI \backslash { \SF } }
    \left( 1 -  \MP \left[ \gamma_i \leq \ell / n \right] \right)
    \prod_{ j \in { \SF } \cup { \SG } } \MP \left[ \gamma_j = \ell / n \right]
    \nonumber \\
& \hspace{0.17 in} \prod_{ \kappa \in \overline{ \SI } \backslash { \SG } }
    \MP\left[\gamma_{ \kappa } \leq ( \ell - 1 ) / n \right],
\end{align}
where $\SI = \{i_1, \ldots, i_k\}$ is the index set of $\{P_{i_1},\ldots,
P_{i_k}\}$, and $\overline{ \SI } \! = \! \{1, \ldots, N\} \backslash \SI $ is
the complement of $\SI$. And $\MP [ \gamma_{i} \!=\! \ell / n]$ is specified in
Eq. (\ref{equation_mass_prob_dis_avr_rating}), and $\MP [ \gamma_{i} \!\leq\!
\ell/n ]$ is specified in Eq. (\ref{equation_cumulative_prob_dis_avr_score}).
\end{theorem}

\noindent {\bf Proof:} Let $\SH \! = \! \{ P_{ i_1 }, \ldots, P_{ i_k } \}$.
Let $ \overline{ \SH } \! = \! \{ P_1 , \ldots , P_N \} \backslash \SH $ be the
complement of $ \SH $. Let $\gamma_{ \text{min} } ( \SH ) \! = \! \mbox{min} \{
\gamma_i , i \in \mathcal{I} \} $ denote the minimum average score of paper set
$ \SH $. Let $ \gamma_{ \text{max} }( \overline{\SH} ) \! = \! \mbox{max} \{
\gamma_i , i \in \overline{ \SI} \}$ denote the maximum average score of paper
set $ \overline{ \SH } $. The probability that the accepted paper set equals to
$ \{ P_{ i_1 } , \ldots , P_{ i_k } \}$ can be divided into the following three
parts:
\begin{align}
\label{equation_prob_partition}
\MP [ \SA ( k ) = \{ P_{ i_1 }, \ldots, P_{ i_k } \} ]
& = \MP [ \SA ( k ) = \SH, \: \gamma_{ \text{min} } ( \SH ) < \gamma_{ \text{max} }
    ( \overline{ \SH } ) ] + \nonumber \\
&\hspace{0.17 in}  \MP [ \SA ( k ) = \SH, \: \gamma_{ \text{min} } ( \SH ) >
  \gamma_{ \text{ max } } ( \overline{ \SH } ) ] + \nonumber \\
&\hspace{0.17 in} \MP[ \SA ( k ) = \SH, \: \gamma_{ \text{ min } } ( \SH ) =
  \gamma_{ \text{ max } } ( \overline{ \SH } ) ].
\end{align}
Let us derive these three terms one by one.

According to our voting rule, the accepted paper set $ \mathcal{ A } ( k ) $
equals to $ \SH $ is 0 conditioned on that $\gamma_{ \text{ min } } ( \SH ) $
is less than $\gamma_{ \text{ max } } ( \overline{ \SH } )$. Thus,
\begin{eqnarray}
\MP[ \SA ( k ) = \SH , \:
     \gamma_{\text{min}}(\SH) < \gamma_{\text{max}}(\overline{\SH})
   ]
& = &
   \MP[ \gamma_{\text{min}}(\SH) < \gamma_{\text{max}}(\overline{\SH})]
   \MP[ \SA ( k ) = \SH \: |\:
        \gamma_{ \text{ min } }( \SH ) < \gamma_{ \text{ max }}(\overline{\SH})
      ] \nonumber \\
&=& 0
\label{equation_min_lequal_max_final}
\end{eqnarray}

According to our voting rule, the accepted paper set $ \mathcal{ A } ( k ) $
equals to $ \SH $ is 1 conditioned on $\gamma_{ \text{ min } } ( \SH ) >
\gamma_{ \text{ max } } ( \overline{ \SH } )$. Thus,
\begin{align}
 \MP[\SA(k) = \SH, \: \gamma_{\text{min}}(\SH)> \gamma_{\text{max}}(\overline{\SH})]
& = \MP[ \gamma_{\text{min}}(\SH) \! > \! \gamma_{\text{max}}(\overline{\SH})]
    \MP[\SA(k) \! = \! \SH\: |\: \gamma_{\text{min}}(\SH) \! > \! \gamma_{\text{max}}(\overline{\SH})] \nonumber \\
& = \MP[ \gamma_{\text{min}}(\SH) \! > \! \gamma_{\text{max}}(\overline{\SH})].
\label{equation_min_gequal_max_med}
\end{align}
Note that $\gamma_i = \sum_{i=1}^n S^i_j / n, \forall i$. Since the scores
$S^i_j$, $\forall i, j$,  are independent random variables, thus the average
scores $\gamma_1, \ldots, \gamma_N$ are also independent random variables.
Based on this fact, we derive the analytical expression of $ \MP[ \gamma_{
\text{ min } }( \SH) \! > \! \gamma_{\text{max}}(\overline{\SH})]$ as
\begin{align}
\MP[\gamma_{\text{min}}(\SH)> \gamma_{\text{max}}(\overline{\SH})]
&= \sum\nolimits_{ \ell = n }^{ nm } \MP \left[ \gamma_{\text{min}}(\SH) = \ell / n \right]
     \MP \left[ \gamma_{\text{max}} (\overline{\SH}) < \ell / n \right] \nonumber  \\
&= \sum\nolimits_{ \ell = n }^{ nm } \!\!\! \left( \prod\nolimits_{ i \in \mathcal{I}} \!
     \MP[\gamma_i \! \leq \! \ell /\! n ] \!-\! \prod\nolimits_{i \in \mathcal{I}} \!
     \MP[ \gamma_i \! \leq ( \ell \!-\! 1) \!/\! n] \right)
     \prod\nolimits_{ j \in \overline{\SI} } \MP [ \gamma_j \leq ( \ell-1) / n].
\label{equation_min_gequal_max_final}
\end{align}

The remaining task is to derive the last term of Eq.
(\ref{equation_prob_partition}). When $\gamma_{ \text{min} }( \SH ) \! = \!
\gamma_{ \text{max} }( \overline{ \SH } )$ occurs, tie breaking will be
performed on the set of papers with the average score equal to $\gamma_{
\text{min} }( \SH )$. Let us provide some notations first. Let $\SF \! = \! \{
i \:|\: \gamma_i \! = \! \gamma_{\text{min}}( \SH ), i \in \SI \}$ be the index
set of the papers that belong to set $ \SH $ and with average scores equal to
minimum average score of $ \SH $. Let $\SG \! = \! \{i \:|\: \gamma_i \! = \!
\gamma_{\text{min}}(\SH), i \in \overline{ \SI } \}$ be the index set of the
papers that belong to set $\overline{ \SH }$ and with average scores equal to
the minimum average score of $\SH$. Thus, tie breaking will perform on the
papers with index set $\mathcal{F} \cup \mathcal{G}$, from which only
$|\mathcal{F}|$ papers will be selected for acceptance. By enumerating all
possible tie breaking paper sets, we can divide the the last term of Eq.
(\ref{equation_prob_partition}) into the following form:
\begin{align}
\MP [\mathcal{A}(k) = \SH , \: \gamma_{\text{min}}(\SH)= \gamma_{\text{max}}(\overline{\SH})]
&= \sum_{ \SF \subseteq \SI, \: \SG \subseteq \overline{ \SI }, \: \SF, \: \SG \neq \emptyset }
   \hspace{-0.2in} \MP [\SF \cup \SG ] \MP[ \SF \:|\: \SF \cup \SG ],
\label{equation_min_equals_max_overall}
\end{align}
where $\MP [ \SF \cup \SG ]$ is the probability that tie breaking is performed
on papers with index set $ \SF \cup \SG $, and $\MP [ \SF | \SF \cup \SG ]$ is
the conditional probability that papers with index set $ \SF $ is selected for
acceptance under the condition that tie breaking is performed on the papers
with index set $\SF \cup \SG$. Since the tie-breaking rule is the random rule,
under which we just randomly pick $| \SF |$ papers, thus
\begin{equation}
    \MP [ \SF | \SF \cup \SG ] = { | \SF \cup \SG | \choose | \SF | }^{-1} .
\label{equation_min_equals_max_tie_breaking}
\end{equation}
Because the average scores $\gamma_1, \ldots, \gamma_N$ are independent random
variables, we can derive $\MP [\SF \cup \SG]$ as:
\begin{align}
\MP [ \SF \cup \SG ]
& = \MP [ \gamma_i > \gamma_{ \text{ min } } ( \SH ) \text{ \:\:\: for all $ i \in
    \SI \: \backslash \SF $ } ] \times \nonumber \\
& \hspace{0.17 in} \MP [ \gamma_i = \gamma_{ \text{ min } } ( \SH ) \text{ \:\:\: for
    all $ i \in \SF \cup \SG $ } ] \times \nonumber \\
& \hspace{0.17 in} \MP [ \gamma_i < \gamma_{ \text{ min } } ( \SH ) \text{ \:\:\: for
    all $ i \in \overline{ \SI } \: \backslash \SG $ } ]  \nonumber \\
& = \sum_{ \ell = n }^{ nm } \prod_{ i \in \SI \backslash {\SF } }
    \left( 1 -  \MP \left[ \gamma_{i}  \leq  \ell / n \right] \right)
    \prod_{ j \in { \SF } \cup {\SG } } \MP \left[ \gamma_{j}
    = \ell / n \right] \prod_{ \kappa \in \overline{ \SI } \backslash
    { \SG } } \MP \left[ \gamma_{ \kappa } \leq ( \ell - 1 ) / n \right].
\label{equation_min_equals_max_prob}
\end{align}
Combining Eq. (\ref{equation_min_equals_max_overall}) $-$
(\ref{equation_min_equals_max_prob}) we obtain
\begin{align}
\MP [ \SA ( k ) = \SH, \: \gamma_{ \text{ min } }( \SH ) = \gamma_{ \text{ max } }
  ( \overline{ \SH} ) ]
& = \sum_{ \SF \subseteq \SI, \: \SG \subseteq \overline{ \SI }, \:
  \SF, \: \SG \neq \emptyset }  \sum_{ \ell = n }^{ nm } { | \SF \cup \SG |
  \choose |\SF | }^{-1}
  \prod_{ i \in \SI \backslash { \SF } } \left( 1 - \MP \left[
  \gamma_{i} \leq  \ell / n \right] \right)   \nonumber \\
&\hspace{0.2 in }
  \prod_{ j \in \SF \cup \SG } \MP \left[ \gamma_{ j } = \ell / n \right]
  \prod_{ \kappa \in \overline{ \SI } \backslash \SG } \MP
     \left[ \gamma_{ \kappa } \leq  ( \ell - 1 ) / n \right].
\label{equation_min_equals_max_final}
\end{align}
Combining Eq. (\ref{equation_prob_partition}) $-$
%(\ref{equation_min_lequal_max_final}),
(\ref{equation_min_gequal_max_final}), and
(\ref{equation_min_equals_max_final}), we obtain Eq.
(\ref{equation_prob_specific_set_accpt}).
%which completes the proof.
%\end{proof}
\done

To illustrate the analytical expression of $\MP[\SA(k) \!=\! \{P_{i_1},\ldots,
P_{i_k}\} ]$, consider a simple example where $n \!=\! 1$, $m \! = \! 2$, $N =
3$ and $k \!=\! 1$. Table \ref{table:examp_accept_specific_set} shows the
analytical expression $\MP[\SA(1) \!=\! \{P_{i_1}\} ]$.

{ \renewcommand{\arraystretch}{1.5}
\begin{table}[htb]
\centering {\renewcommand{\tabcolsep}{0.12cm}
\begin{tabular}{|c|c|}
\hline
$\{P_{i_1}\}$ & $ \MP[\SA(1) = \{P_{i_1}\}]$ \\ \hline \hline
              & $ \MP[\gamma_1 = 2]\MP[\gamma_2 = 1]\MP[\gamma_3 = 1] +
                  \MP[\gamma_1 = 2]\MP[\gamma_2 = 2]\MP[\gamma_3 = 1]/2+$   \\
$\{ P_1 \}$   & $ \MP[\gamma_1 = 2]\MP[\gamma_3 = 2]\MP[\gamma_2 = 1]/2 +
                  \MP[\gamma_1 = 1]\MP[\gamma_3 = 1]\MP[\gamma_2 = 1]/3 $  \\
              &  $ \MP[\gamma_1 = 2]\MP[\gamma_3 = 2]\MP[\gamma_2 = 2]/3$  \\  \hline
              & $ \MP[\gamma_2 = 2]\MP[\gamma_1 = 1]\MP[\gamma_3 = 1] +
                  \MP[\gamma_2 = 2]\MP[\gamma_1 = 2]\MP[\gamma_3 = 1]/2+$   \\
$\{ P_2 \}$   & $ \MP[\gamma_2 = 2]\MP[\gamma_3 = 2]\MP[\gamma_1 = 1]/2 +
                  \MP[\gamma_2 = 1]\MP[\gamma_3 = 1]\MP[\gamma_1 = 1]/3 $  \\
              &  $ \MP[\gamma_2 = 2]\MP[\gamma_3 = 2]\MP[\gamma_1 = 2]/3$  \\  \hline
              & $ \MP[\gamma_3 = 2]\MP[\gamma_2 = 1]\MP[\gamma_1 = 1] +
                  \MP[\gamma_3 = 2]\MP[\gamma_2 = 2]\MP[\gamma_1 = 1]/2+$   \\
$\{ P_3 \}$   & $ \MP[\gamma_3 = 2]\MP[\gamma_1 = 2]\MP[\gamma_2 = 1]/2 +
                  \MP[\gamma_3 = 1]\MP[\gamma_1 = 1]\MP[\gamma_2 = 1]/3 $  \\
             &  $ \MP[\gamma_3 = 2]\MP[\gamma_1 = 2]\MP[\gamma_2 = 2]/3$  \\  \hline
\end{tabular}
} \caption{Examples of the analytical expression of $\MP[\SA(1) \!=\!
\{P_{i_1}\} ]$, where $i_1 \! = \! 1, \ldots, 3$.  }
\label{table:examp_accept_specific_set}
\end{table}
}

Up to now, we have derived the probability of a specific set of accepted
papers. Let us derive the pmf of a general set of $k$ papers:
\[
    \MP[|\SA^I(k) \cap \SA(k)| = i], \hspace{0.2in}
    \mbox{for $i=0,1,\ldots,k$,}
\]
which is shown in the following theorem.
\begin{theorem}
The pmf of $|\SA^I(k) \cap \SA(k)|$ is:
\begin{align}
\label{equation_X_mass_prob}
\MP[|\SA^I(k) \cap \SA(k)| = i]
& = \sum\nolimits_{ \begin{subarray}{c}
            \SF \subseteq \mathcal{A}^I(k),  \\
            |\SF| =i
            \end{subarray}
        }
    \sum\nolimits_{\begin{subarray}{c}
            \SG \subseteq \overline{\SA^I}(k),  \\
            |\SG| =k-i
            \end{subarray}
        }
    \MP[\mathcal{A}(k) = \SF \cup \SG], \nonumber \\
& \hspace{1.2in} \mbox{for all $i=0,1,\ldots,k$},
\end{align}
where $\overline{\SA^I}(k) = \{P_1, \ldots, P_N\}/\SA^I(k)$ is the complement
of $\SA^I(k)$ and  $\MP[\mathcal{A}(k) = \SF \cup \SG]$ is given in Eq.
(\ref{equation_prob_specific_set_accpt}).
\end{theorem}
\noindent {\bf Proof:} The paper set $\SA(k)$ can be divided into two disjoint
subsets of which one is $\SA(k) \cap \SA^I(k)$ and the other one is $\SA(k)
\cap \overline{\SA^I}(k)$. Note that we have derived the pmf that a specific
set of accepted papers in Eq. (\ref{equation_prob_specific_set_accpt}). Then by
enumerating subsets of $\SA(k)$ with cardinality $i$ and all the subsets of
$\overline{\SA^I}(k)$ with cardinality $k-i$ we can obtain probability $ \MP
[|\SA^I(k) \cap \SA(k)| = i]$, or
\begin{align*}
\MP[|\SA^I(k) \cap \SA(k)| = i]
&= \sum\nolimits_{ \begin{subarray}{c}
            \SF \subseteq \mathcal{A}^I(k),  \\
            |\SF| =i
            \end{subarray}
        }
    \sum\nolimits_{\begin{subarray}{c}
            \SG \subseteq \overline{\SA^I}(k),  \\
            |\SG| =k-i
            \end{subarray}
        }
    \MP[\mathcal{A}(k) = \SF \cup \SG], \nonumber \\
&\hspace{1.2in} \mbox{for all $i=0,1,\ldots,k$}, \nonumber
\end{align*}
where $\MP[\mathcal{A}(k) = \SF \cup \SG]$ is given in Eq.
(\ref{equation_prob_specific_set_accpt}). \done

To illustrate the analytical expression of $\MP[|\SA^I(k) \cap \SA(k)| \!=\!
i]$, let us consider the same example with Table
\ref{table:examp_accept_specific_set}. Table \ref{table:examp_pmf_accuracy}
shows the analytical expression $\MP[\SA^I(1) \cap \SA(1) \!=\! i ]$, where $i
= 0, 1$.

{ \renewcommand{\arraystretch}{1.5}
\begin{table}[htb]
\centering {\renewcommand{\tabcolsep}{0.12cm}
\begin{tabular}{|c|c|}
\hline
$i$ & $ \MP[ \SA^I(1) \cap \SA(1) = i ]$ \\ \hline \hline
    & $ \MP[\gamma_1 = 2]\MP[\gamma_2 = 1]\MP[\gamma_3 = 1] +
        \MP[\gamma_1 = 2]\MP[\gamma_2 = 2]\MP[\gamma_3 = 1]/2+$   \\
1   & $ \MP[\gamma_1 = 2]\MP[\gamma_3 = 2]\MP[\gamma_2 = 1]/2 +
        \MP[\gamma_1 = 1]\MP[\gamma_3 = 1]\MP[\gamma_2 = 1]/3 $  \\
    & $ \MP[\gamma_1 = 2]\MP[\gamma_3 = 2]\MP[\gamma_2 = 2]/3$  \\  \hline
    & $ \MP[\gamma_2 = 2]\MP[\gamma_1 = 1]\MP[\gamma_3 = 1] +
        \MP[\gamma_3 = 2]\MP[\gamma_2 = 1]\MP[\gamma_1 = 1] + $  \\
0   & $ \MP[\gamma_2 = 2]\MP[\gamma_1 = 2]\MP[\gamma_3 = 1]/2+
        \MP[\gamma_3 = 2]\MP[\gamma_1 = 2]\MP[\gamma_2 = 1]/2 $   \\
    & $ \MP[\gamma_2 = 2]\MP[\gamma_3 = 2]\MP[\gamma_1 = 1] +
        \MP[\gamma_2 = 1]\MP[\gamma_3 = 1]\MP[\gamma_1 = 1] \times 2/3  $ \\
    & $ \MP[\gamma_2 = 2]\MP[\gamma_3 = 2]\MP[\gamma_1 = 2] \times 2/3$ \\  \hline
\end{tabular}
} \caption{Examples of the analytical expression of $\MP[\SA^I(1) \cap \SA(1)
\!=\! i ]$, where $i = 0, 1$. } \label{table:examp_pmf_accuracy}
\end{table}
}

Now we have derived the analytical expression of the pmf of $|\SA^I(k) \cap
\SA(k)|$, then it is easy to obtain the analytical expression of $E[|\SA^I(k)
\cap \SA(k)|]$ and $\MV[|\SA^I(k) \cap \SA(k)|]$ according to the definition of
expectation and variance.

Examining Eq. (\ref{equation_X_mass_prob}), we can see that $\MP[\SA (k) \! =
\! \SF \cup \SG]$ is an essential part of the analytical expression of the pmf.
The analytical expression of $\MP [ \SA (k) \! = \! \SF \cup \SG]$ is given in
Eq. (\ref{equation_prob_specific_set_accpt}), which is quite complicated, and
these analytical expressions cannot be reduced to a simple form. Thus, it is
not easy to use these analytical results to gain some insight of a conference
recommendation system. An alternative way is to compute the numerical results
of the pmf of $|\SA^I(k) \cap \SA(k)|$ based on the analytical expressions in
Eq. (\ref{equation_X_mass_prob}). After we obtain the numerical results of the
pmf of $|\SA^I(k) \cap \SA(k)|$, we can then compute its expectation and
variance. Unfortunately, computing the numerical results of Eq.
(\ref{equation_X_mass_prob}) is computationally expensive, which is shown in
the following theorem.

\begin{theorem}
The computational complexity in calculating the numerical results of the pmf of
$|\SA^I(k) \cap \SA(k)|$ based on Eq. (\ref{equation_X_mass_prob}) is
exponential, or $\Theta(2^N)$.
\end{theorem}

\noindent {\bf Proof:} Examining Eq. (\ref{equation_X_mass_prob}), we can see
that the calculation of $\MP[ \SA(k) = \SF \cup \SG ]$ is the core part on the
calculation of the the pmf of $|\SA^I(k) \cap \SA(k)|$. Assume the running time
of calculating $\MP[\SA(k) = \SF \cup \SG]$ is $t$, then by Equation
(\ref{equation_X_mass_prob}), the running time of calculating the pmf is
\begin{equation}
   \Theta\left(\sum_{i=0}^k {k \choose i} {N-k \choose k-i} t \right) =\Theta\left( {N \choose k}t\right).
\label{complexity_prob}
\end{equation}
In the following we analyze the running time of calculating the numerical
result of $\MP[ \SA(k) \! = \! \SF \cup \SG ]$ based on its analytical
expression derived by Eq. (\ref{equation_prob_specific_set_accpt}). Examining
Eq. (\ref{equation_prob_specific_set_accpt}), we can see that there are two
basic computations of Eq. (\ref{equation_prob_specific_set_accpt}), of which
the first one is
\begin{equation*}
\left( \prod_{ i \in \SI } \! \MP \! \left[ \! \gamma_i \! \leq \!  \frac{\ell}{n} \! \right]
\! - \! \prod_{ i \in \SI } \! \MP \! \left[ \! \gamma_i \! \leq \! \frac{\ell  \!- \! 1}{n} \! \right] \right)
\prod_{ j \in \overline{\SI} } \! \MP \! \left[ \! \gamma_j \! \leq \! \frac{\ell \! - \! 1}{n} \! \right],
\end{equation*}
let us assume the running time of calculating this basic part is $t_1$. The
second basic computation is
\begin{equation*}
\prod_{ i \in \SI \backslash { \SF }} \!\! \left( \! 1 \!-\! \MP \! \left[ \! \gamma_{i}
\! \leq \! \frac{\ell}{n} \! \right] \! \right) \! \prod_{ j \in { \SF } \cup \SG }
 \! \MP \! \left[ \! \gamma_{ j } \! = \! \frac{\ell}{n} \! \right]
 \prod_{ \kappa \in \overline{ \SI } \backslash { \SG } }
  \! \MP \! \left[ \! \gamma_{ \kappa } \! \leq \!  \frac{\ell \! - \! 1}{n} \! \right],
\end{equation*}
let us assume the running time of calculating this basic part is $t_2$. The
running time of computing $\MP[ \SA(k) = \SF \cup \SG ]$ by Eq.
(\ref{equation_prob_specific_set_accpt}) is
\begin{align*}
   t &= \Theta (npt_1 + (2^{k} - 1)(2^{N-k} - 1)npt_2) \\
     & = \Theta(npt_1 + 2^{k-1} 2^{N-k-1} npt_2) \\
     & = \Theta( npt_1 + 2^{N-2}npt_2)
       = \Theta( 2^Nnpt_2/4)  .
\end{align*}
By letting $t = \Theta( 2^Nnpt_2/4)$ in Eq. (\ref{complexity_prob}), we can
obtain the result stated in this theorem. \done

To illustrate the complexity, consider the number of mathematical operations
(i.e., addition, subtraction, multiplication and division) that we need in the
computation of the numerical results of the pmf of $|\SA^I(k) \cap \SA(k)|$.
Table \ref{table:examp_complexity} illustrates the computational complexity of
three conferences: Recsys'11, Sigcomm'11, WWW'11 and
IEEE Infocom'11. Since the expectation and
variance are based on these operations, so their computational complexity are
also exponential.
%We use three conferences, Recsys, Sigcomm and WWW, as examples to
%illustrate the computational complexity. These examples are shown in Table \ref{table:examp_complexity}.
{ \renewcommand{\arraystretch}{1.3}
\begin{table}[htb]
\centering {\renewcommand{\tabcolsep}{0.12cm}
\begin{tabular}{|c|c|c|c|c|c|c|}
\hline
{\bf Conference} & $N$ & $k$ & {\bf Acceptance \%} & {\bf Complexity}
\\ \hline \hline
ACM Recsys'11 & 110 & 22 & 20\%  & $\sim 2^{161}$  \\  \hline
ACM Sigcomm'11 & 223 & 32 & 13\% &  $\sim 2^{312}$  \\  \hline
WWW'11 & 658 & 81 & 12\% &  $\sim 2^{902}$   \\  \hline
IEEE Infocom'11 & 1823 & 291 &16\% & $\sim 2^{2593}$ \\ \hline
\end{tabular}
} \caption{Examples of computational complexity} \label{table:examp_complexity}
\end{table}
}

In summary, we have the following conclusions in analyzing the above conference
recommendation system:
\begin{itemize}
   \item We can analytically derive the pmf of:
    \[
        \MP[|\SA^I(k) \cap \SA(k)| = i], \hspace{0.2in}
        \mbox{for $i=0,1,\ldots,k$,}
    \]

   \item The analytical expression is complex and it is not easy to obtain
	insights of the underlying recommendation system.

   \item Computing the numerical results based on these analytical results
       is computational expensive.
\end{itemize}

\subsection{Derivation for the General Case}
For the general case, we can derive the analytical experssions of the pmf,
expectation and variance of $|\SA^I(k) \cap \SA(k)|$ with similar methods used
in the special case. The analytical expressions for the general case will have
a similar form compared with the analytical expressions derived in the special
case. Furthermore, it is reasonable to expect that for the general case, there
can be different types of paper and that reviewers are not homogeneous (e.g,
they may have different topics preference). Also, tie breaking rules will be
more complicated than the random rule. Hence we expect the analytical
expressions for the general case will be more complicated. Thus, one may not
easily obtain insight by examining the analytical expression of the general
case. Instead, let us focus on finding a practical approach to solve the
general case, and that it should be {\em computational inexpensive} to obtain
numerical results of:
\[
    \MP[|\SA^I(k) \cap \SA(k)| = i], \hspace{0.2in}
    \mbox{for $i=0,1,\ldots,k$.}
\]
In the following section, we present this practical approach, and to show that
not only we can have a computational efficient approach to compute all
probability measures, but more importantly, provides performance guarantees on
our operation.

\section{Randomized Algorithm}
\label{section: monte carlo}

In this section, we present a randomized algorithm to evaluate the pmf,
expectation and variance of $|\SA^I(k) \cap \SA(k)|$. Our randomized algorithm
is computationally efficient with performance guarantee. We will use the
following notations to describe our algorithm. We define $ I(k) \! = \! |
\SA^I(k) \cap \SA(k) |$. Let $\HP[I(k)=i]$, $\HE [I(k)]$ and $\HV [I(k)]$
denote the approximate value of $\MP[ I(k) = i ]$, $E[I(k)]$ and $\MV [I(k)]$
respectively. We first present the algorithm, then show its performance
guarantee.

\subsection{Randomized Algorithm}
Our algorithm is stated in Algorithm 1. The main idea of this randomized
algorithm is that we first approximate the pmf $\MP[ I(k) = i ]$, then we use
the approximate value $\HP[ I(k) = i ]$ to compute $\HE [ I(k) ]$ and $\HV
[I(k)]$.
\begin{algorithm}
\caption{{\bf: Randomized Algorithm}} \label{algorithm_monte_carlo_algorithm}
\begin{algorithmic}[1]
    \STATE for all $i =0, \ldots, k$, $\ell_i \leftarrow 0$
    \FOR{$j =1$ to $K$}
        \STATE for all $i = 1, \ldots, N$, generate the intrinsic quality for
               paper $P_i$ by simulating the paper submission process according
               to a {\em self-selectivity} type.
        \STATE produce the intrinsic top-$k$ paper set $\mathcal{A}^I(k)$.
        \STATE assign papers to reviewers, generate the degree of criticality
            based on the paper-reviewer matching degree.
        \STATE for $i = 1, \ldots, N$, generate score set $\mathcal{S}^i$ for paper $P_i$ by simulating the scoring process.
        \STATE simulate the decision making process, i.e., applying
          the voting rule $\mathcal{V}$ and the tie breaking rule $\mathcal{T}$ to produce
               the set $\mathcal{A}(k)$ based on the score
                sets $\{\mathcal{S}^1, \ldots, \mathcal{S}^N\}$
        \STATE if the cardinality of the intersection of $\mathcal{A}^I(k)$ and $\mathcal{A}(k)$ is equal to $i$, then $\ell_i \leftarrow \ell_i + 1$.
    \ENDFOR
    \STATE for all $i =0, \ldots, k$, $\HP [I(k) = i ] \leftarrow \ell_i / K $
    \STATE $\HE [I(k)]  \leftarrow \sum_{i=0}^k i \HP [I(k) = i ] $
    \STATE $\HV [I(k)] \leftarrow \sum_{i=0}^k \left(i - \HE [I(k)]\right)^2 \HP [I(k) = i ] $
\end{algorithmic}
\end{algorithm}
We can state two properties of this algorithm. The first one is its running
time complexity and the other one is its theoretical performance guarantee. The
following theorem states its running time complexity.
\begin{theorem}
The computational complexity of our randomized algorithm is $\Theta(K N \log N
)$, where $K$ is the number of simulation round and $N$ is the number of
submitted papers.
\end{theorem}

\noindent {\bf Proof:} We prove this theorem by examining the complexity of
each step of our Algorithm \ref{algorithm_monte_carlo_algorithm}. The
complexity of step 10 $-$ 12 and 1 are the same, say $\Theta ( k )$. The
complexity of step 3 $-$ 8 are $\Theta ( N )$, $\Theta ( N \log N  )$, $\Theta
( \sum_{i = 1}^N n_i )$, $\Theta ( \sum_{i = 1}^N n_i )$, $\Theta ( N \log N
)$, and $\Theta (1)$ respectively. Since each reviewer only review a small
subset of the submitted papers, namely $n_i \ll N$, thus we have $\Theta
(\sum_{i = 1}^N n_i) \! = \! \Theta ( N )$. By adding the complexity of step 1
$-$ 12 up we could obtain the theorem.  \done

The remaining technical issue is how to set the parameter $K$. Specifically,
how many simulation rounds $K$ can produce a good approximation of the pmf,
expectation and variance? Let us proceed to answer this question by deriving
the theoretical performance guarantee for Algorithm
\ref{algorithm_monte_carlo_algorithm}.

\subsection{Theoretical Performance Guarantee}
First, we derive a loose bound on the number of simulation rounds $K$ needed
but have good performance guarantee. Then we show how one can have a tight
bound on $K$, and tradeoff between $K$ and its performance guarantee. The
following theorem states the loose bound on $K$.
\begin{theorem}
\label{theorem_performance_guarantee} When the following condition holds:
\begin{equation}
K \geq  \max\nolimits_{ \begin{subarray}{c}
                        i = 0, 1, \ldots,k,   \\
                        \text{Pr}[I(k)=i]\neq 0
                     \end{subarray}
                    }
\frac{3\ln(2( k+1) / \delta )}{ \MP[I(k)=i] \epsilon^2} ,
\label{Inequality_loose_bound_simu_rounds}
\end{equation}
then Algorithm \ref{algorithm_monte_carlo_algorithm} guarantees:
\begin{eqnarray*}
\left|\HP[I(k) = i] - \MP[I(k) = i] \right| & \leq &
       \epsilon \MP[I(k) = i], \quad \forall i = 0,\ldots,k, \\
\left|\HE[I(k)]-E[I(k)] \right| & \leq &  \epsilon E[I(k)],  \\
\left|\HV[I(k)] - \MV[I(k)] \right| & \leq &   \epsilon (1+\epsilon)\MV[I(k)],
\end{eqnarray*}
with probability at least $1-\delta$.
\end{theorem}

\noindent {\bf Proof:} By applying Lemma \ref{lemma:joint_prob_loose_round}, we
obtain that
\[
    \left|\HP[I(k) =i] - \MP[I(k) = i] \right| \leq \epsilon \MP[I(k) =i],
\]
holds for all $x =0, 1, \ldots, k$, with probability at least $1-\delta$. By
applying Lemma \ref{lemma:expectation_loose_round} and
\ref{lemma:joint_prob_loose_round} we have
\[
    \left|\HE[I(k)]-E[I(k)] \right| \leq \epsilon E[I(k)]
\]
holds with probability at least $1-\delta$. By applying Lemma
\ref{lemma:variance_loose_round} and \ref{lemma:joint_prob_loose_round} we have
\[
   \left| \HV[I(k)] - \MV[I(k)] \right| \leq \epsilon (1+\epsilon)\MV[I(k)]
\]
holds with probability at least $1-\delta$. Please refer appendix for the
proofs of these lemmas. \done

When one examines the Inequality (\ref{Inequality_loose_bound_simu_rounds}),
one can see that the bound of $K$ is useful when $\MP[I(k) = i]$ is not small
for $i = 0, 1, \ldots, k$. Consider the case when $\MP[I(k) = i] \leq 2^N$ for
some $i = 0, 1, \ldots, k $, then we have $K \geq 2^N$. For such cases, $K$ is
too large. In the following theorem, we show a tight bound on $K$.

\begin{theorem}
When $ K \geq 3\ln(2( k+1) / \delta ) / \epsilon^2$, Algorithm
\ref{algorithm_monte_carlo_algorithm} guarantees:
\begin{eqnarray*}
\left|\HP[I(k) = i] - \MP[I(k) = i] \right| & \leq & \epsilon \sqrt{\!\MP[I(k) = i]},\quad \forall i\!=\!0,\ldots,k, \\
\left|\HE[I(k)]-E[I(k)] \right|  & \leq &  \epsilon \sqrt{k(k+1) E[I(k)] / 2} \\
\left|\HV[I(k)] - \MV[I(k)]\right| & \leq &  \epsilon ( k + 1) \left( \epsilon \MV[I(k)]+ \sqrt{ (2k + 1)  \MV[I(k)]/6} \right)
\end{eqnarray*}
with probability at least $1-\delta$.
\label{theorem_performance_guarantee_sqrt}
\end{theorem}

\noindent {\bf Proof:} By applying Lemma \ref{lemma:joint_prob_tight_round}, we
obtain that
\[
    \left|\HP[I(k) = i] - \MP[I(k) = i] \right| \leq \epsilon \sqrt{\MP[I(k) = i]}
\]
holds for all $i = 0, 1, \ldots, k$, with probability at least $1-\delta$. By
applying Lemma \ref{lemma:expectation_tight_round}, and
\ref{lemma:joint_prob_tight_round} we have
\[
    \left| \HE[I(k)]-E[I(k)] \right| \leq \epsilon \sqrt{k(k+1) E[I(k)] / 2}
\]
holds with probability $1-\delta$. By applying Lemma
\ref{lemma:variance_tight_round} and \ref{lemma:joint_prob_tight_round} we have
\[
    \left| \HV[I(k)] - \MV[I(k)] \right| \leq \epsilon ( k + 1)
    \left(\epsilon \MV[I(k)]+ \sqrt{ (2k + 1)  \MV[I(k)]/6}\right)
\]
hold with probability at least $1-\delta$. Detailed proof of these lemmas are
in the appendix. \done

\section{Evaluation of Peer Review System}
\label{section: simulations}

In this section we evaluate the accuracy of conference recommendation systems.
We consider a conference with two hundred submissions, or $N\!=\!200$, and only
$k\!=\!30$ submissions will be accepted. In other words, a 15\% acceptance
rate. In consistent with realistic conference review systems, we set $m\!=\!5$,
or the rating set is $\{1,\ldots,5\}$. The simulation rounds $K$ in Algorithm
\ref{algorithm_monte_carlo_algorithm} is set to $10^9$. In the following, we
start our evaluation from a simple case, then we extend it step by step and
evaluate the impact of various factors on the overall accuracy in the final
recommendation.

\subsection{Probability distribution, expectation and variance of $|\SA^I(k) \cap \SA(k)|$}
\label{section:prob_expec_var}

Here we consider a homogenous conference recommendation system, in which papers
and reviewers are homogeneous and each paper is reviewed by the {\em same
number} of reviewers, or $n_i\! =\! n$, $\forall i$. Specifically, each
reviewer is non-biased with the same critical degree, or $c^i_j \! = \! c$,
$\forall i, j$. Hence the function $\sigma(c^i_j)$, $\forall i, j$, within
Equation (\ref{equation_rating_distribution_general}) has the same value
$\sigma(c)$ and further we set $\sigma(c) \! = \! 1$. We select one type of
{\em self-selectivity}, say medium {\em self-selectivity} specified by Equation
(\ref{equation:medium_self_select}), to study here. The voting rule $\SV$ is
the {\em average score rule} and the tie breaking rule is the {\em least
variance rule} that selects the paper with the least variance. If there is
still tie, paper is randomly selected.
\begin{definition}
  The reviewing workload of a conference recommendation system is the sum
of all reviews, or
  \[
    W = \sum\nolimits_{i=1}^N n_i.
  \]
\label{definition_workload}
\end{definition}

\begin{definition}
 Let $I_i$ be the random variable indicating
\begin{equation}
  I_i  = | \SA^I(i) \cap \SA(k) |.
\label{equation: measure}
\end{equation}
In other words, if $i=1$, $I_1$ reflects the event that the {\em best} paper is
accepted by this conference recommendation system. When $i=5$, $I_5$ reflects
the event that the top five submitted papers are finally accepted.
\label{definition:I_i}
\end{definition}

\begin{figure}[thb]
\centering
\includegraphics[width=0.42\textwidth]{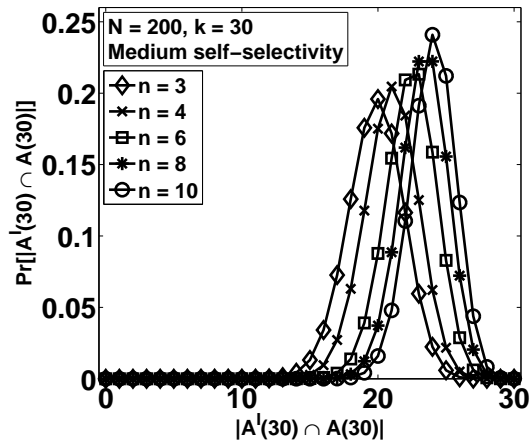}
\caption{pmf of $|\SA^I(30)\! \cap\! \SA(30)|$ when $n\! =\! 3, 4, 6, 8, 10$ and
          papers are submitted with medium {\em self-selectivity}.}
\label{fig:sec_prob_exp_var:prob}
\end{figure}

The numerical results of the pmf of $|\SA^I(30) \cap \SA(30)|$ are shown in
Fig. \ref{fig:sec_prob_exp_var:prob}. The numerical results of the expectation
and variance of $I_1$, $I_5$, $I_{10}$ and $I_{30}$ are shown in Table
\ref{table:sec_prob_exp_var:exp_var}.

In Fig. \ref{fig:sec_prob_exp_var:prob}, the horizontal axis represents the
number of top 30 papers that got finally accepted, or $|\SA^I(30) \cap
\SA(30)|$. The vertical axis shows the corresponding probability.  From Fig.
\ref{fig:sec_prob_exp_var:prob} we could see that when we increase the number
of reviews per paper, or $n$, the probability mass function shifts toward the
right. In other words, the more reviews each paper received, the higher the
accuracy of the conference recommendation system.  From Table
\ref{table:sec_prob_exp_var:exp_var}, we have the following observations. When
each paper is reviewed by three reviewers, approximately 19.8 papers from the
top 30 papers will get accepted. It is interesting to note that the chance of
accepting the best paper is invariant of the reviewing workload, since $E[I_1]
\!=\!0.98$ when $n \!=\! 3$ and it improves to $0.9999$ when $n \!=\! 10$. This
statement also holds for the top ten papers.  From Table
\ref{table:sec_prob_exp_var:exp_var}, we can see that as we increase $n$, we
decrease the variance, which reflects that the conference recommendation system
is more accurate.

\noindent {\bf Lessons learned:} If reviewers are non-biased, fair and
critical, and the quality of submitted papers is of medium {\em
self-selectivity} as specified by Eq. (\ref{equation:medium_self_select}), we
have a pretty accurate conference recommendation system. There are number of
interesting questions to explore further, i.e., are these results dependent on
the distribution of intrinsic quality, or the {\em self-selectivity} type of
papers? Are these results sensitive to any voting rule? Let us continue to
explore.

% figure for mass probability distribution of $|\SA^I(30) \cap \SA(30)|$

%table for the expectation $\mbox{Var}[X_1]$, $\mbox{Var}[X_5]$,
%$\mbox{Var}[X_10]$ and $\mbox{Var}[X]$
\begin{table}[htb]
\centering
\begin{tabular}{|c||c|c|c|c|c|} \hline
            & $n=3$  & $n=4$  & $n=6$  & $n=8$  & $n=10$ \\ \hline \hline
$E[I_{1}]$  & 0.9832 & 0.9931 &0.9986  & 0.9996 & 0.9999 \\  \hline
$E[I_{5}]$  & 4.6854 & 4.8284 & 4.9352 & 4.9723 & 4.9869 \\  \hline
$E[I_{10}]$ & 8.7763 & 9.1643 & 9.5576 & 9.7446 & 9.8426 \\  \hline
$E[I_{30}]$     & 19.8270 & 20.960 & 22.400 & 23.328 & 23.980 \\ \hline \hline
$\mbox{Var}[I_{1}]$  & 0.0165 & 0.0067 & 0.0014 & 0.0004 & 0.0001 \\  \hline
$\mbox{Var}[I_{5}]$  & 0.2942 & 0.1701 & 0.0654 & 0.0280 & 0.0132 \\  \hline
$\mbox{Var}[I_{10}]$ & 1.0793 & 0.7793 & 0.4370 & 0.2586 & 0.1609 \\  \hline
$\mbox{Var}[I_{30}]$      & 4.1210 & 3.7710 & 3.2934 & 2.9580 & 2.7121 \\
\hline\end{tabular} \caption{Expectation and variance of $I_1$, $I_5$,
$I_{10}$, and $I_{30}$ when we increase number of reviews per paper, $n \! = \!
3, 4, 6, 8, 10$, and papers are submitted with medium {\em self-selectivity}.}
\label{table:sec_prob_exp_var:exp_var}
\end{table}

\subsection{Effect of Intrinsic Quality}
\label{section:self_select}

Here, we explore the effect of {\em intrinsic quality} (or {\em
self-selectivity}) of papers on the conference recommendation system.
Specifically, we consider four representative types of {\em self-selectivity}
of papers: high, medium, low and random {\em self-selectivity} specified by Eq.
(\ref{equation:high_self_select}) $-$ (\ref{equation:random_self_select})
respectively. We explore the effect of {\em self-selectivity} of papers on the
homogeneous conference recommendation system specified in Section
\ref{section:prob_expec_var} with each paper receiving three reviews, or $n
\!=\! 3$. We use the following notations to present our results:

\noindent {\bf \em H-S-S:} high {\em self-selectivity} specified in Eq.
(\ref{equation:high_self_select}).

\noindent {\bf \em M-S-S:} medium {\em self-selectivity} specified in Eq.
(\ref{equation:medium_self_select}).

\noindent {\bf \em L-S-S:} low {\em self-selectivity} specified in Eq.
(\ref{equation:low_self_select}).

\noindent {\bf \em R-S-S:} random {\em self-selectivity} specified in Eq.
(\ref{equation:random_self_select}).

\begin{figure}[thb]
\centering
\includegraphics[width = 0.42\textwidth ]{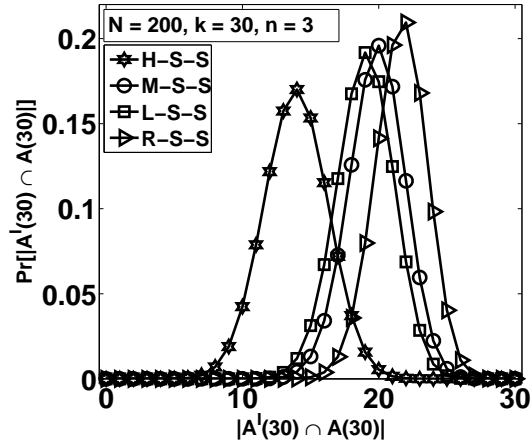}
\caption{pmf of $|\SA^I(30)\! \cap\! \SA(30)|$ when $n\! =\! 3$ and papers are
submitted with high, medium, low or random {\em self-selectivity}.}
\label{fig:sec_self_select:prob}
\end{figure}

\noindent The numerical results of the pmf of $|\SA^I(30) \cap \SA(30)|$ is
shown in Fig. \ref{fig:sec_self_select:prob}. The numerical results of the
expectation and variance of $I_1$, $I_5$, $I_{10}$ and $I_{30}$ are shown in
Table \ref{table:sec_self_select:exp_var}.

In Fig. \ref{fig:sec_self_select:prob}, the horizontal axis represents the
number of top 30 papers that got finally accepted, or $|\SA^I(30) \cap
\SA(30)|$. The vertical axis shows the corresponding probability. From Fig.
\ref{fig:sec_self_select:prob} we could see that as the {\em self-selectivity}
type varies in the order of high, low, medium, random {\em self-selectivity},
the corresponding mass probability distribution curve moves towards right. In
other words, the accuracy of the conference recommendation system corresponding
to the random {\em self-selectivity} is the highest followed by medium, low and
high {\em self-selectivity}. From Table \ref{table:sec_self_select:exp_var} we
have the following observations. When papers are submitted with medium, low or
random {\em self-selectivity}, around 20 papers from the top 30 papers will be
accepted. It is interesting to note that the chance of accepting the best paper
is invariant to these three {\em self-selectivity} types, since the
corresponding three values of $E[I_1]$ are all around 0.98. This statement also
holds for top ten papers. But when papers are submitted with high {\em
self-selectivity}, the accuracy of the conference recommendation system is {\em
remarkably lower} than that the other three {\em self-selectivity} types.
Specifically, only a small number, around 13.2, of papers from the top 30
papers will be accepted. Even the best paper will get rejected with high
probability, around 0.46. This statement also holds for top ten or top five
papers. The variance corresponding to the high {\em self-selectivity} is the
highest among those four, which reflects that the results of the conference
recommendation system is the least accurate and more likely to depart from the
expectation. Hence, for a prestigious conference, i.e., SIGCOMM, assume papers
are submitted with high {\em self-selectivity}, and if the conference insists
to have a small technical program committee with reviewers having moderate
reviewing workload, say $n\!=\!3$, the final recommendation may not be
accurate.

\noindent {\bf Lessons learned:} When reviewers are non-biased fair and
critical, the above simple conference review system is quite accurate except
when papers are of high {\em self-selectivity}. In that case, one may explore
other means to improve the accuracy. Again, there are number of interesting
questions to explore, i.e., how large the reviewing workload do we need to have
an accurate recommendation?

% figure for probability distribution comparison of four self-selectivity types

%table for expectation and variance of X_1, X_5, X_10 and X_30 under four types of submissions.
\begin{table}[htb]
\centering
\begin{tabular}{|c||c|c|c|c|c|} \hline
            & {\em H-S-S} & {\em M-S-S} & {\em L-S-S} & {\em R-S-S} \\ \hline \hline
$E[I_{1}]$  & 0.5401 & 0.9832 & 0.9788 & 0.9846  \\  \hline
$E[I_{5}]$  & 2.6250 & 4.6950 & 4.6082 & 4.7599  \\  \hline
$E[I_{10}]$ & 5.0687 & 8.7736 & 8.5162 & 9.0810  \\  \hline
$E[I_{30}]$ & 13.2258 & 19.8270 & 18.9798 & 21.5821  \\ \hline \hline
$\mbox{Var}[I_{1}]$ & 0.2437 & 0.0165 & 0.0208 & 0.0151 \\ \hline
$\mbox{Var}[I_{5}]$ & 1.2236 & 0.2942 & 0.3698 & 0.2314 \\ \hline
$\mbox{Var}[I_{10}]$ & 2.3933 & 1.0793 & 1.2491 & 0.8275 \\ \hline
$\mbox{Var}[I_{30}]$ & 5.4420 & 4.1210 & 4.2888 & 3.5473 \\
\hline\end{tabular} \caption{Expectation and Variance of $I_1$, $I_5$,
$I_{10}$, and $I_{30}$ when $n \!=\! 3$ and papers are submitted with high,
medium, low or random {\em self-selectivity}. }
\label{table:sec_self_select:exp_var}
\end{table}

\subsection{Effect of Number of Reviews per Paper}
\label{section:workload}
%Let us explore the effect of reviewing workload $W$ on a
%conference recommendation system.

\begin{figure}[htb]
\centering
\subfigure[\# of top 1 papers get in]{
\includegraphics[width = 0.42\textwidth]{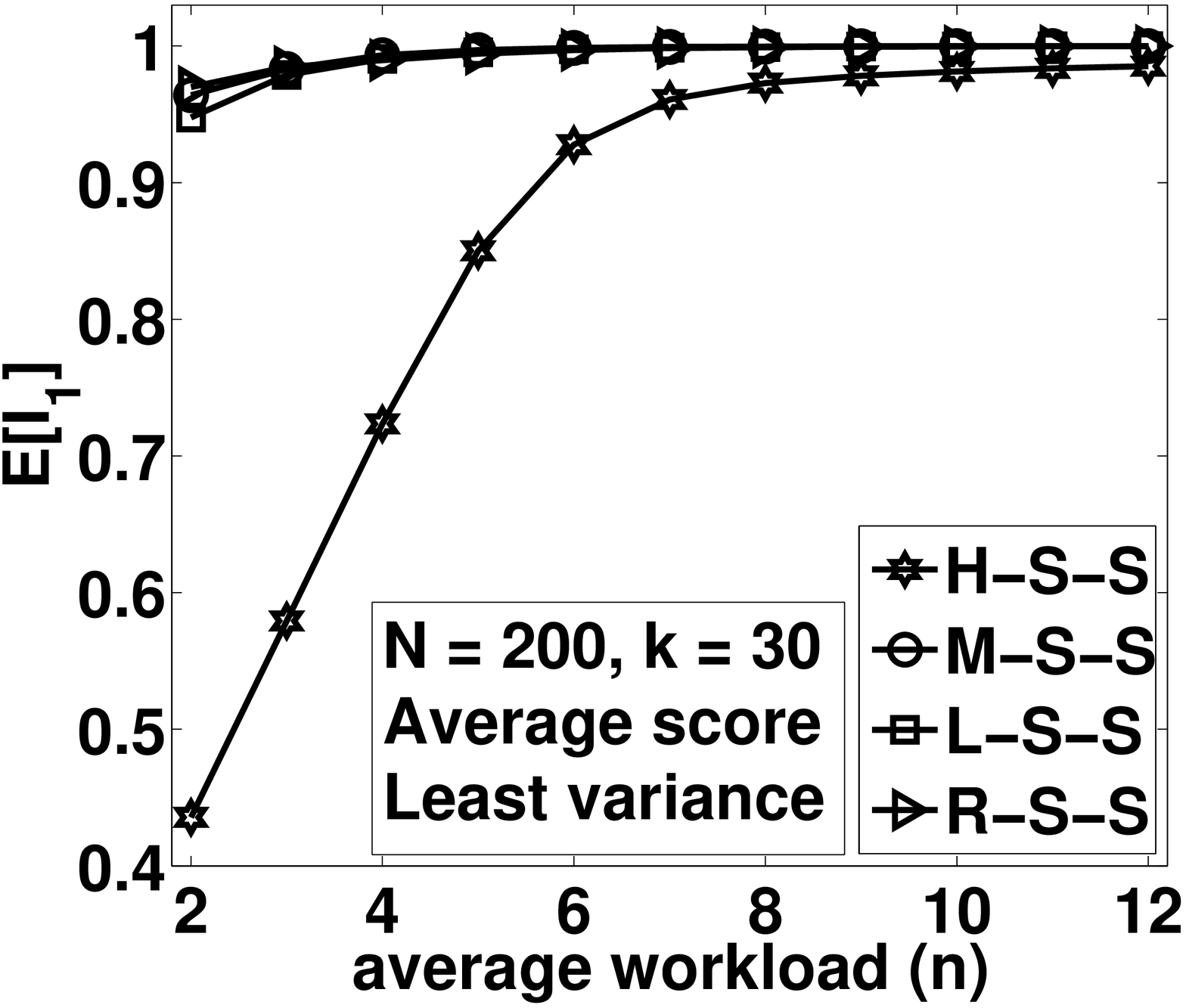}
\label{fig:sec_workload:exp_top_1}
}
\subfigure[\# of top 5 papers get in]{
 \includegraphics[width = 0.42\textwidth]{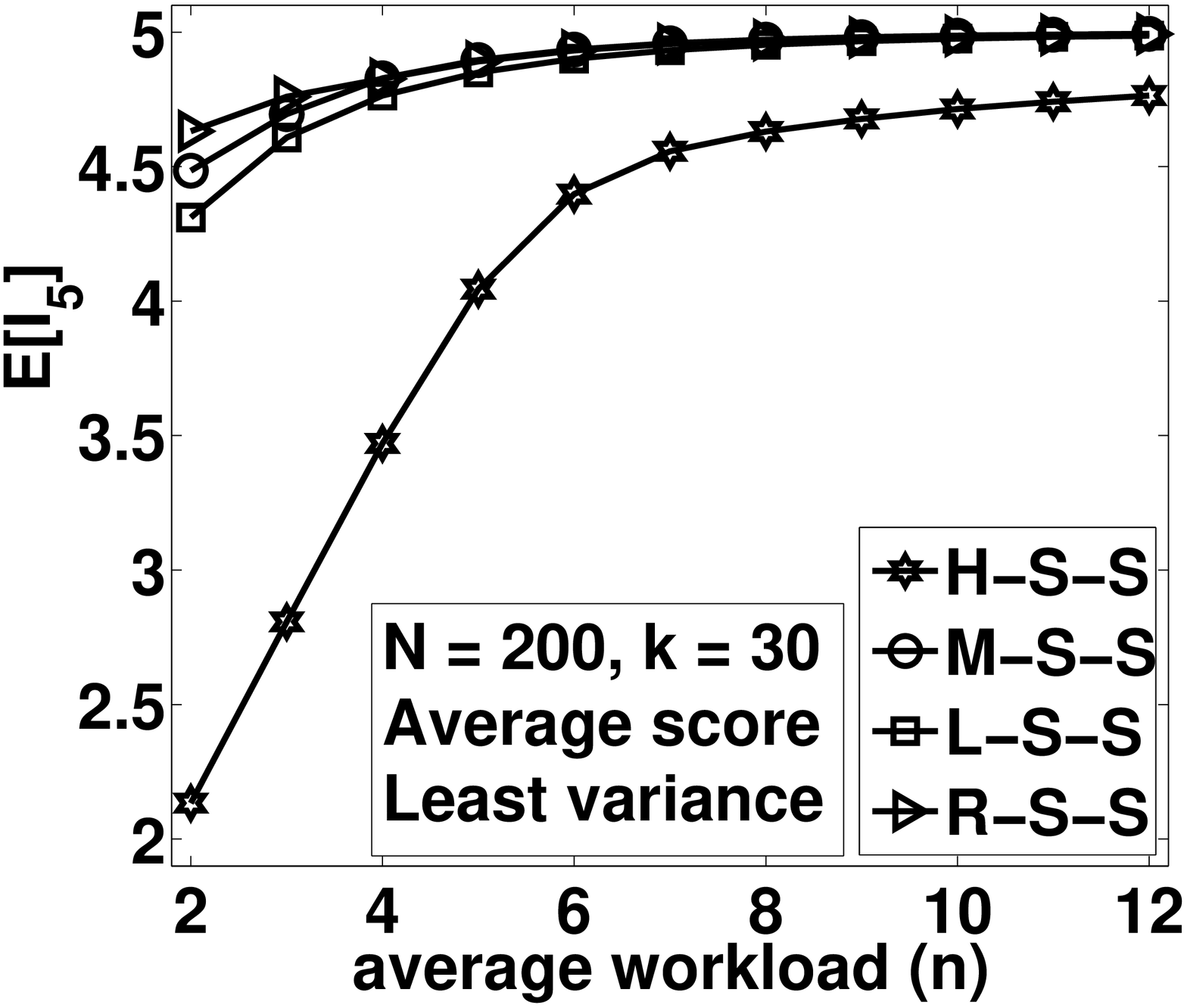}
 \label{fig:sec_workload:exp_top_5}
 }
\\
\subfigure[\# of top 10 papers get in]{
\includegraphics[width = 0.42\textwidth]{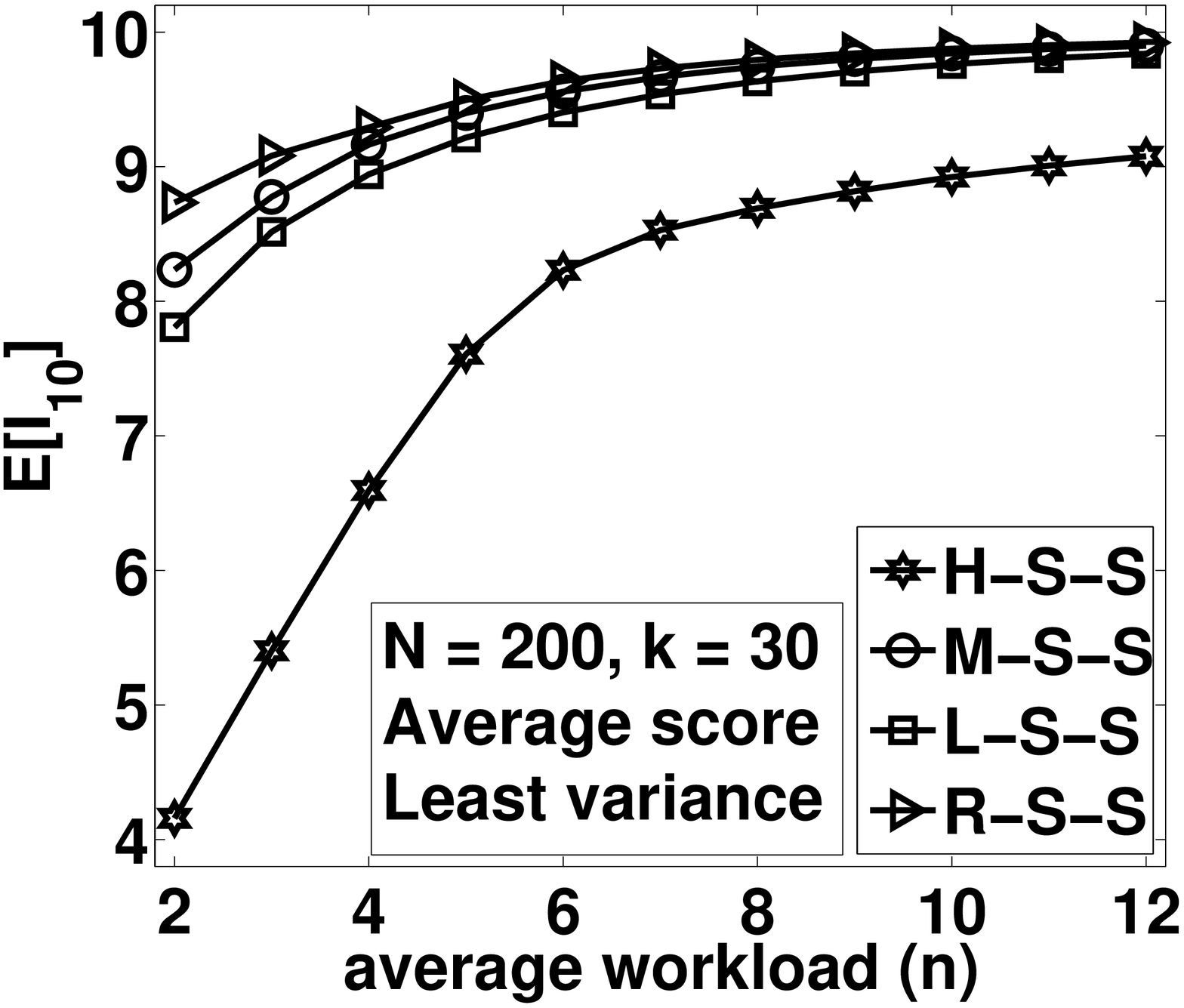}
\label{fig:sec_workload:exp_top_10}
}
\subfigure[\# of top 30 papers get in]{
 \includegraphics[width = 0.42\textwidth]{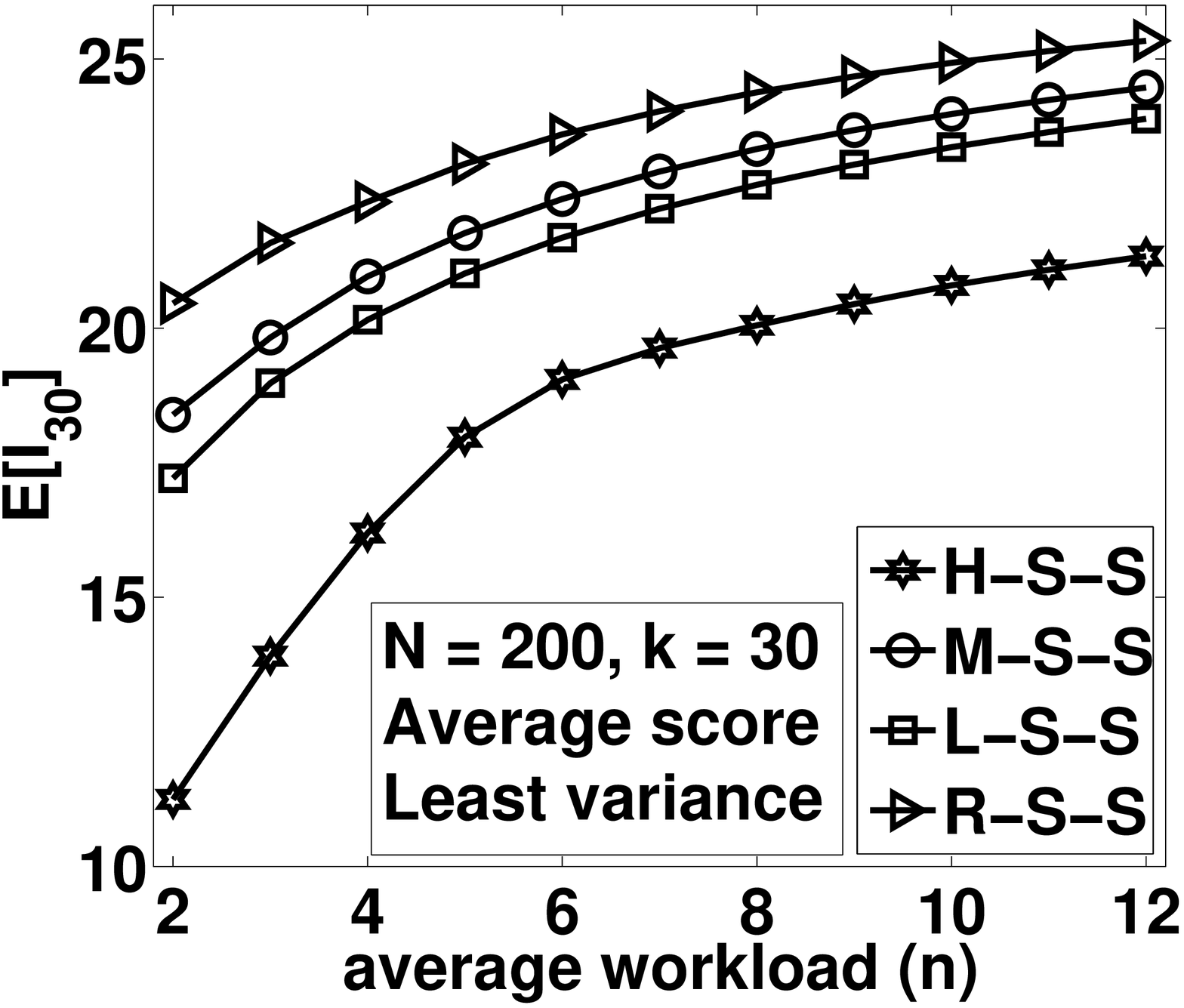}
 \label{fig:sec_workload:exp_top_30}
 }
\caption{Expectation of $I_1$, $I_5$, $I_{10}$, and $I_{30}$ when
we vary number of reviews per paper $n \!=\! 2, \ldots, 12$,
with papers are from high, low, medium or random {\em self-selectivity}. }
\label{fig:sec_workload:exp}
\end{figure}

Here we consider a homogeneous conference recommendation system specified in
Section \ref{section:prob_expec_var}. We select one probability measure,
expectation of acceptance, to study. The numerical results of $E[I_1]$,
$E[I_5]$, $E[I_{10}]$ and $E[I_{30}]$ are shown in Fig.
\ref{fig:sec_workload:exp}.

In Fig. \ref{fig:sec_workload:exp}, the horizontal axis represents the number
of reviews per paper, or $n$. The vertical axis shows the corresponding
expectation.  From Fig. \ref{fig:sec_workload:exp}, we have the following
observations. When we increase the number of reviews per paper, or $n$, the
expectation increased, which reflects the improvement in accuracy of the
conference recommendation system. As the {\em self-selectivity} type varies in
the order of high, low, medium, random {\em self-selectivity}, the expectation
curve shifts toward up. In other words, the accuracy corresponding to the
random {\em self-selectivity} is the highest followed by medium, low, high {\em
self-selectivity}. It is interesting to note that the best paper is invariant
of the workload, except for papers with high {\em self-selectivity}. This
statement also holds for the top five or ten papers. When papers are highly
{\em self-selective}, the accuracy of the conference recommendation system is
remarkably lower than the other three {\em self-selectivity} types. Especially,
when the reviewing workload is low, say $n \!=\! 3$, with less than 15 papers
from the top 30 papers will get accepted. This holds even for the best paper,
which only has a probability of less than 0.6 of being accepted when $n \! = \!
3$. Same can be said for the top five or top ten papers. In closing, for papers
with high {\em self-selectivity}, and we may have to increase the reviewing
workload to at least $n \!\geq \!7$ such that we have a strong guarantee that
the best paper will be accepted.

\noindent {\bf Lessons learned:} If reviewers are non-biased and fair, using
the above simple conference review system achieves relatively high accuracy
except when papers are of high {\em self-selectivity}. In that case, we have to
increase the workload to $n \!\geq\!7$ to improve the system. Again, there are
number of interesting questions to explore, i.e., is increasing the workload
the only way to improve the conference recommendation system? Can we improve it
by using different voting rule or tie breaking rule?

% figure for workload exploration

\subsection{Effect of Voting Rules}
\label{section:voting_rule}

In this section we explore the effect of voting rules on the accuracy of
conference recommendation system. Specifically, we will evaluate the
performance of the following three representative voting rules:

\noindent {\bf \em Average score rule ($\SV_{as}$)}: specified by
\begin{equation}
    \gamma_i = \sum\nolimits_{ S \in \mathcal{ S }^i } S / |\mathcal{S}^i|.
\label{equation:average_score_rule}
\end{equation}

\noindent {\bf \em Eliminate the highest \& lowest score rule ($\SV_{ehl}$)}:
eliminate the highest and lowest score of each paper, and calculate the average
score of each paper by the remaining scores, or
\begin{equation}
\gamma_i = \left( \sum\nolimits_{ S \in \mathcal{ S }^i } S- \mbox{ max }
           \{ \mathcal{ S }^i \} - \mbox{ min }\{ \mathcal{ S }^i \} \right) /
           ( | \mathcal{ S }^i | - 2 ).
\label{equation_eliminate_highest_lowest}
\end{equation}

\noindent {\bf \em Punish low scores rule ($\SV_{pl}$)}: punish the low scores.
Specifically, a low score, or 1, brings the an extra punishment of decreasing
its score by $\eta$, or
\begin{equation}
\gamma_i = \sum\nolimits_{ S \in \mathcal{ S }^i } S / | \mathcal{ S }^i |
           - \eta | \{ S \: | \: S = 1, S \in \mathcal{ S }^i \} |.
\label{equation:punish_low_score}
\end{equation}
Let us set the punishment $\eta$ to be 0.5 throughout this paper.

\begin{figure}
\centering
\subfigure[High {\em self-selectivity}]{
\includegraphics[width=0.42\textwidth]{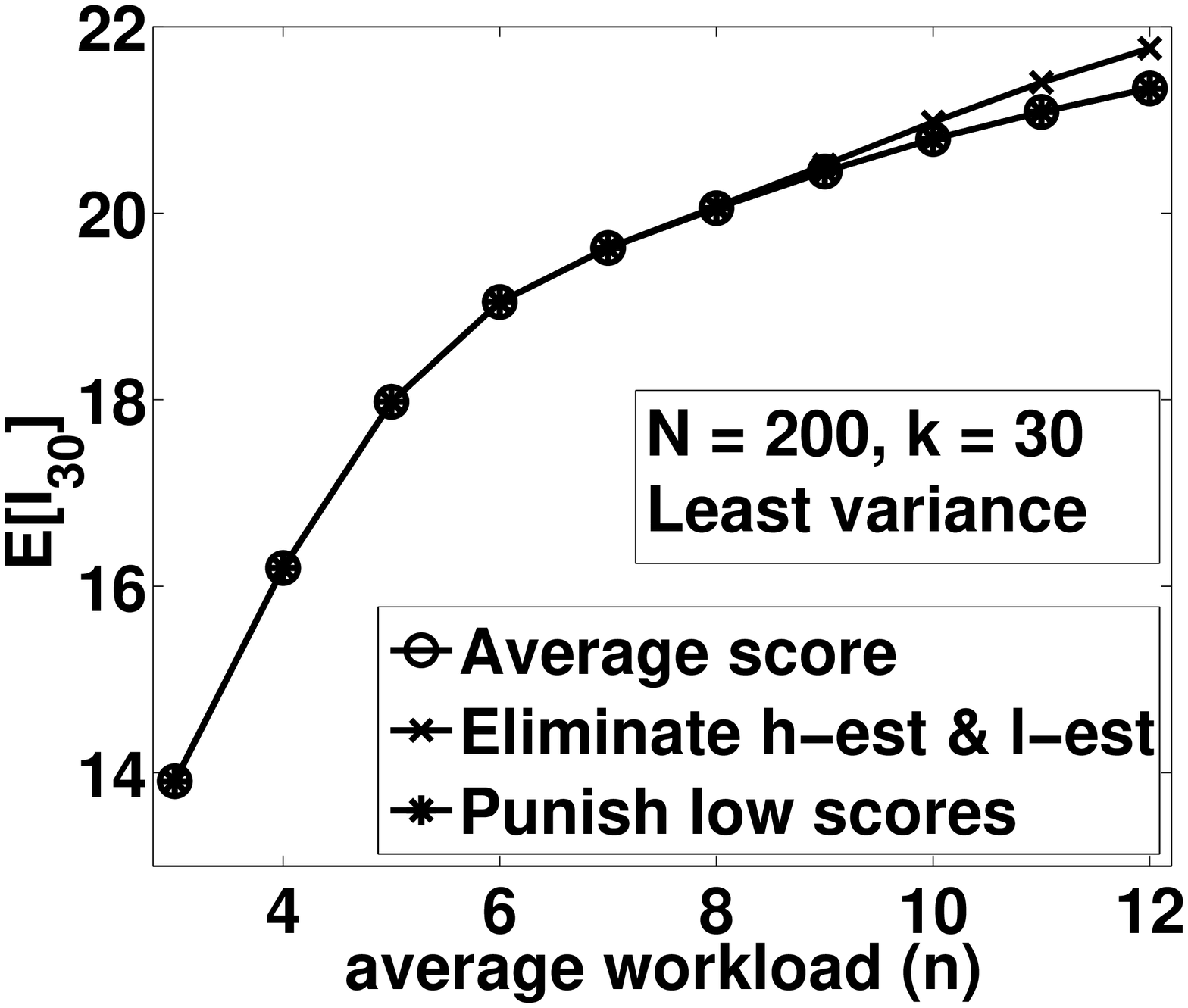}
\label{fig:sec_vot_rule:exp_top_30_high}
}
\subfigure[Medium {\em self-selectivity}]{
 \includegraphics[width=0.42\textwidth]{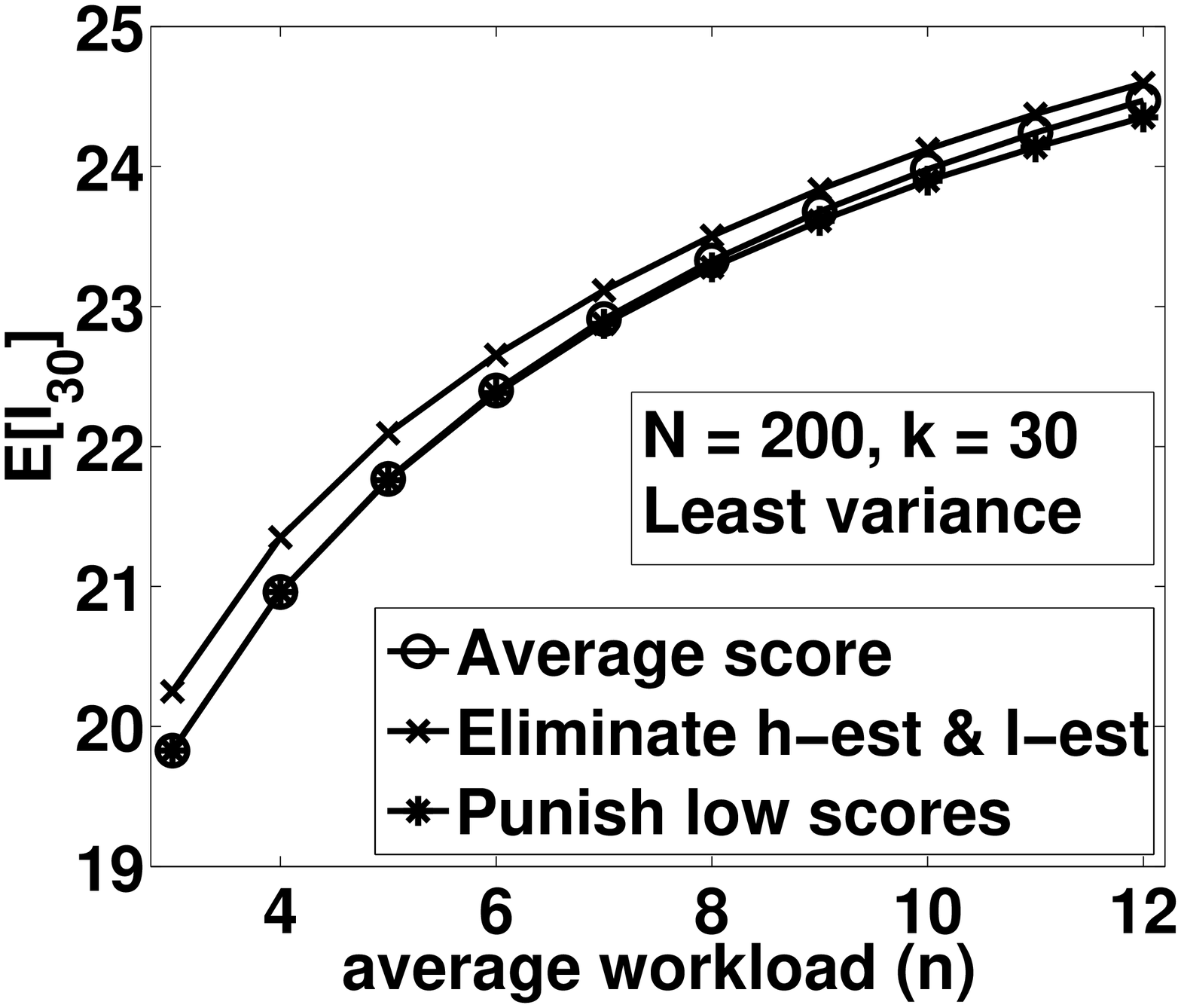}
 \label{fig:sec_vot_rule:exp_top_30_medium}
 }
 \\
\subfigure[Low {\em self-selectivity}]{
\includegraphics[width=0.42\textwidth]{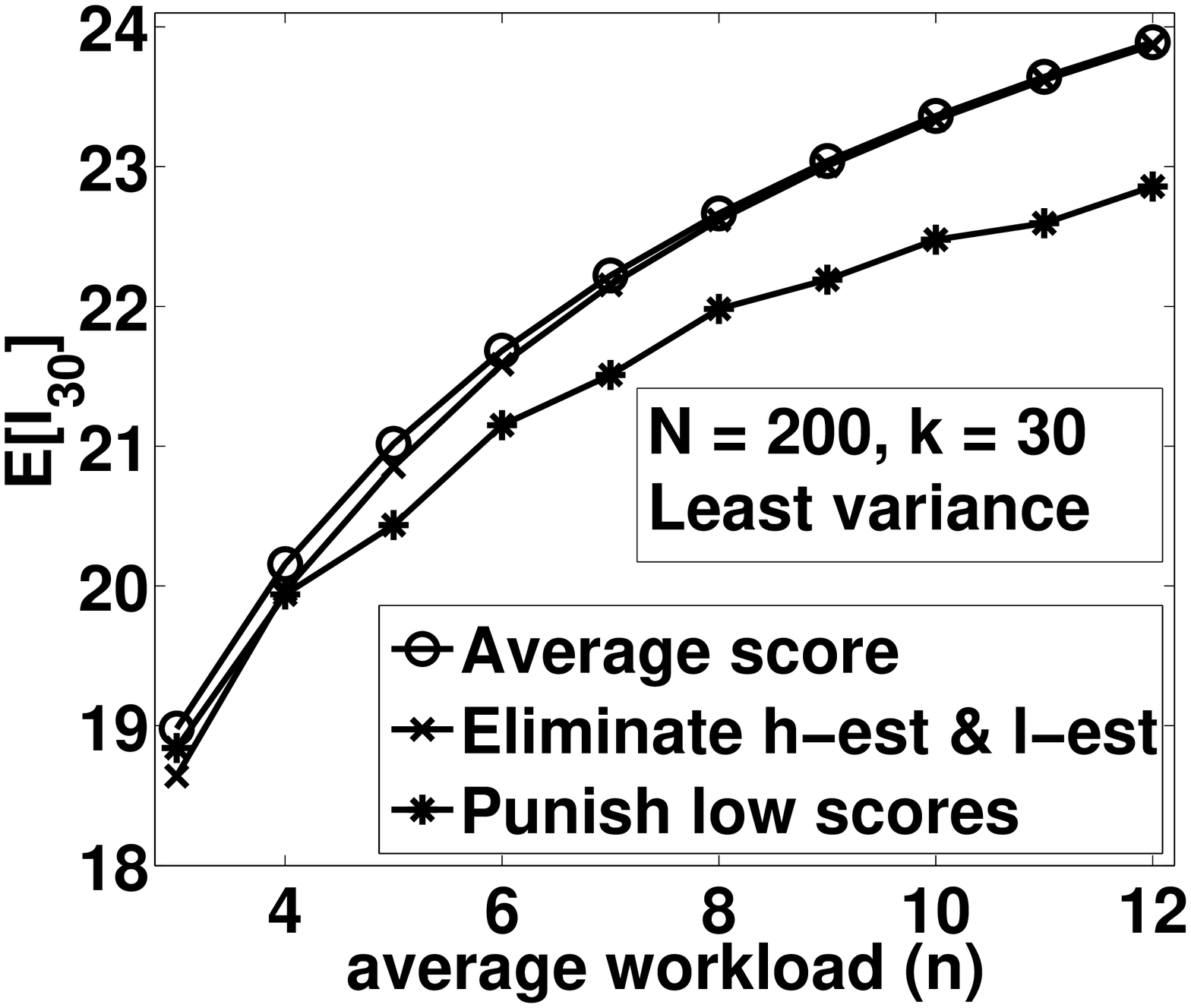}
\label{fig:sec_vot_rule:exp_top_30_low}
}
\subfigure[Random {\em self-selectivity}]{
 \includegraphics[width=0.42\textwidth]{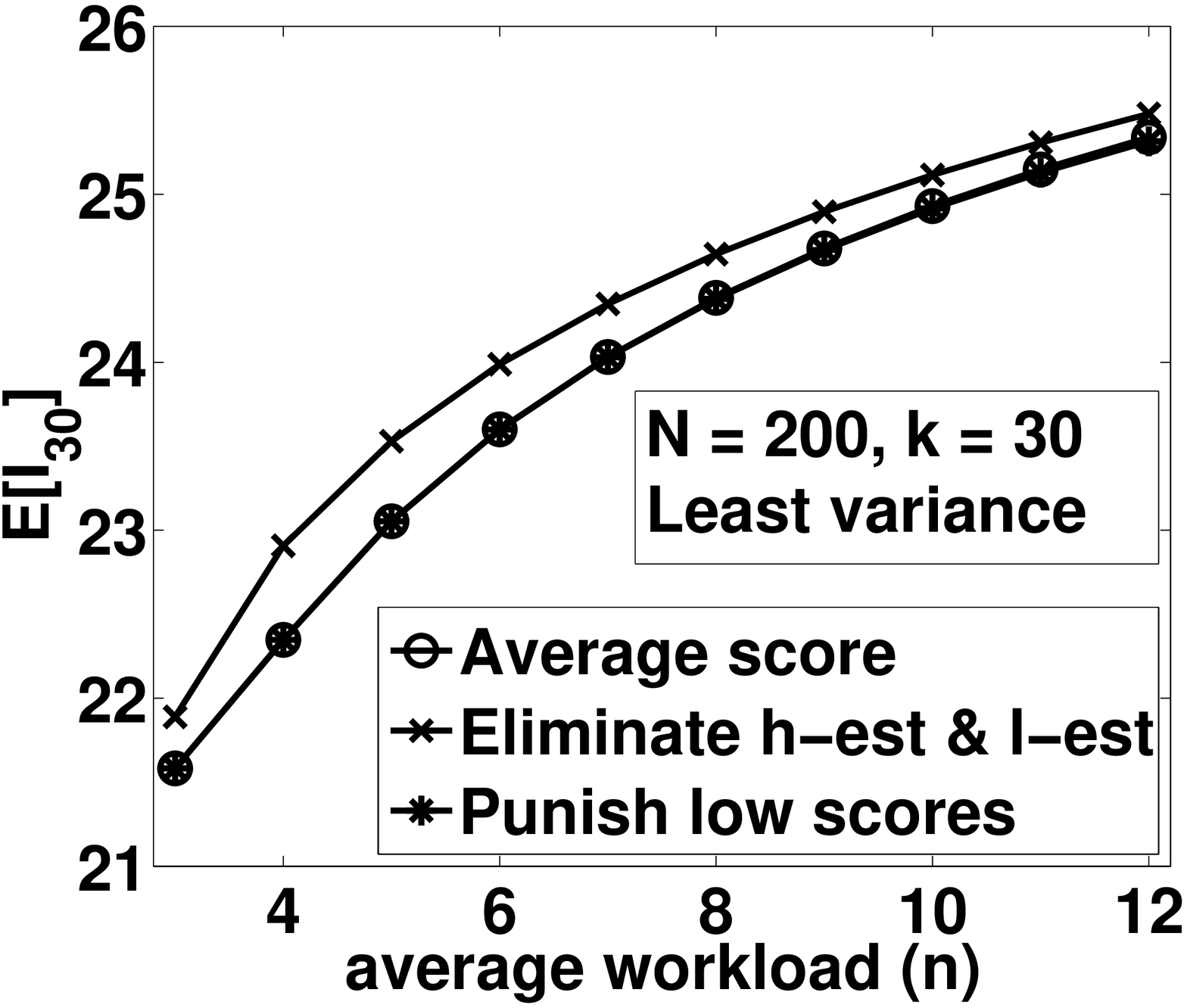}
 \label{fig:sec_vot_rule:exp_top_30_uniform}
 }
\caption{$E[I_{30}]$ for three voting rules: $\SV_{as}$, $\SV_{ehl}$, and
$\SV_{pl}$.}
\label{fig:sec_vot_rule:exp_top_30}
\end{figure}

\begin{figure}
\centering
\subfigure[\# of top 1 paper get in]{
\includegraphics[width=0.42\textwidth]{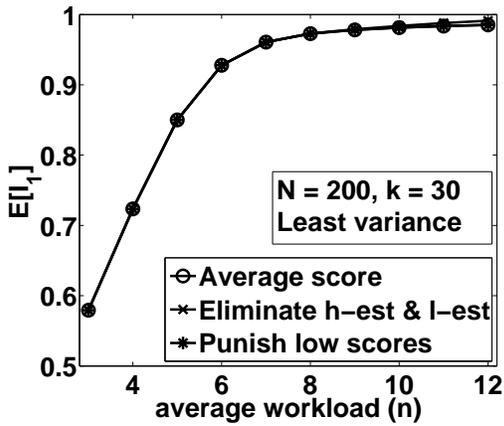}
\label{fig:sec_vot_rule:exp_top_1_high}
}
\subfigure[\# of top 5 papers get in]{
 \includegraphics[width=0.42\textwidth]{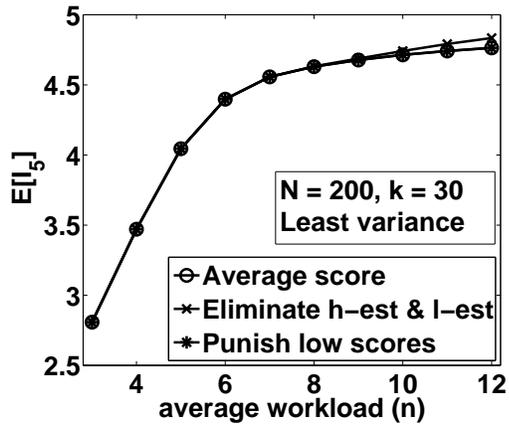}
 \label{fig:sec_vot_rule:exp_top_5_high}
 }
\caption{ Expectation of $E[I_1]$, $E[I_5]$ for three voting rules:
$\SV_{as}$, $\SV_{ehl}$, and $\SV_{pl}$ for high {\em self-selectivity} submission.}
\label{fig:sec_vot_rule:exp_top_1_5_high}
\end{figure}

We use the {\em least variance} rule for tie breaking and we choose expectation
as our performance measure. We evaluate the accuracy of these three voting
rules on the conference recommendation system specified in Section
\ref{section:prob_expec_var}. The numerical results of expectation of $I_{30}$
are shown in Fig. \ref{fig:sec_vot_rule:exp_top_30}. The numerical results of
expectation of $I_1$ and $I_5$ for high {\em self-selectivity} submissions are
shown in Fig. \ref{fig:sec_vot_rule:exp_top_1_5_high}. The numerical results of
$E[I_1]$, $E[I_5]$, and $E[I_{10}]$ when $n = 3$ are shown in Table
\ref{table:sec_vot_rule:exp_top1_5_10_n_3}.

In Fig. \ref{fig:sec_vot_rule:exp_top_30} and
\ref{fig:sec_vot_rule:exp_top_1_5_high} the horizontal axis represents the
number of reviews per paper, or $n$. The vertical axis shows corresponding
expectation.  From Fig. \ref{fig:sec_vot_rule:exp_top_30}, we have the
following observations.  When we increase $n$, the expectation $E[I_{30}]$
corresponding to each voting rule increased, which reflects that the accuracy
of the conference recommendation system increased.  From Fig.
\ref{fig:sec_vot_rule:exp_top_30_high}, we see that when submitted papers are
of high {\em self-selectivity}, the expectation curves overlapped together. In
other words, for high {\em self-selectivity} papers, these three voting rules
are similar.  From Fig. \ref{fig:sec_vot_rule:exp_top_30_medium} we see that
for submitted papers with medium {\em self-selectivity}, the {\em average score
rule} and {\em punish low scores rule } have the same degree of accuracy, and
the {\em eliminate highest and lowest score rule } has slightly higher accuracy
than the other two voting rules. This statement also holds for the random {\em
self-selectivity} submissions as shown in Fig.
\ref{fig:sec_vot_rule:exp_top_30_uniform}.  From Fig.
\ref{fig:sec_vot_rule:exp_top_30_low}, we see that when submitted papers are of
low {\em self-selectivity}, all three voting rules have nearly the same
accuracy but the {\em punish low scores rule } has slightly lower accuracy.
From Table \ref{table:sec_vot_rule:exp_top1_5_10_n_3}, we observe that when $n
\!=\! 3$, the chance of accepting the best paper is invariant of these voting
rules, unless when submitted papers are of high {\em selectivity}, since
$E[I_1]$ is around 0.98 for medium, low and random {\em self-selectivity}
papers. This statement also holds for the top five and the top ten papers. When
papers are submitted with high {\em self-selectivity}, the chance of accepting
the best paper is low, in fact, it is less than 0.6. The same statements holds
for the top five and top ten papers. From Fig.
\ref{fig:sec_vot_rule:exp_top_1_5_high}, we see that when papers submitted with
high {\em self-selectivity}, the expectation curves overlapped together. And
for each voting rule we have to increase the reviewing workload to at least
seven such that we have a strong guarantee that the best paper or the top five
papers will get accepted.

\noindent {\bf Lessons learned:} These three voting rules have comparable
accuracy, no rule can outperform others remarkably. Thus the improvement of
conference recommendation system by voting rules is limited. For each voting
rule, we till have to increase the average workload to at least seven to have a
strong guarantee that the best paper will get in. Again, there are number of
interesting questions to explore, i.e., how about the tie-breaking rules?

% voting rule comparisons

%voting rule comparison top-1 papers
\begin{table}
\centering
\begin{tabular}{|c||c|c|c|} \hline
           & $\SV_{as}$ & $\SV_{ehl}$ & $\SV_{pl}$  \\ \hline \hline
\multicolumn{4}{|c|}{ \bf High self-selectivity }  \\ \hline
$E[I_{1}]$ & 0.5793 & 0.5793 & 0.5793  \\  \hline
$E[I_{5}]$ & 2.8078 & 2.8078 & 2.8077  \\  \hline
$E[I_{10}]$ & 5.4028 & 5.4026 & 5.4026   \\  \hline \hline
\multicolumn{4}{|c|}{ \bf Medium self-selectivity }  \\ \hline
$E[I_{1}]$ & 0.9832 & 0.9911 & 0.9832   \\  \hline
$E[I_{5}]$ & 4.6950 & 4.7733 & 4.6950   \\  \hline
$E[I_{10}]$ & 8.7763 & 8.9641 & 8.7763   \\  \hline \hline
\multicolumn{4}{|c|}{ \bf Low self-selectivity }  \\ \hline
$E[I_{1}]$ & 0.9788 & 0.9741 & 0.9746   \\  \hline
$E[I_{5}]$ & 4.6082 & 4.5616 & 4.5691   \\  \hline
$E[I_{10}]$ & 8.5162 & 8.4032 & 8.4297   \\  \hline \hline
\multicolumn{4}{|c|}{ \bf Random self-selectivity }  \\ \hline
$E[I_{1}]$ & 0.9846 & 0.9873 & 0.9846   \\   \hline
$E[I_{5}]$ & 4.7599 & 4.7937 & 4.7599    \\ \hline
$E[I_{10}]$ & 9.0810 & 9.1791 & 9.0810    \\  \hline
\end{tabular}
\caption{ Expectation of $E[I_1]$, $E[I_5]$, and $E[I_{10}]$ for three voting
rules: $\mathcal{V}_{as}$, $\mathcal{V}_{ehl}$, and $\mathcal{V}_{pl}$ when $n
= 3$} \label{table:sec_vot_rule:exp_top1_5_10_n_3}
\end{table}

% voting rule comparisons for high self-selectivity

\subsection{Effect of Tie Breaking Rules}

In this section we explore the effect of tie breaking rules on conference
recommendation system. Specifically, we evaluate the performance of the
following four representative tie breaking rules:

\noindent {\bf \em Least variance ($\ST_{var}$):} select one with least
variance, if there is still tie, perform random selection.

\noindent {\bf \em Largest max score ($\ST_{maxs}$):} select one with the
largest max score, if there is still tie, perform random selection.

\noindent {\bf \em Largest min score ($\ST_{mins}$):} select one with the
largest min score, if there is still tie, perform random selection.

\noindent {\bf \em Largest medium score ($\ST_{meds}$):} select one with the
largest medium score, if there is still tie, perform random selection.

\begin{figure}
\centering
\subfigure[High {\em self-selectivity}]{
\includegraphics[width=0.42\textwidth]{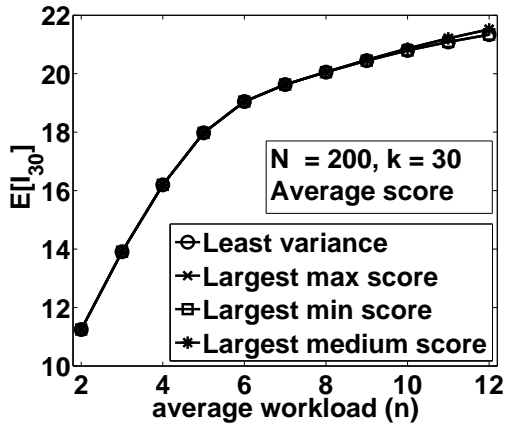}
\label{fig:sec_tie_break_rule:exp_top_30_high}
}
\subfigure[Medium {\em self-selectivity}]{
 \includegraphics[width=0.42\textwidth]{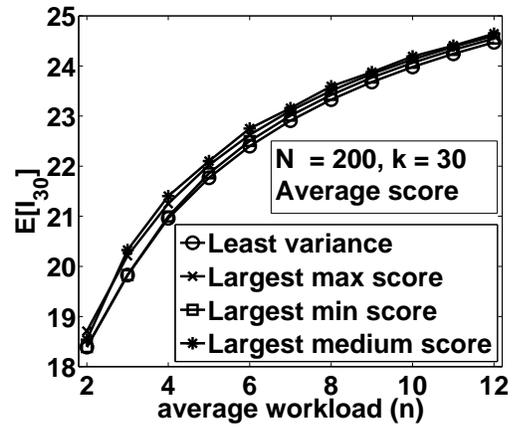}
 \label{fig:sec_tie_break_rule:exp_top_30_medium}
 }
\\
\subfigure[Low {\em self-selectivity}]{
\includegraphics[width=0.42\textwidth]{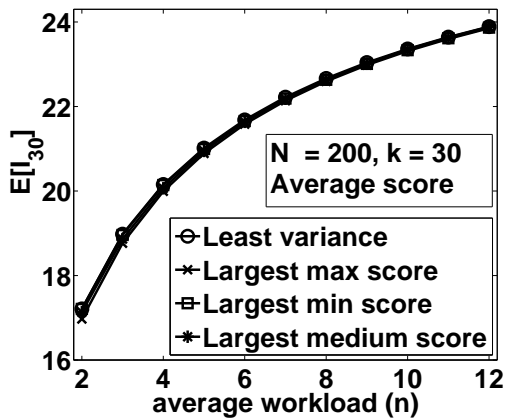}
\label{fig:sec_tie_break_rule:exp_top_30_low}
}
\subfigure[Random {\em self-selectivity}]{
 \includegraphics[width=0.42\textwidth]{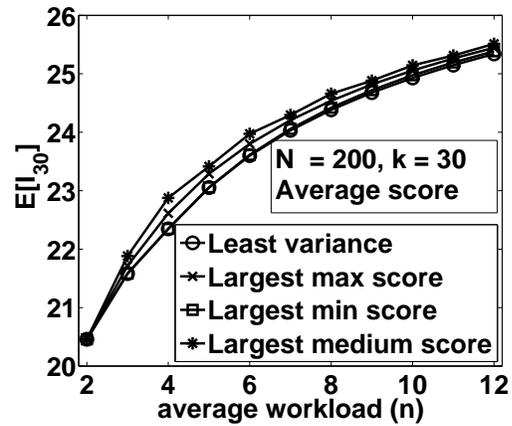}
 \label{fig:sec_tie_break_rule:exp_top_30_uniform}
 }
\caption{ Expectation of $E[I_{30}]$ for four tie breaking rules: $\ST_{var}$,
          $\ST_{maxs}$, $\ST_{mins}$, and $\ST_{meds}$.}
\label{fig:sec_tie_break_rule:exp_top_30}
\end{figure}

\begin{figure}
\centering
\subfigure[\# of top 1 paper get in]{
\includegraphics[width=0.42\textwidth]{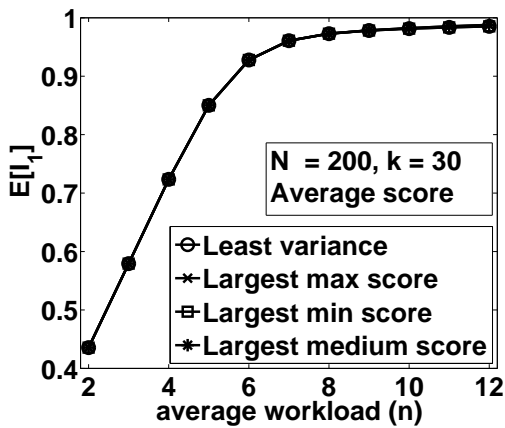}
\label{fig:sec_tie_break_rule:exp_top_1_high}
}
\subfigure[\# of top 5 papers get in]{
 \includegraphics[width=0.42\textwidth]{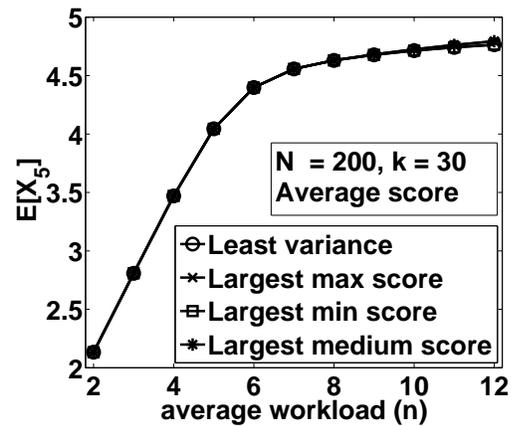}
 \label{fig:sec_tie_break_rule:exp_top_5_high}
 }
\caption{ Expectation of $E[I_1]$, $E[I_5]$ for four tie breaking rules:
$\ST_{var}$, $\ST_{maxs}$, $\ST_{mins}$, and $\ST_{meds}$ for high
{\em self-selectivity} submission.}
\label{fig:sec_tie_break_rule:exp_top_1_5_high}
\end{figure}

To evaluate the performance of these four tie breaking rules, let us select a
voting rule: {\em average score rule} and we use expectation as our performance
measure. We evaluate the performance of these four tie breaking rules on the
conference recommendation system specified in Section
\ref{section:prob_expec_var}. The numerical results of $E[I_{30}]$ are shown in
Fig. \ref{fig:sec_tie_break_rule:exp_top_30}. The numerical results of $E[I_1]$
and $E[I_5]$ for high {\em self-selectivity} papers are shown in Fig.
\ref{fig:sec_tie_break_rule:exp_top_1_5_high}. The numerical results of
$E[I_1]$, $E[I_5]$, and $E[I_{10}]$ when $n = 3$ are shown in Table
\ref{table:sec_tie_break_rule:exp_top_1_5_10_n_3}.

In Fig. \ref{fig:sec_tie_break_rule:exp_top_30} and
\ref{fig:sec_tie_break_rule:exp_top_1_5_high}, the horizontal axis represents
the number of reviews per paper, or $n$. The vertical axis shows the
corresponding expectation. From Fig. \ref{fig:sec_tie_break_rule:exp_top_30},
we could have the following observations. When we increase the reviewing
workload, the expectation $E[I_{30}]$ corresponding to each tie breaking rule
increased, which reflects that the accuracy of the conference recommendation
system increased. From Fig. \ref{fig:sec_tie_break_rule:exp_top_30_high} and
\ref{fig:sec_tie_break_rule:exp_top_30_low}, we could see that when submitted
papers are of high or low {\em self-selectivity}, the expectation curves
corresponding to these four tie breaking rules overlapped together. In other
words, these four rules have the same accuracy.  From Fig.
\ref{fig:sec_tie_break_rule:exp_top_30_medium} and
\ref{fig:sec_tie_break_rule:exp_top_30_uniform}, we see that when submitted
papers are of medium or random {\em self-selectivity}, the expectation curves
corresponding to these four tie breaking rules bunched together and the {\em
largest medium score rule} has a slightly higher accuracy than others.  From
Table \ref{table:sec_tie_break_rule:exp_top_1_5_10_n_3}, we see that when $n
\!=\! 3$, the chance of accepting the best paper is invariant of tie breaking
rules, unless when submitted papers are of high {\em self-selectivity}, since
values of $E[I_1]$ corresponding to medium, low or random {\em
self-selectivity} are all around 0.98. This statement also holds for the top
five and top ten papers. When submitted papers are of high {\em
self-selectivity}, the chance of the best paper being accepted is low, or less
than 0.6. The same statements holds for the top five and top ten papers. From
Fig. \ref{fig:sec_tie_break_rule:exp_top_1_5_high}, we see that for each tie
breaking rule, we still have to increase the reviewing workload $n$ to at least
7 so as to have a strong guarantee that the best paper or the top five papers
will get accepted.

\noindent {\bf Lessons learned:} These four tie breaking rules have comparable
accuracy, no rule can outperform others remarkably. Thus, the improvement of
conference recommendation system by tie breaking rules is limited. For each tie
breaking rule, we till have to increase the reviewing workload $n$ to at least
seven to have a strong guarantee that the best paper will get in.

%tie breaking rule comparison

%tie breaking comparison for top-1 paper
\begin{table}
\centering
\begin{tabular}{|c||c|c|c|c|} \hline
    & $\ST_{var}$ & $\ST_{maxs}$ & $\ST_{mins}$ & $\ST_{meds}$  \\ \hline \hline
\multicolumn{5}{|c|}{ \bf High self-selectivity }  \\ \hline
$E[I_{1}]$ & 0.5793 & 0.5793 & 0.5793 & 0.5793  \\  \hline
$E[I_{5}]$ & 2.8078 & 2.8077 & 2.8077 & 2.8078  \\  \hline
$E[I_{10}]$ & 5.4026 & 5.4026 & 5.4026 & 5.4026 \\  \hline \hline
\multicolumn{5}{|c|}{ \bf Medium self-selectivity }  \\ \hline
$E[I_{1}]$ & 0.9832 & 0.9880 & 0.9832 & 0.9907  \\  \hline
$E[I_{5}]$ & 4.6951 & 4.7514 & 4.6951 & 4.7755  \\  \hline
$E[I_{10}]$ & 8.7764 & 8.9311 & 8.7765 & 8.9831  \\  \hline \hline
\multicolumn{5}{|c|}{ \bf Low self-selectivity }  \\ \hline
$E[I_{1}]$ & 0.9788 & 0.9787 & 0.9788 & 0.9886  \\  \hline
$E[I_{5}]$ & 4.6083 & 4.5989 & 4.6074 & 4.6040  \\  \hline
$E[I_{10}]$ & 8.5164 & 8.4723 & 8.5102 & 8.3990  \\  \hline \hline
\multicolumn{5}{|c|}{ \bf Random self-selectivity }  \\ \hline
$E[I_{1}]$ & 0.9846 & 0.9858 & 0.9846 & 0.9872  \\  \hline
$E[I_{5}]$ & 4.7599 & 4.7747 & 4.7599 & 4.7926  \\   \hline
$E[I_{10}]$ & 9.0810 & 9.1239 & 9.0810 & 9.1761  \\  \hline
\end{tabular}
\caption{ Expectation of $E[I_1]$, $E[I_5]$, and $E[I_{10}]$ for four tie
breaking rules: $\ST_{var}$, $\ST_{maxs}$, $\ST_{mins}$, and $\ST_{meds}$ when
$ n \! = \! 3$. } \label{table:sec_tie_break_rule:exp_top_1_5_10_n_3}
\end{table}

\subsection{Effect of Reviewers Type -- two types case}
\label{section:two_type}

\begin{figure}[thb]
\centering
\subfigure[\# of top 1 papers get in]{
\includegraphics[width=0.42\textwidth]{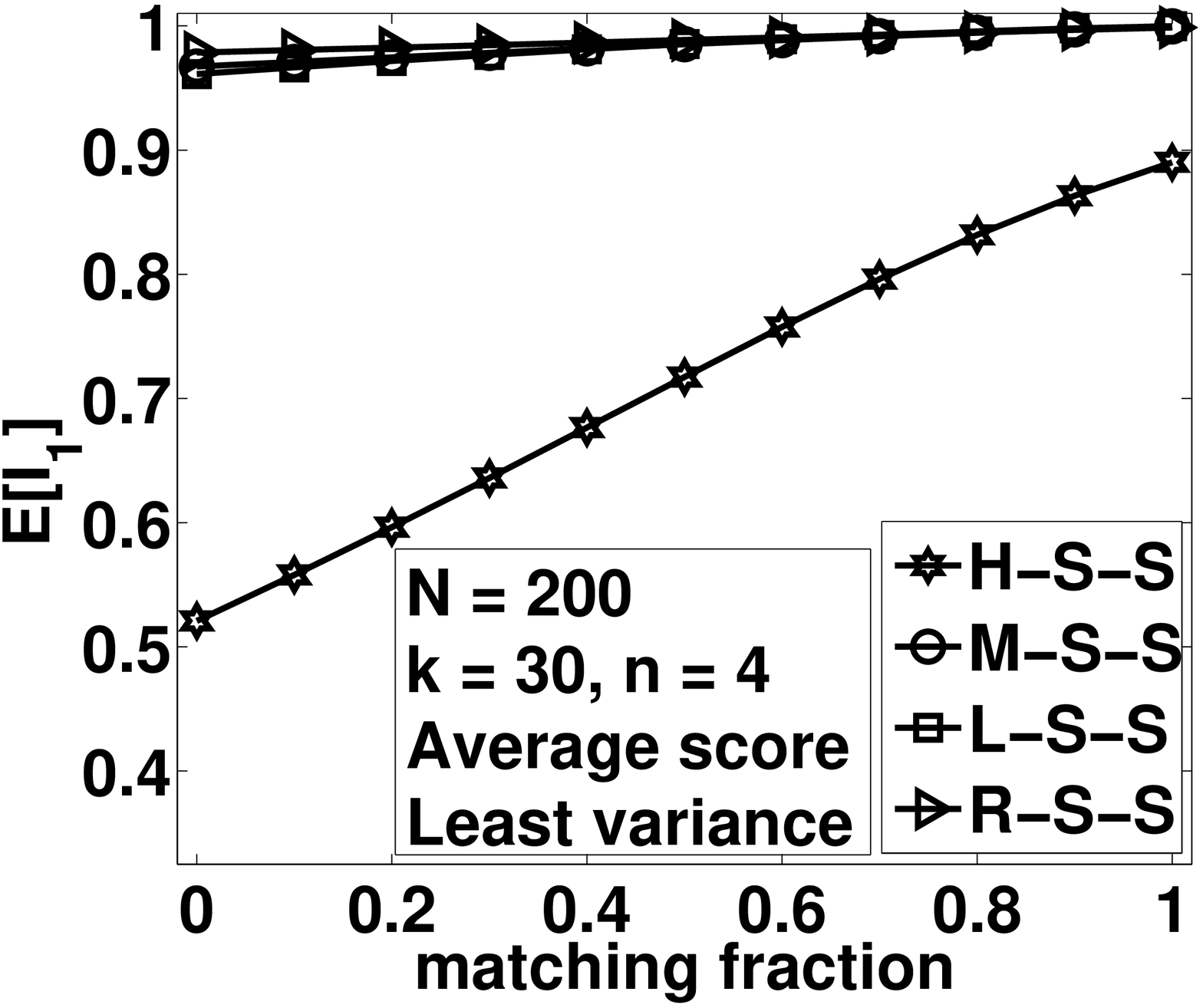}
\label{fig:sec_two_types:exp_top1}
}
\subfigure[\# of top 5 papers get in]{
 \includegraphics[width=0.42\textwidth]{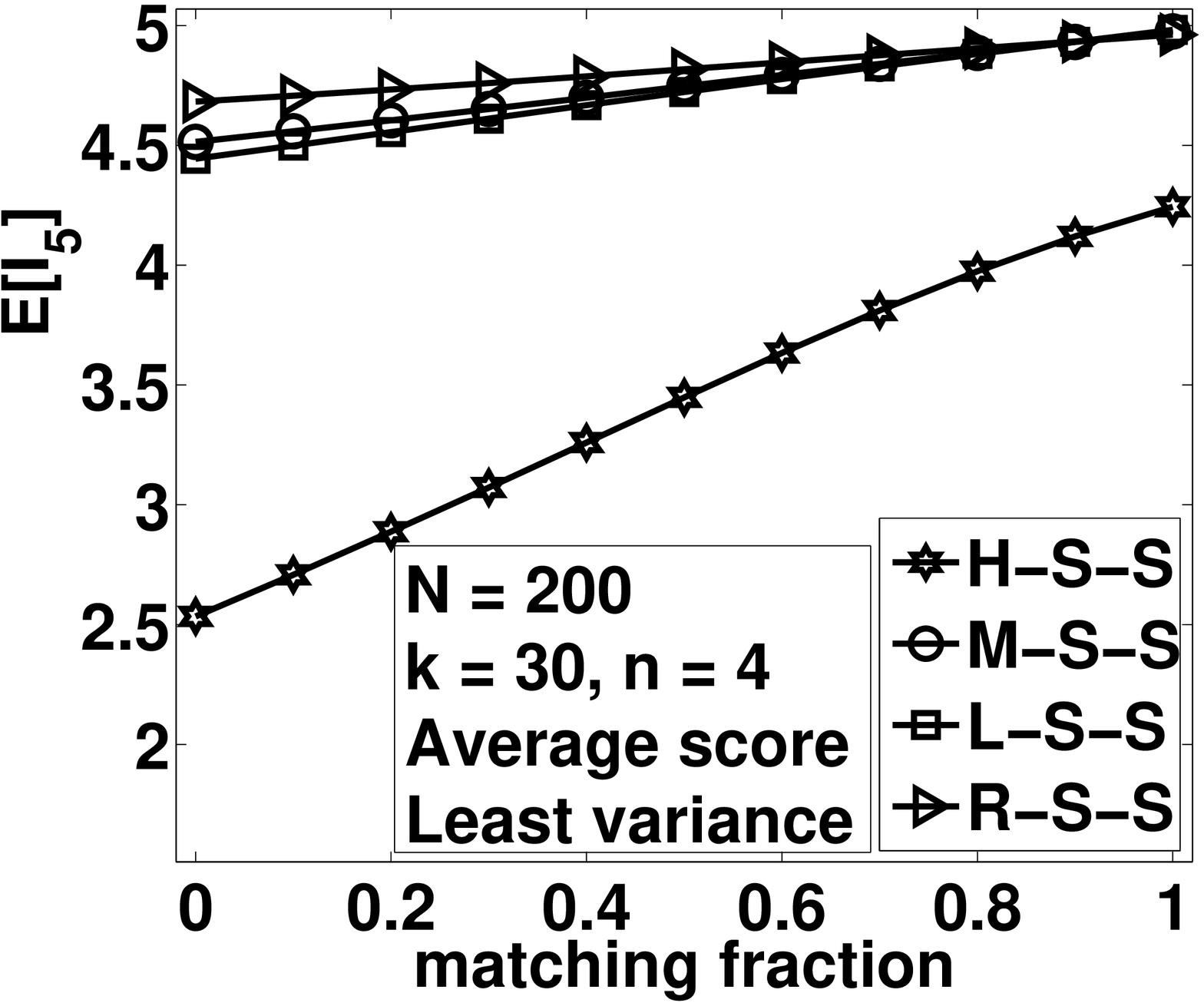}
 \label{fig:sec_two_types:exp_top5}
 }
\\
\subfigure[\# of top 10 papers get in]{
\includegraphics[width=0.42\textwidth]{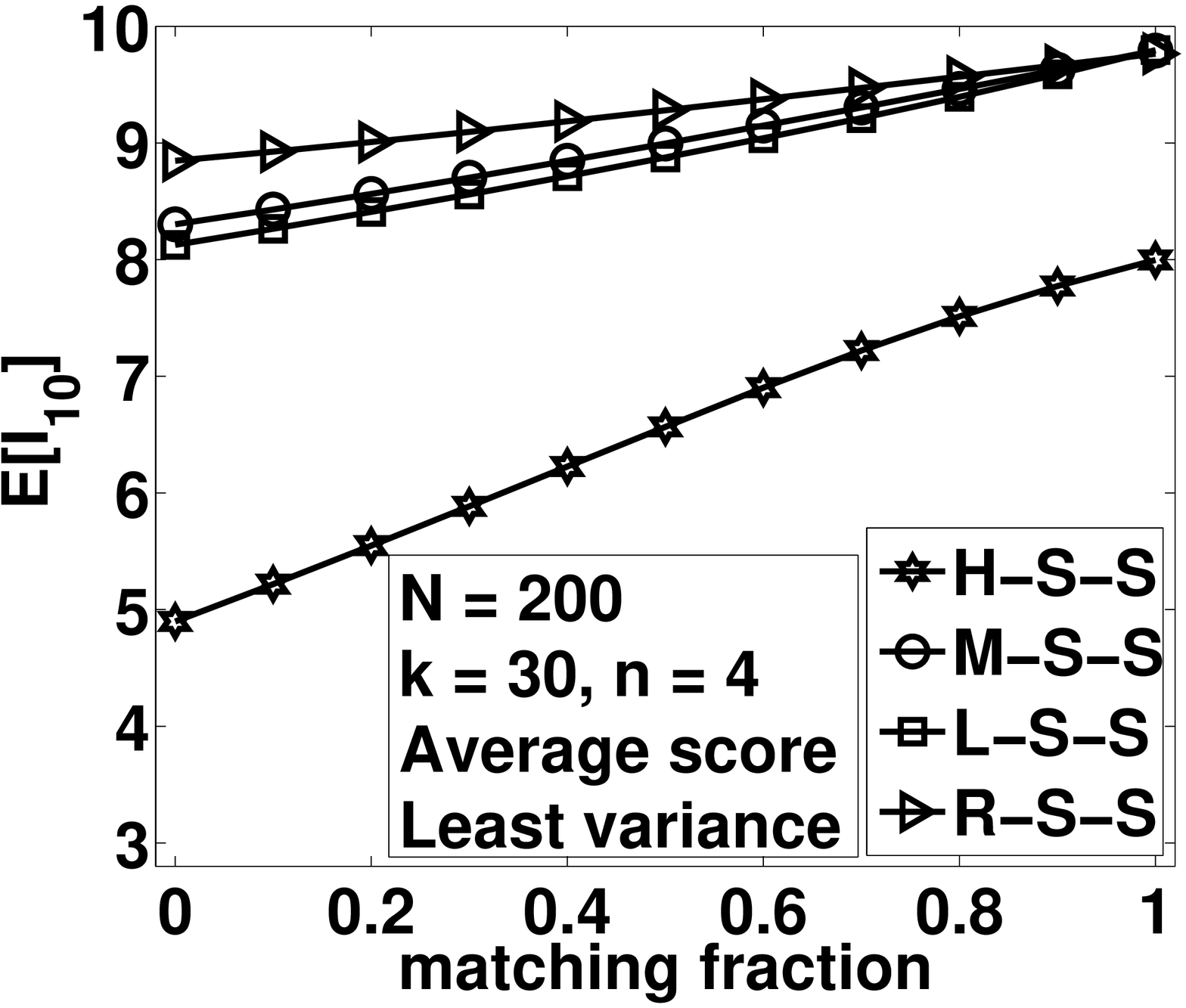}
\label{fig:sec_two_types:exp_top10}
}
\subfigure[\# of top 30 papers get in]{
 \includegraphics[width=0.42\textwidth]{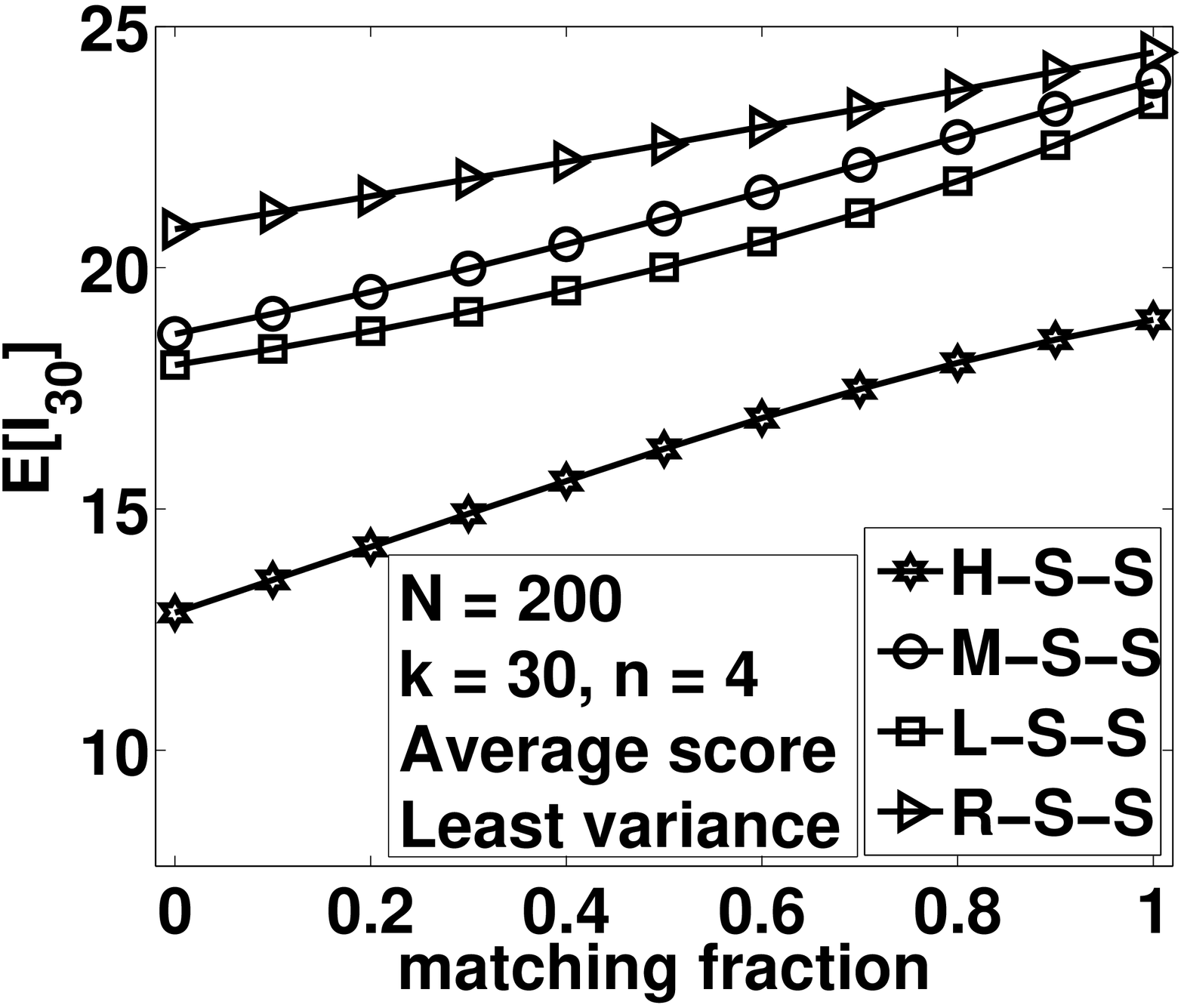}
 \label{fig:sec_two_types:exp_top30}
 }
\caption{Impact of paper-reviewing matching}
%Exploring the fraction of paper-reviewer matched in the system type}
\label{fig:sec_two_types:exp}
\end{figure}

We extend the homogeneous conference recommendation system specified in Section
\ref{section:prob_expec_var} to a heterogeneous conference recommendation
system, for which papers and reviewers can be of different types. Specifically,
we consider papers and reviewers are of two types (e.g. system or theory). If
paper-reviewer matches in the same type, the critical degree is high, else the
critical degree is low. Here we explore the effect of the fraction that
paper-reviewer matches in the same type on the accuracy of the recommendation.
We use expectation as our performance measure and we set the number of review
per paper in this heterogeneous recommendation system be four, or $n \! = \!
4$. We vary the fraction of paper-reviewer that matches in the same type from
0.1 to 1. Before showing our results, let us specify $\sigma(c^i_j)$ derived in
Eq. (\ref{equation_rating_distribution_general}): if papers-reviewer matches in
the same type, $\sigma(c^i_j) \!=\! 0.5$, otherwise $\sigma(c^i_j) \!=\! 2$.
The numerical results of $E[I_1]$, $E[I_5]$, $E[I_{10}]$, and $E[I_{30}]$ are
shown in Fig. \ref{fig:sec_two_types:exp}.

In Fig. \ref{fig:sec_two_types:exp}, the horizontal axis represents the the
fraction that paper-reviewer matches in the same type. The vertical axis shows
the corresponding expectation. From Fig. \ref{fig:sec_two_types:exp}, we have
the following observations. When we increase the matching fraction, the
expectation $E[I_1]$, $E[I_5]$, $E[I_{10}]$, and $E[I_{30}]$ increased. In
other words, the accuracy of the conference recommendation system increased
when more paper-reviewer matches in the same type. It is interesting to see
that the expectation $E[I_1]$, $E[I_5]$, $E[I_{10}]$ and $E[I_{30}]$ increase
in a linear rate and the rate corresponding to the high {\em self-selectivity}
is the highest, and the random {\em self-selectivity} is the lowest. When
papers are submitted with medium, low or random {\em self-selectivity}, the
chance of accepting the best paper is invariant of the matching fraction, since
the corresponding values of $E[I_1]$ are approximately equal to 1 when the
matching fraction is 0 and increased slightly when matching faction increased
to 1. This statement also holds for the top five papers. When submitted papers
are highly {\em self-selective}, the accuracy of the system is very sensitive
to the matching fraction, this is because $E[I_1] \!\approx\! 0.5$, $E[I_{10}]
\!\approx\! 5$ and $E[I_{30}] \!\approx\! 13$ when matching fraction is 0, and
$E[I_1]\!\approx \! 0.9$, $E[I_{10}] \!\approx\! 8$ and $E[I_{30}] \!\approx\!
20$ when matching fraction is 1.

\noindent {\bf Lessons learned:} If reviewers' preference and papers match, it
can significantly affect the accuracy of the conference recommendation system.
This is especially true for submitted papers which are of high {\em
self-selectivity} (or prestigious conferences). To achieve this, it is
especially important to select the appropriate technical program committee
members to match the research topics since it can greatly influence the
accuracy of the final recommendation.

% two types of reviewers, vary the fraction

\subsection{Effect of Reviewers type -- many types case}

\begin{figure}[thb]
\centering
\subfigure[High {\em self-selectivity}]{
\includegraphics[width=0.42\textwidth]{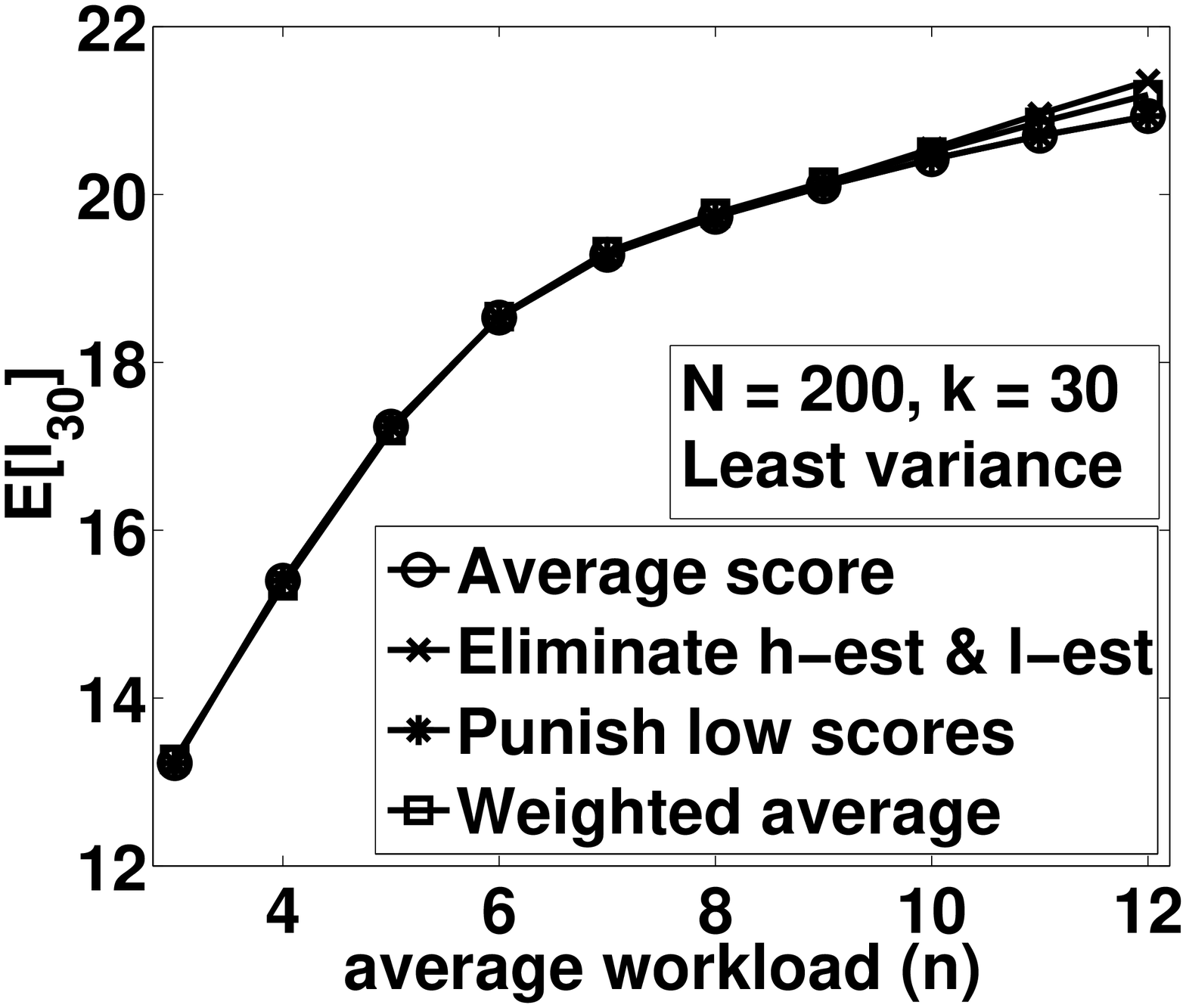}
\label{fig:sec_many_types:exp_top_30_high}
}
\subfigure[Medium {\em self-selectivity}]{
 \includegraphics[width=0.42\textwidth]{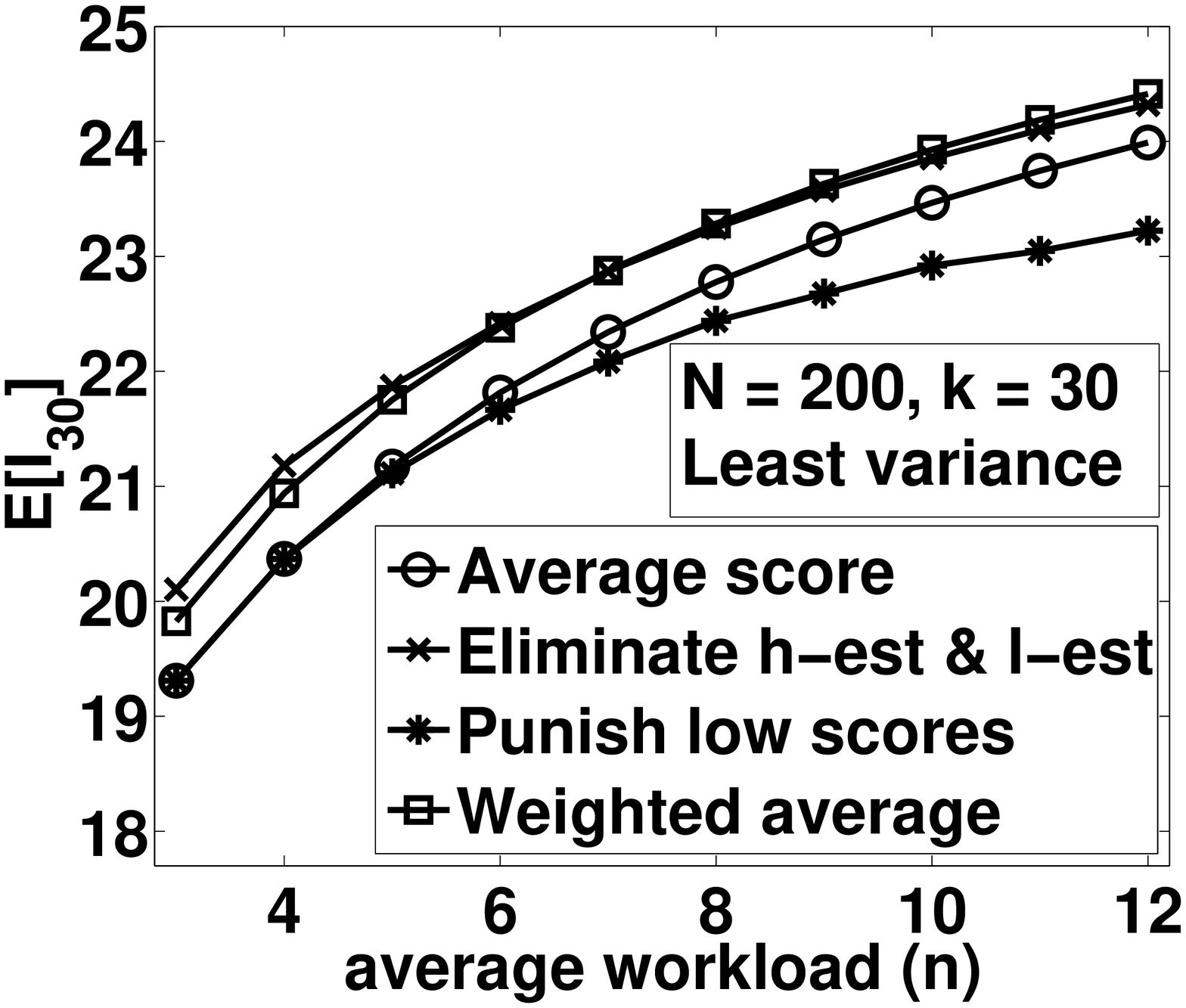}
 \label{fig:sec_many_types:exp_top_30_medium}
 }
\\
\subfigure[Low {\em self-selectivity}]{
\includegraphics[width=0.42\textwidth]{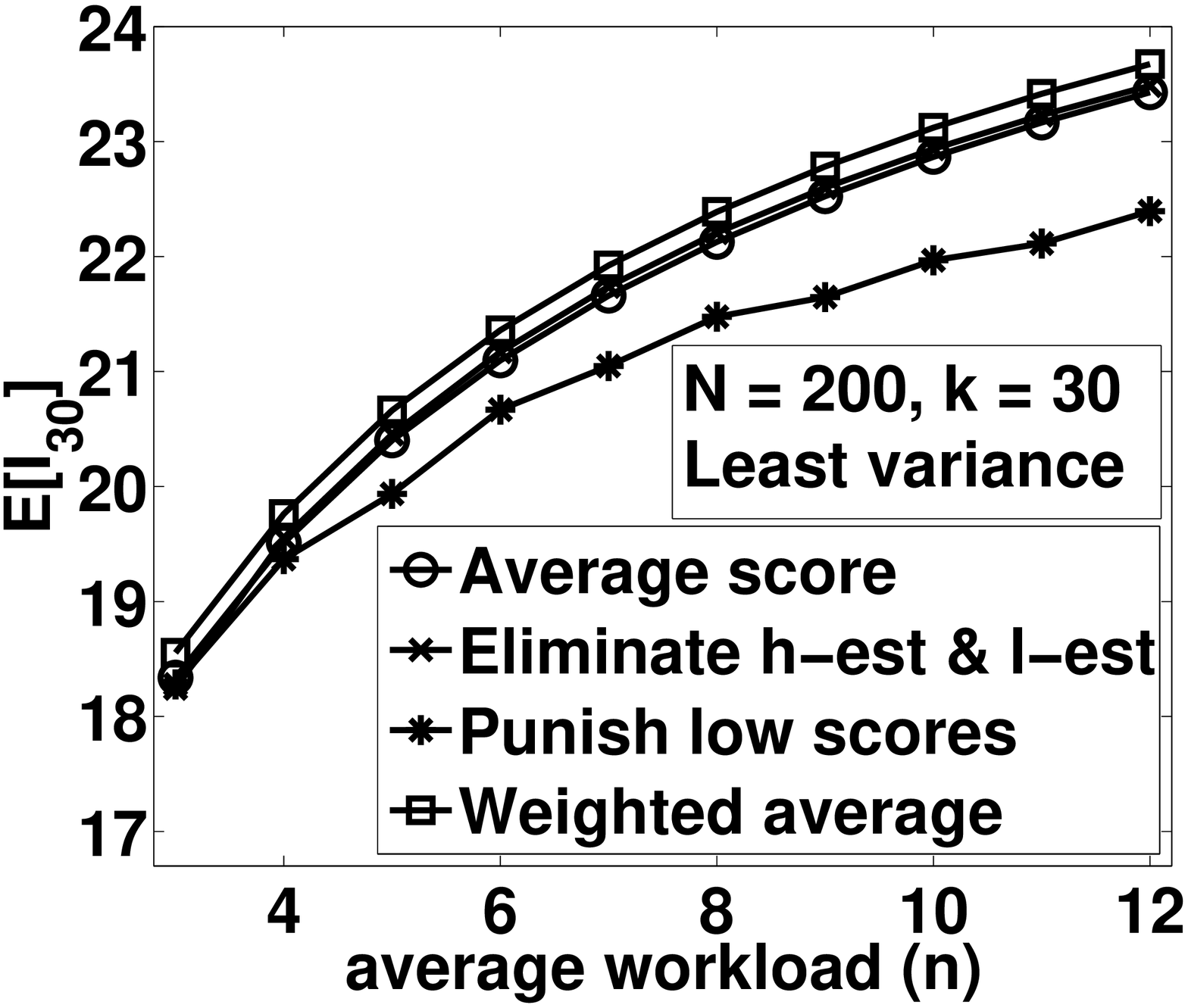}
\label{fig:sec_many_types:exp_top_30_low}
}
\subfigure[Random {\em self-selectivity}]{
 \includegraphics[width=0.42\textwidth]{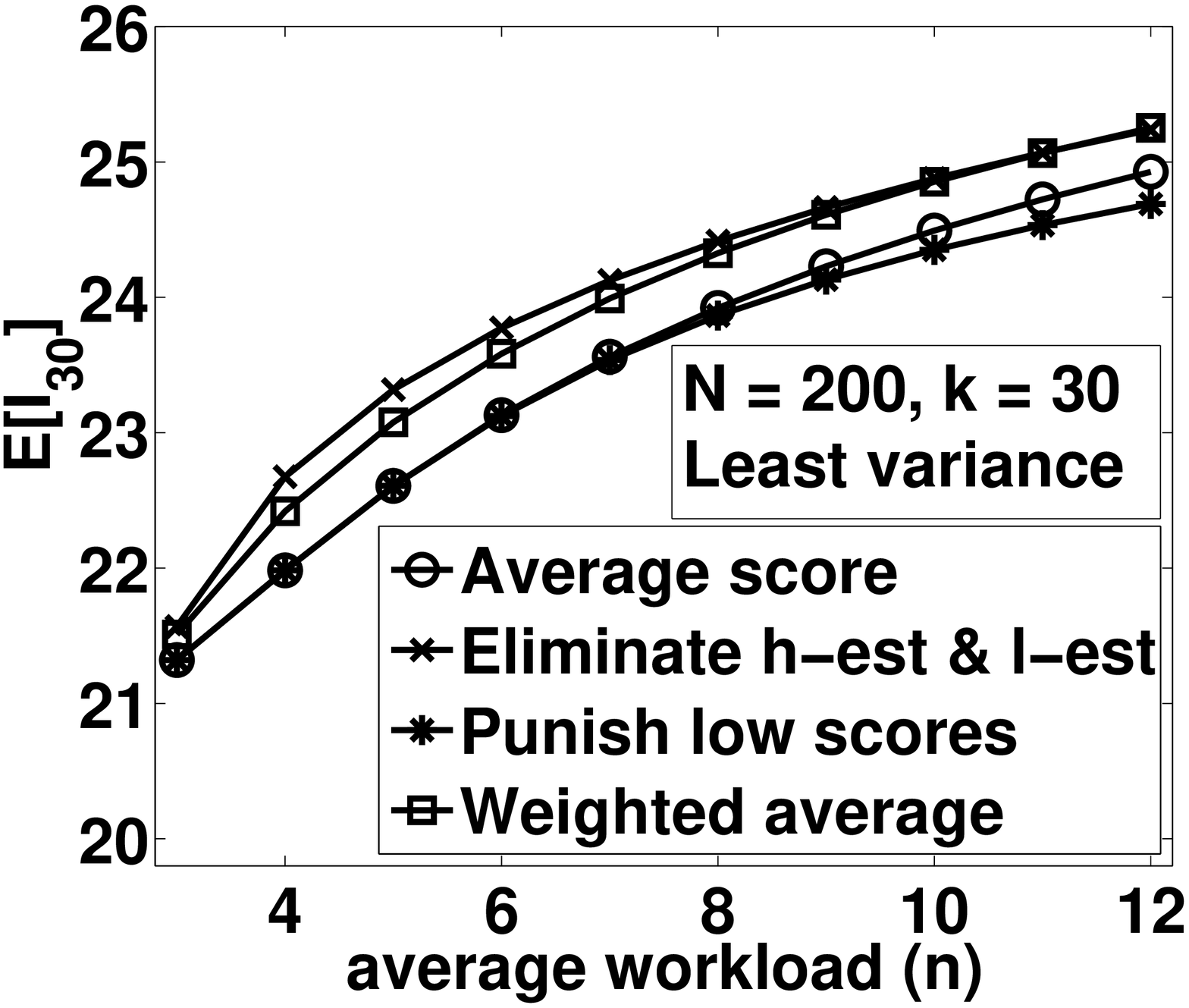}
 \label{fig:sec_many_types:exp_top_30_uniform}
 }
\caption{ Expectation of $I_{30}$ for four voting rules: $\SV_{as}$,
          $\SV_{ehl}$, $\SV_{pl}$ and $\SV_{wa}$.}
\label{fig:sec_many_types:exp_top_30}
\end{figure}

We now generalize the types of papers or reviewers of a conference
recommendation system specified in Section \ref{section:two_type} to many
types. We explore the effect of reviewer types on the accuracy of different
voting rules. Specifically, we consider the following four representative
voting rules: {\bf \em Average score ($\SV_{as}$), Eliminate the highest \&
lowest ($\SV_{ehl}$) , Punish low scores ($\SV_{pl}$)} as given by Equation
(\ref{equation:average_score_rule}) $-$ (\ref{equation:punish_low_score}). We
also consider the {\bf \em Weighted average ($\SV_{wa}$) rule:} The average
score of paper $P_i$ is weighted on $e^i_j$, where $j \! = \! 1, \ldots, n_i $,
the declared expertise level of reviewer $\SR(S^i_j)$ who selects topics
related to $P_i$, or
\begin{equation}
\gamma_i = \sum\nolimits_{j = 1}^{n_i} S^i_j e^i_j
  \left[\sum\nolimits_{j = 1}^{n_i} e^i_j \right]^{-1}.
\label{equation:weighted_average}
\end{equation}
We consider the least variance rule for tie breaking and we use expectation as
our performance measure. In this evaluation, papers and reviewers are randomly
matched. Before showing our results, let us specify two functions that related
to model for reviewers types: the first one is $\sigma(c^i_j)$ within the
probability distribution for score $S^i_j$ derived in Eq.
(\ref{equation_rating_distribution_general}), which is specified by the
following linear function
\[
    \sigma(c^i_j) = 0.5 + 1.5 [1 - c^i_j],
\]
the second one is a monotonic increasing function $f(\mu)$ within Eq.
(\ref{equation_critical_function}), which is specified by the following linear
function
\[
    f(\mu) = \mu.
\]
We set $l$ to be 3, thus $e^i_j \in \{1,2,3\}$. The numerical results of
$E[I_{30}]$ are shown Fig. \ref{fig:sec_many_types:exp_top_30}.

In Fig. \ref{fig:sec_many_types:exp_top_30}, the horizontal axis represents the
number of reviews per paper, or $n$. The vertical axis shows the corresponding
expectation. From Fig. \ref{fig:sec_many_types:exp_top_30}, we have the
following observations. When submitted papers are of high {\em self-selectivity
}, the expectation curves corresponding to these four voting rules overlapped
together. In other words, these four rules have the same accuracy for high {\em
self-selectivity} submissions. When submitted papers of low {\em
self-selectivity}, the {\em punish low scores rule } has the lowest accuracy
than the other three voting rules which have nearly the same accuracy. When
submitted papers of medium {\em self-selectivity}, the weighted average scoring
rule and the {\em eliminate highest and lowest score rule } have nearly the
same accuracy, and the weighted average scoring rule has slightly higher
accuracy than the average scoring rule and {\em punish low scores rule }. This
statement also holds for the submitted papers with random {\em
self-selectivity}.

\noindent {\bf Lessons learned:} When papers and reviewers are of many types
and papers and reviewers are randomly matched, {\em weighted average score
rule} can have slightly higher accuracy than {\em average score rule} and {\em
punish low scores rule }, and it has nearly the same accuracy with eliminate
highest and lowest score rule. We have the following similar observations in
Section \ref{section:voting_rule}: these four voting rules have comparable
accuracy, no rule can outperform others, thus the improvement of conference
recommendation system by voting rules is limited. Again, there are number of
interesting questions to explore, i.e., is the accuracy sensitive to anomaly
behavior?

%voting rule evaluation when paper review of types

\subsection{Effect of Anomaly Behavior -- Random Scoring}

\begin{figure}[thb]
\centering
\subfigure[\# of top 1 papers get in]{
\includegraphics[width=0.42\textwidth]{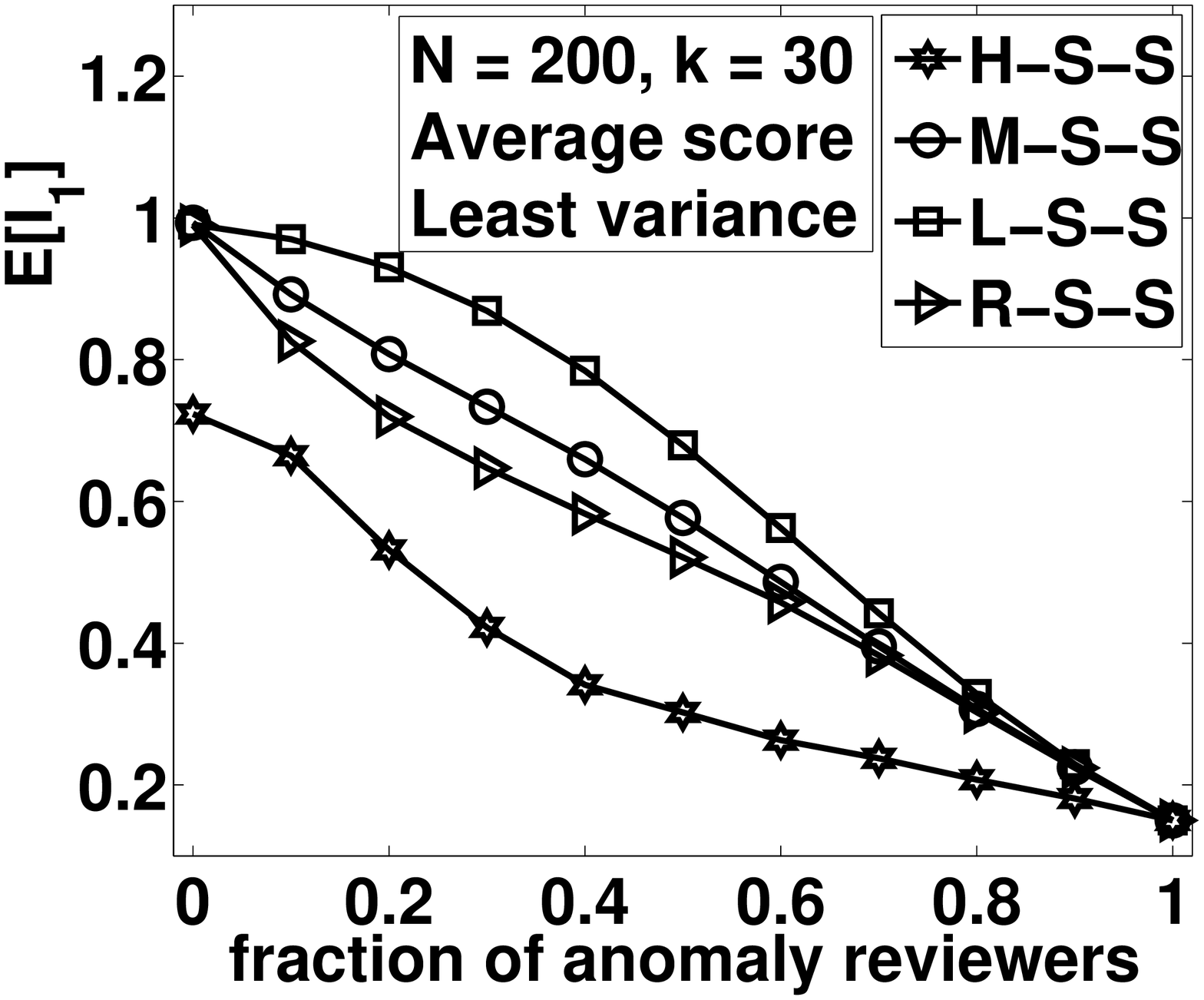}
\label{fig:sec_anomaly_behav:exp_top_1}
}
\subfigure[\# of top 5 papers get in]{
 \includegraphics[width=0.42\textwidth]{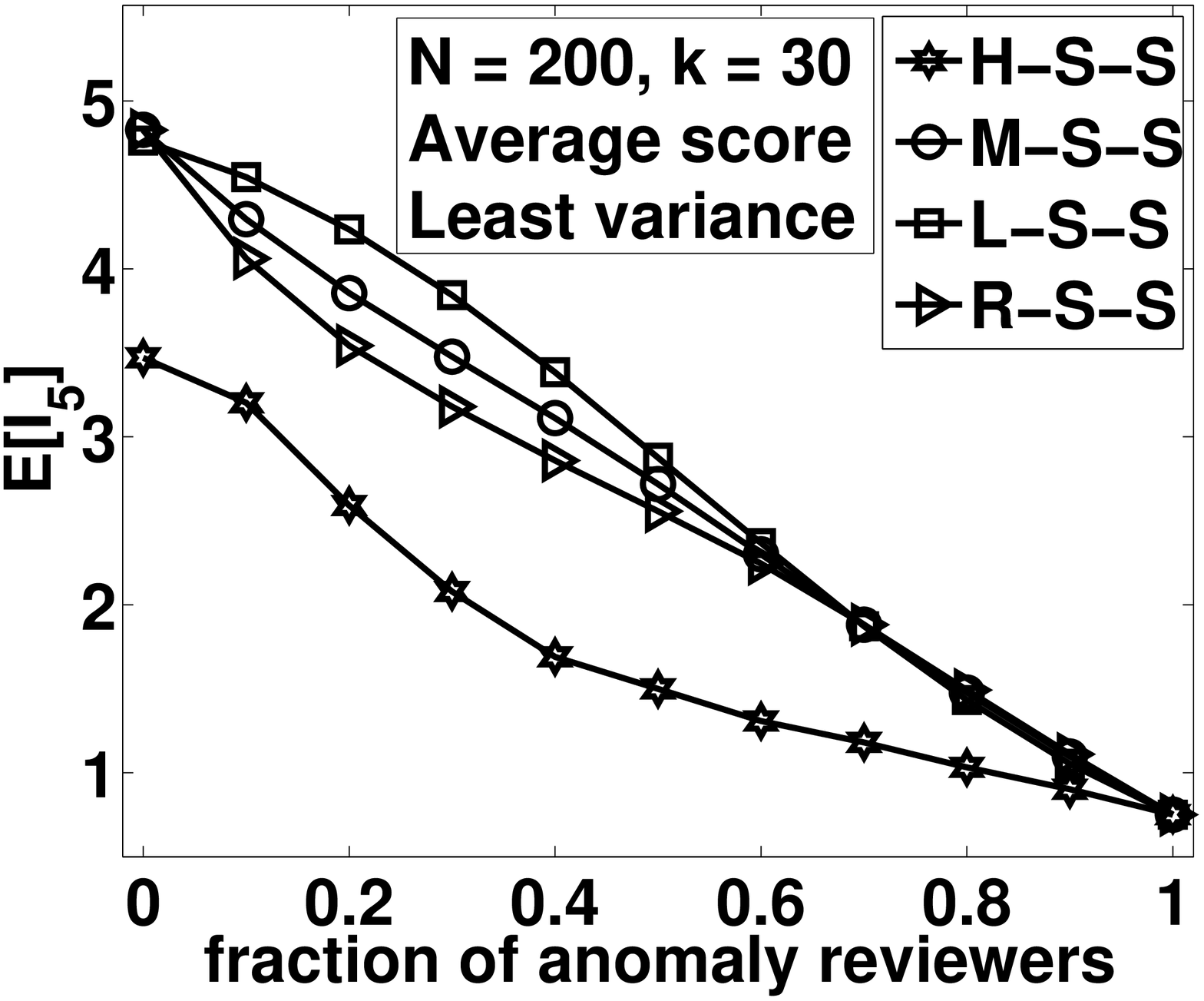}
 \label{fig:sec_anomaly_behav:exp_top_5}
 }
\\
\subfigure[\# of top 10 papers get in]{
\includegraphics[width=0.42\textwidth]{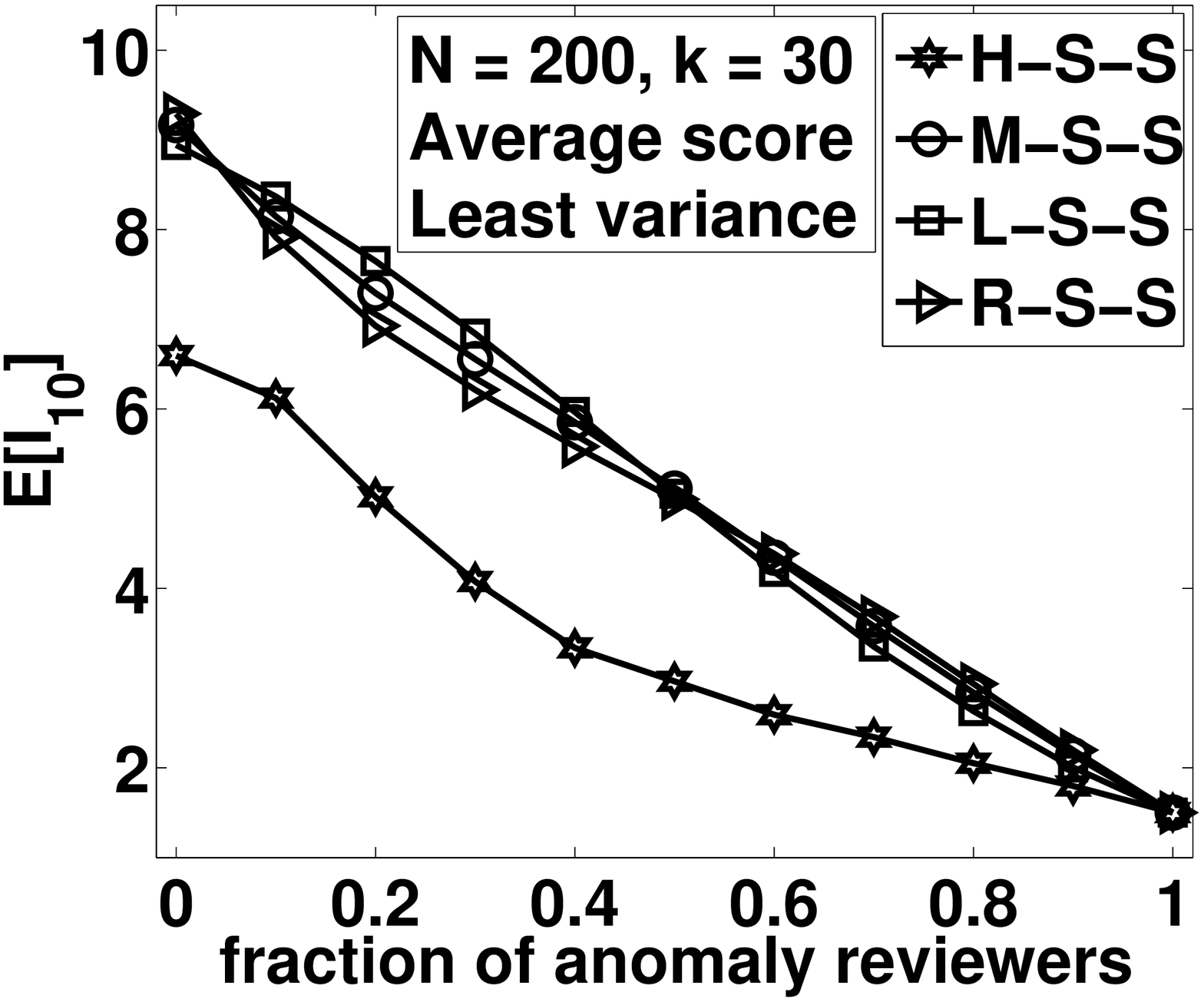}
\label{fig:sec_anomaly_behav:exp_top_10}
}
\subfigure[\# of top 30 papers get in]{
 \includegraphics[width=0.42\textwidth]{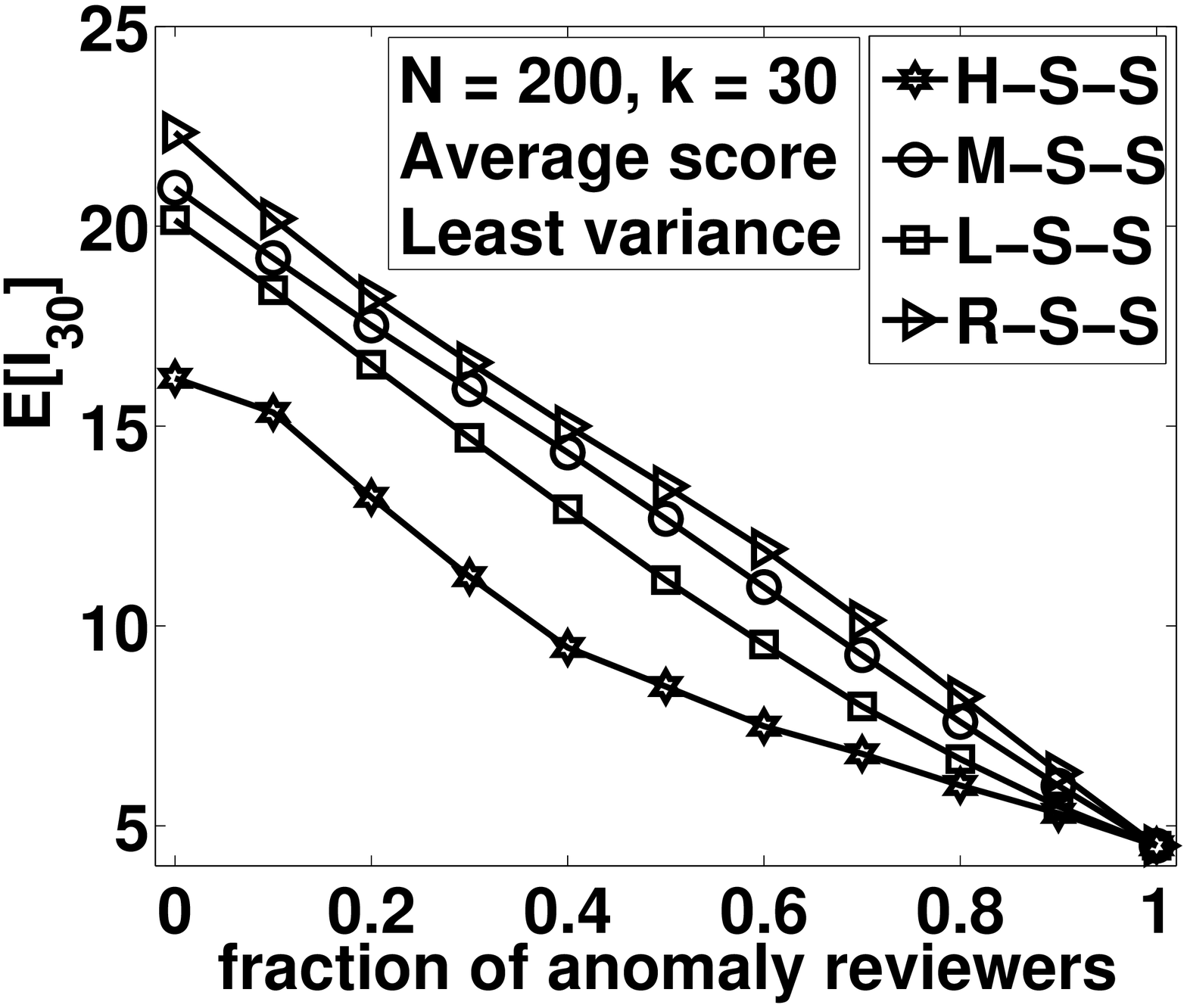}
 \label{fig:sec_anomaly_behav:exp_top_30}
 }
\caption{Impact of random anomaly behavior when $n \! = \! 4$.}
\label{fig:sec_anomaly_behav:exp}
\end{figure}

Let us explore the effect of an anomaly behavior on the accuracy of a
conference recommendation system. Specifically, we consider one potential
anomaly behavior: {\em random scoring behavior}, under which a misbehaving
reviewer gives a random score to any paper she reviews regardless of the
quality of that paper.
%To see the effect of this kind of anomaly behavior
%on the accuracy of conference recommendation system,
We vary the fraction of anomaly behavior from 0.1 to 1 and we use expectation
as our performance measure. We explore the effect of anomaly behavior on the
conference recommendation system specified in Section
\ref{section:prob_expec_var} with number of reviews per paper $ n \! = \! 4 $.
The numerical results of $E[I_1]$, $E[I_5]$, $E[I_{10}]$, and $E[I_{30}]$ are
shown in Fig. \ref{fig:sec_anomaly_behav:exp}.

In Fig. \ref{fig:sec_anomaly_behav:exp}, the horizontal axis represents the
fraction of these misbehaving reviewers. The vertical axis shows the
corresponding expectation. From Fig. \ref{fig:sec_anomaly_behav:exp}, we have
the following observations. When we increase the fraction of anomaly reviewers,
the expectation decreased. In other words, the more anomaly reviewers, the
lower is the accuracy of the conference recommendation system. It is
interesting to note that the accuracy of the conference recommendation system
decreases in a nearly linear rate.  From Fig.
\ref{fig:sec_anomaly_behav:exp_top_1}, we see that for low {\em
self-selectivity } papers, the chance of accepting the best paper can withstand
a small fraction of misbehaving reviewers, say around 10\%. This statement
holds for the top five papers. When submitted papers are of high {\em
self-selectivity}, around 40\% of misbehaving reviewers can drastically disrupt
the accuracy of a conference recommendation system. Because for this case, less
than ten papers from the top 30 papers will get accepted, and less than four
papers from the top ten papers will get accepted. Even the best paper can only
be accepted with the probability of less than 0.4. When papers are submitted
with medium, low or random {\em self-selectivity}, around 60\% misbehaving
reviewers can drastically disrupt the accuracy of a conference recommendation
system.

\begin{figure}[htb]
\centering
\subfigure[High {\em self-selectivity}]{
\includegraphics[width=0.42\textwidth]{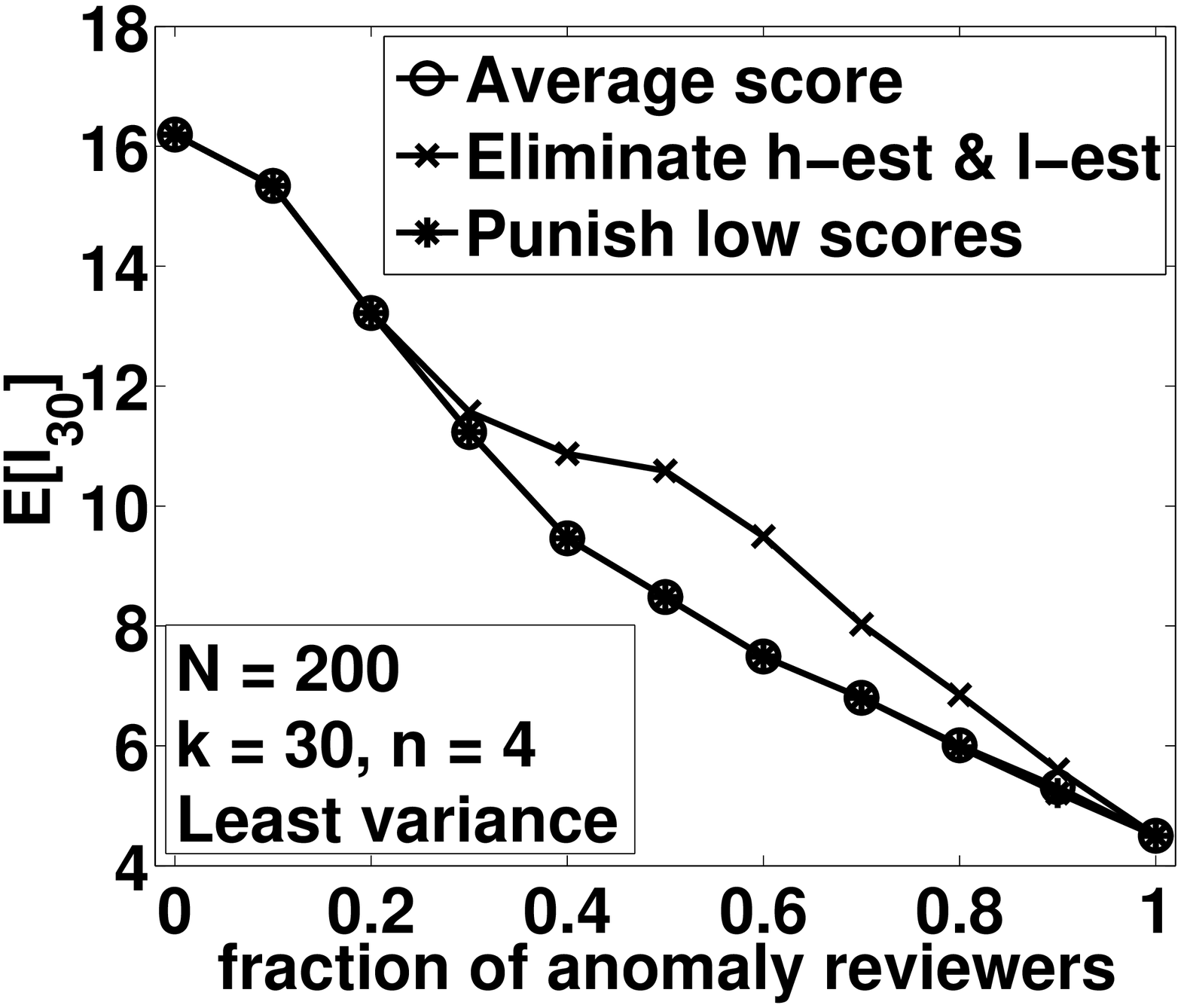}
\label{fig:sec_anomaly_behav:voting_rule_com_exp_high}
}
\subfigure[Medium {\em self-selectivity}]{
 \includegraphics[width=0.42\textwidth]{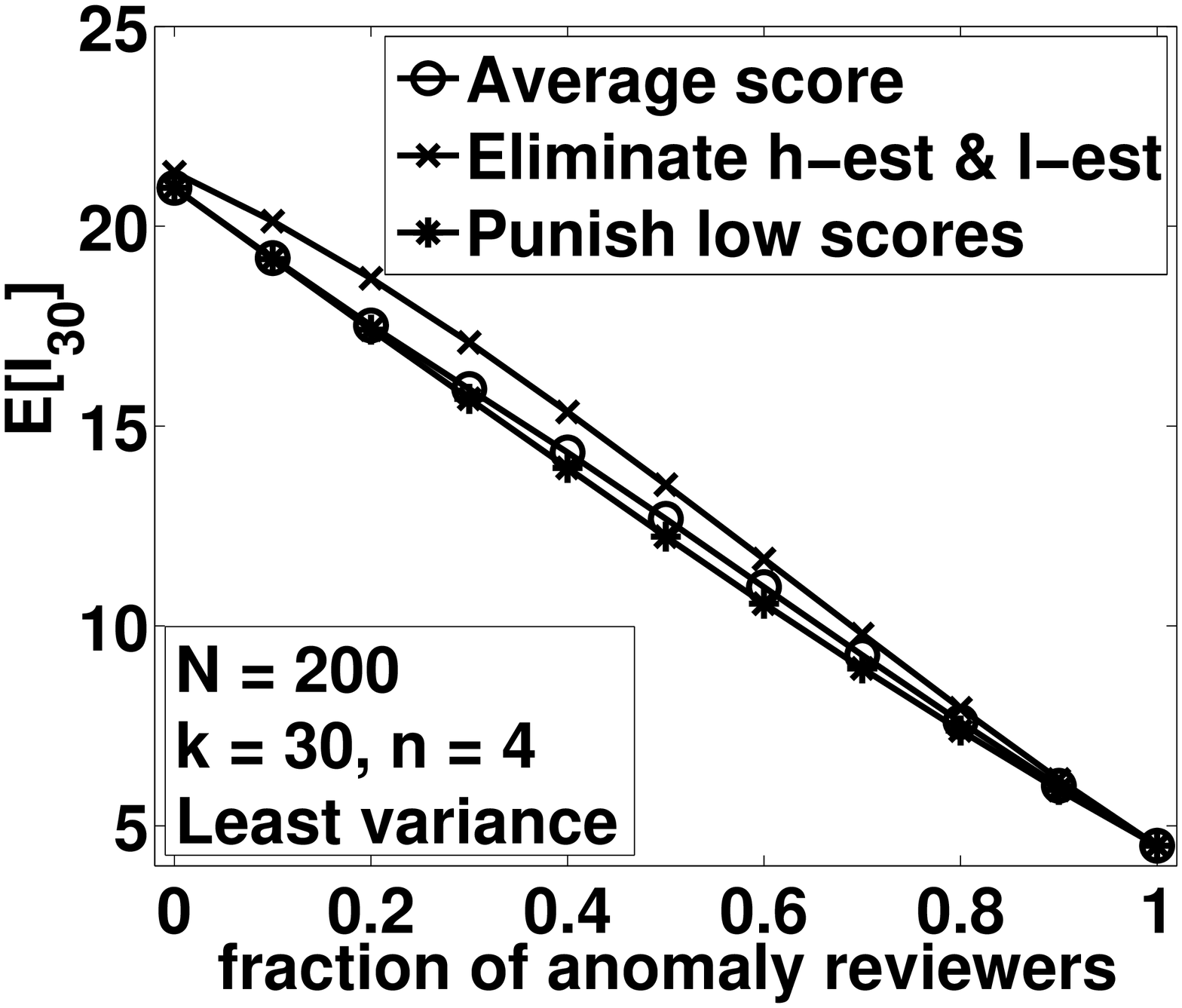}
 \label{fig:sec_anomaly_behav:voting_rule_com_exp_medium}
 }
\\
\subfigure[Low {\em self-selectivity}]{
\includegraphics[width=0.42\textwidth]{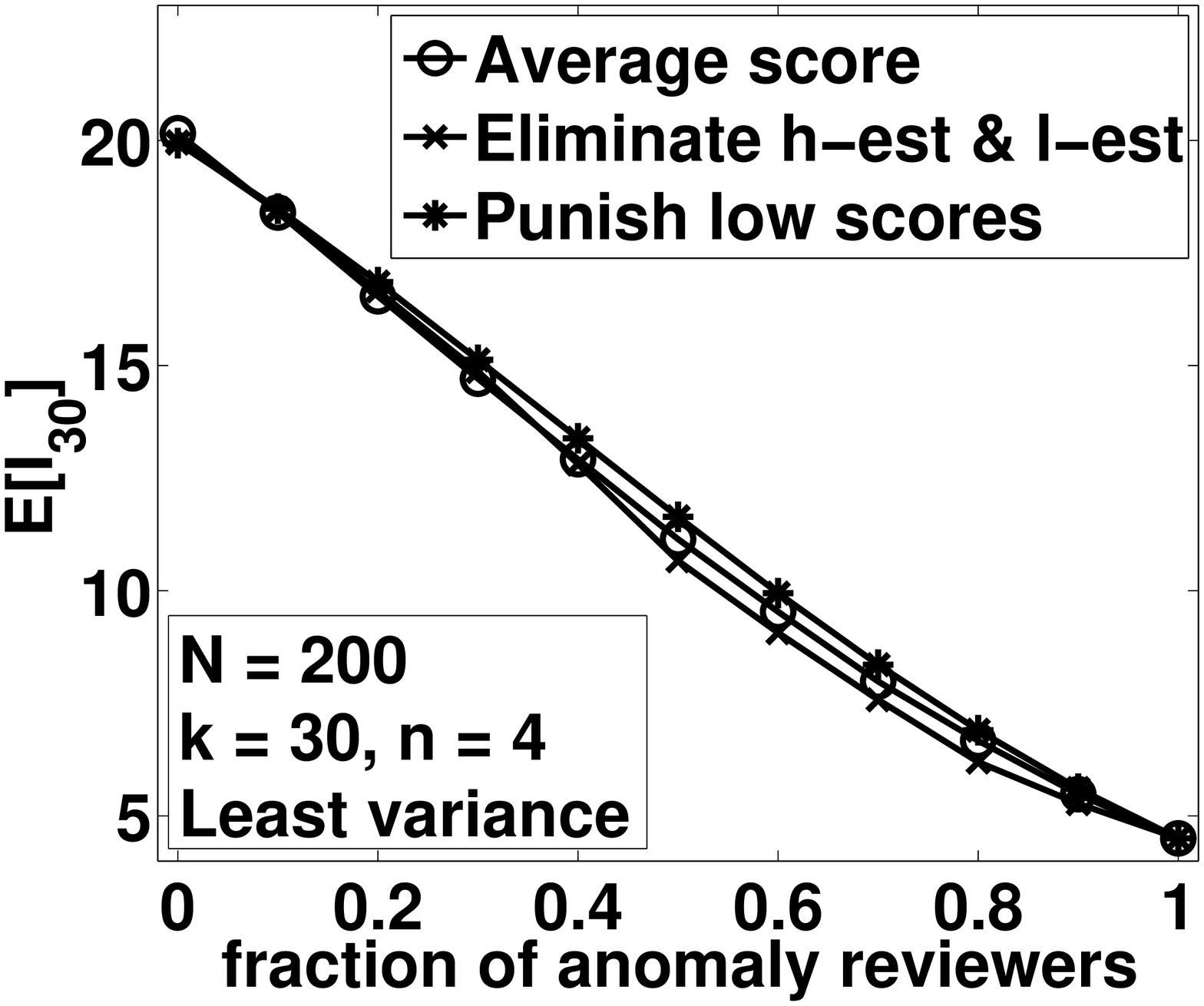}
\label{fig:sec_anomaly_behav:voting_rule_com_exp_low}
}
\subfigure[Random {\em self-selectivity}]{
 \includegraphics[width=0.42\textwidth]{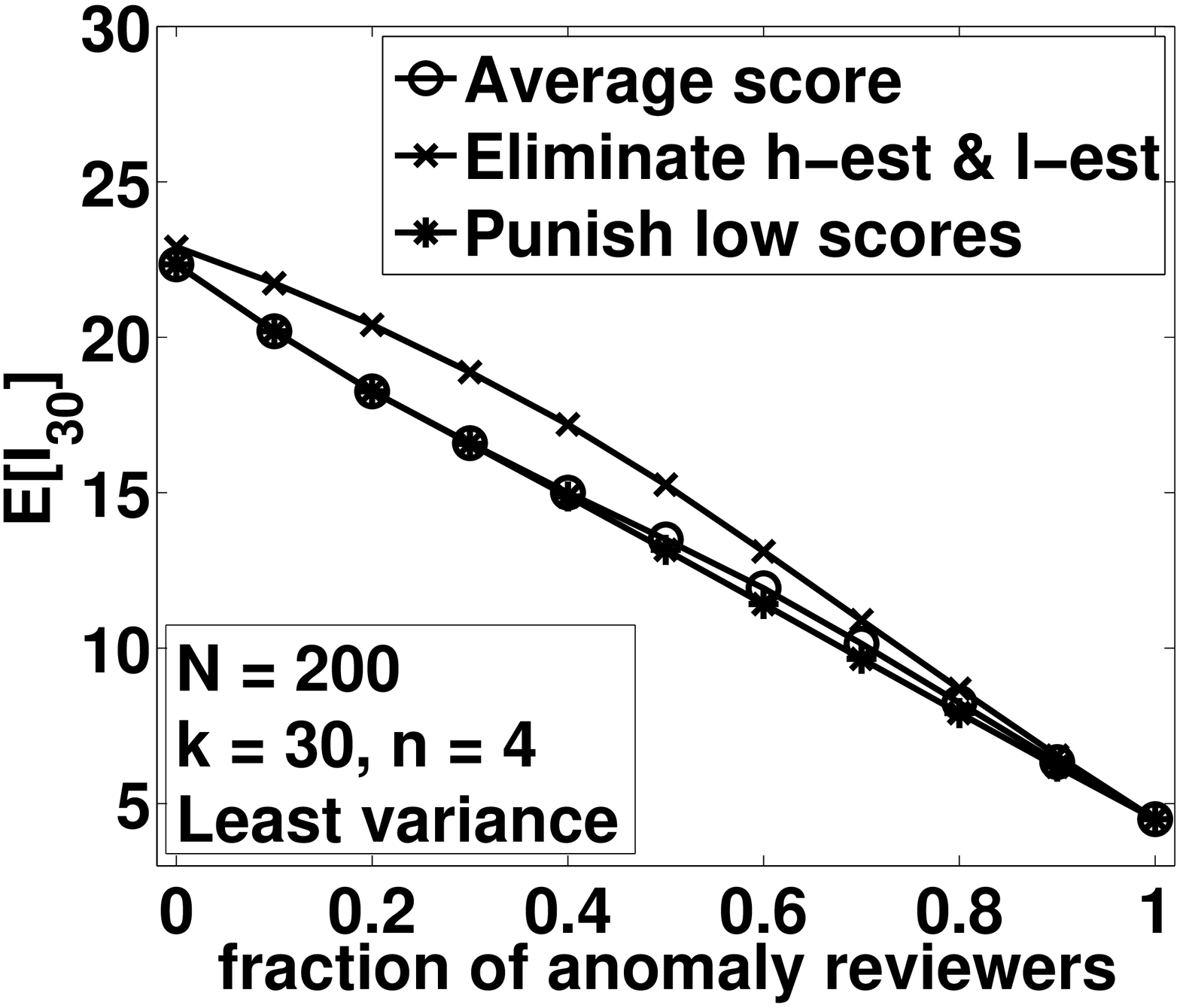}
 \label{fig:sec_anomaly_behav:voting_rule_com_exp_uniform}
 }
\caption{Robustness of three voting rules: $\SV_{as}$, $\SV_{ehl}$, and
$\SV_{pl}$ when $n \! = \! 4$.}
\label{fig:sec_anomaly_behav:voting_rule_com_exp}
\end{figure}

An interesting question is that which voting rule is more robust against this
type of misbehaving reviewers? Here we evaluate three voting rules: {\em
average score rule}, {\em eliminate highest and lowest} and {\em punish low
scores} given by Equation (\ref{equation:average_score_rule}) $-$
(\ref{equation:punish_low_score}). We use expectation as our performance
measure. The numerical results of $E[I_{30}]$ are shown in Fig.
\ref{fig:sec_anomaly_behav:voting_rule_com_exp}.

In Fig. \ref{fig:sec_anomaly_behav:voting_rule_com_exp}, the horizontal axis
represents of the fraction of misbehaving reviewers. The vertical axis shows
the corresponding expectation.  From Fig.
\ref{fig:sec_anomaly_behav:voting_rule_com_exp}, we have the following
observations. When we increase the fraction of anomaly reviewers, the
expectation decreased. From Fig.
\ref{fig:sec_anomaly_behav:voting_rule_com_exp_low}, we see that when submitted
papers are of low {\em self-selectivity}, the expectation curves corresponding
to these three voting rules overlapped together. In other words, for the low
{\em self-selectivity} papers, these three voting rules have the same
robustness. From Fig. \ref{fig:sec_anomaly_behav:voting_rule_com_exp_medium},
\ref{fig:sec_anomaly_behav:voting_rule_com_exp_uniform}, we see that when
submitted papers are of medium or random {\em self-selectivity}, eliminate
highest and lowest score rule are slightly more robust than the other two
rules. From Fig. \ref{fig:sec_anomaly_behav:voting_rule_com_exp_high}, we
observe that when submitted papers of high {\em self-selectivity}, these three
voting rules have the same robustness when the fraction of anomaly reviewer is
less than 30\%, and when it is higher than 30\%,  eliminate highest and lowest
voting rule are more robust than the other two rules.

\noindent {\bf Lessons learned:} Random anomaly behavior can significantly
affect the accuracy of the conference recommendation system. This is especially
true for submitted papers which are of high { \em self-selectivity} (or
prestigious conferences). The conference recommendation system suffers
significantly from this kind of anomaly behavior, say and 20\% of misbehaving
reviewers will reduce the probability of the best paper to be accepted to
around 0.5.
%most 60\% misbehaving reviewer can drastically disrupt the accuracy of the
%conference recommendation system.
These four voting rules have comparable robustness, no rule can outperform
others remarkably, thus to defend this kind of anomaly behavior by voting rules
may not be effective. Again, there are number of interesting questions to
explore, i.e., how to improve the accuracy of the conference recommendation
system?

\subsection{Effect of Anomaly Behavior -- Bias Scoring}

\begin{figure}[thb]
\centering
\subfigure[\# of top 1 papers get in]{
\includegraphics[width=0.40\textwidth]{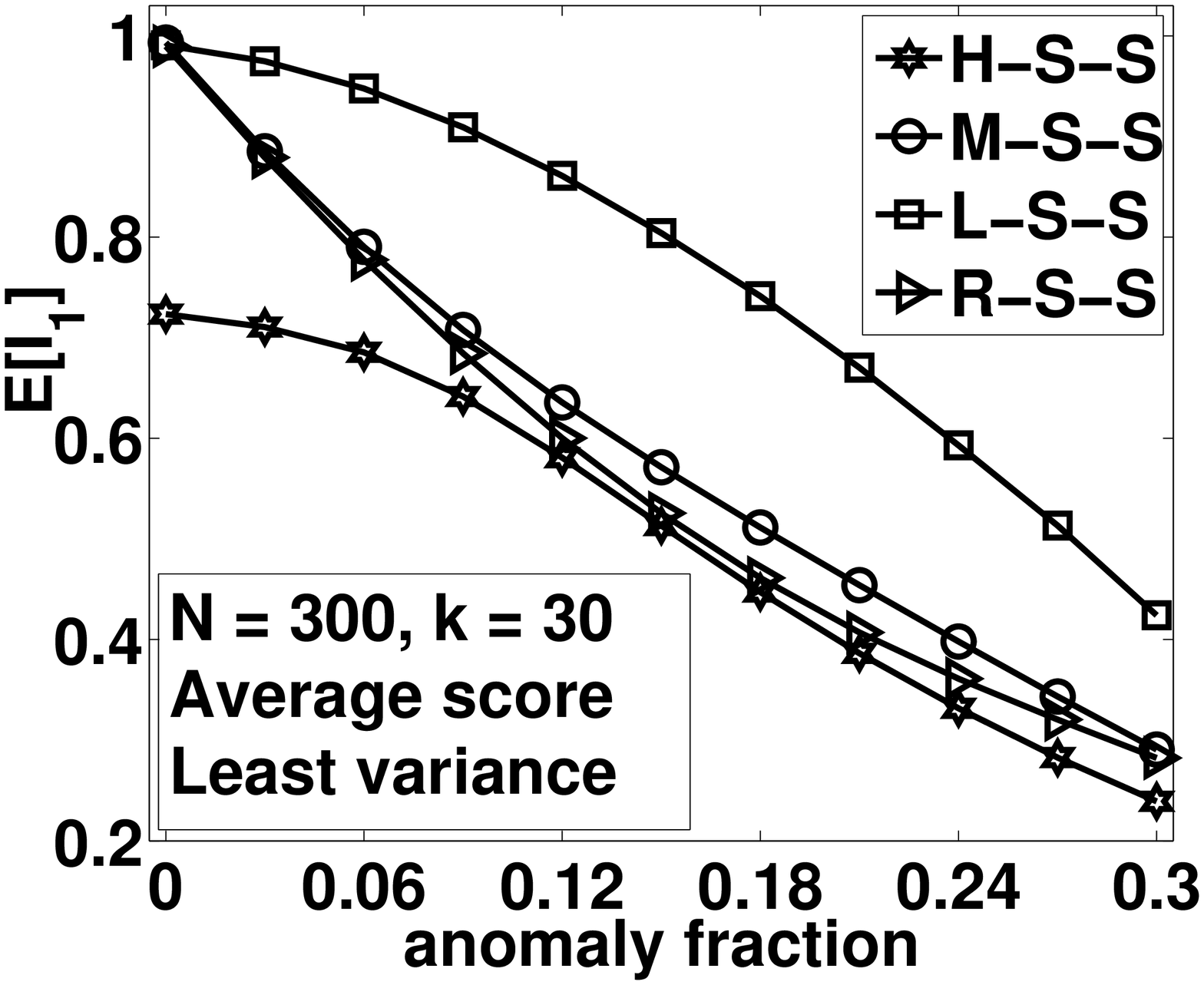}
\label{fig:sec_anomaly_behav_crazy:exp_top_1}
}
\subfigure[\# of top 5 papers get in]{
 \includegraphics[width=0.40\textwidth]{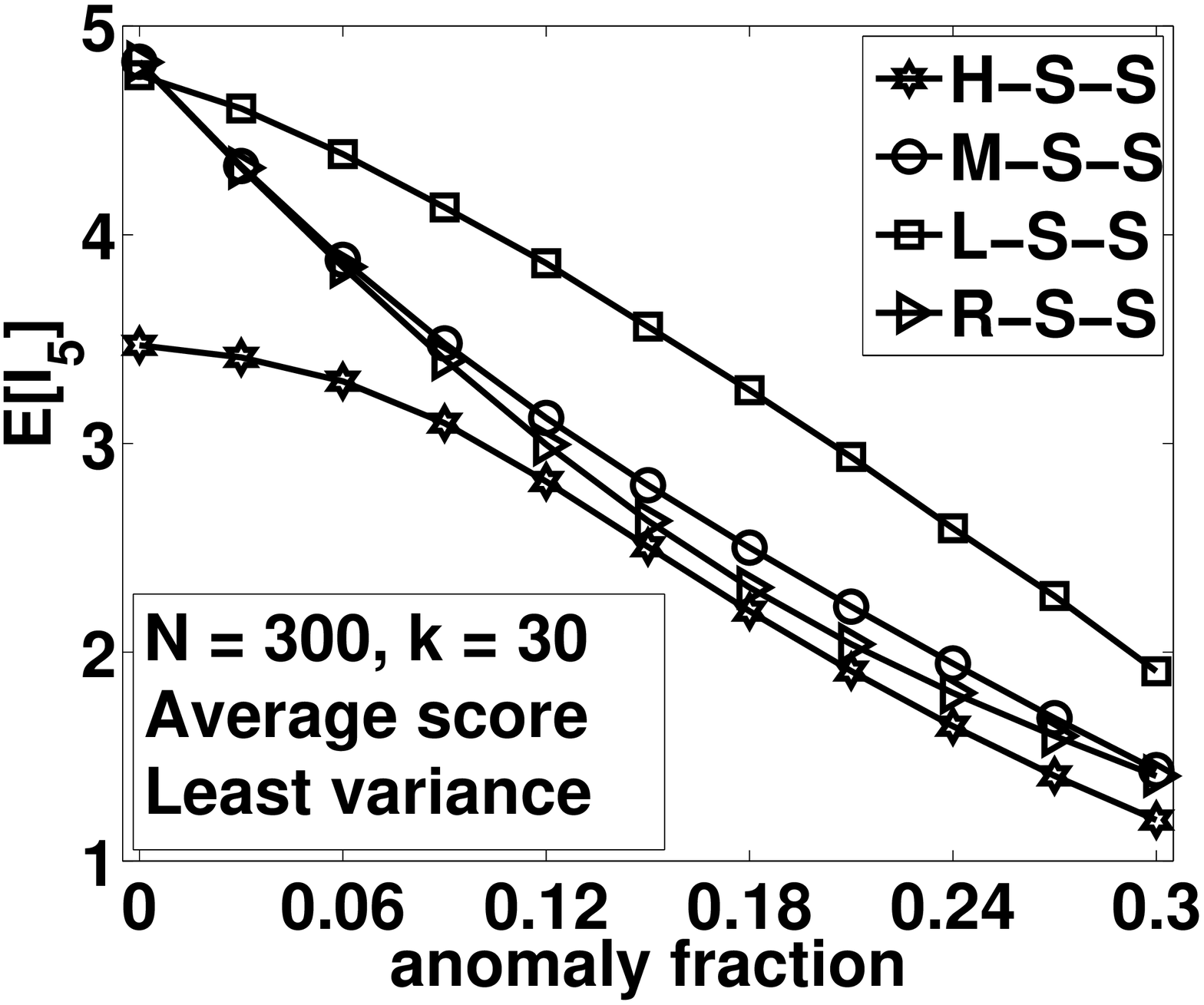}
 \label{fig:sec_anomaly_behav_crazy:exp_top_5}
 }
\\
\subfigure[\# of top 10 papers get in]{
\includegraphics[width=0.40\textwidth]{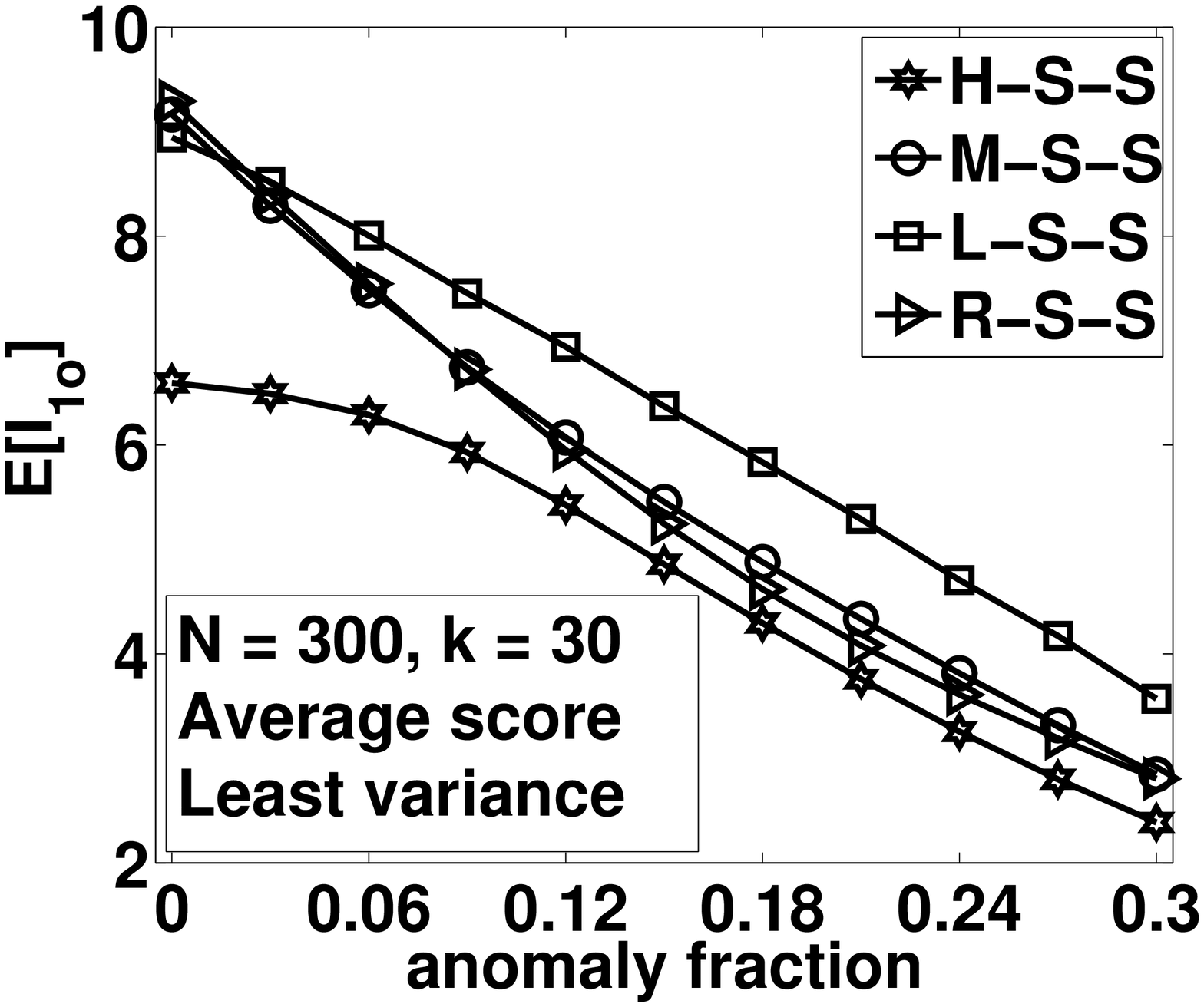}
\label{fig:sec_anomaly_behav_crazy:exp_top_10}
}
\subfigure[\# of top 30 papers get in]{
 \includegraphics[width=0.40\textwidth]{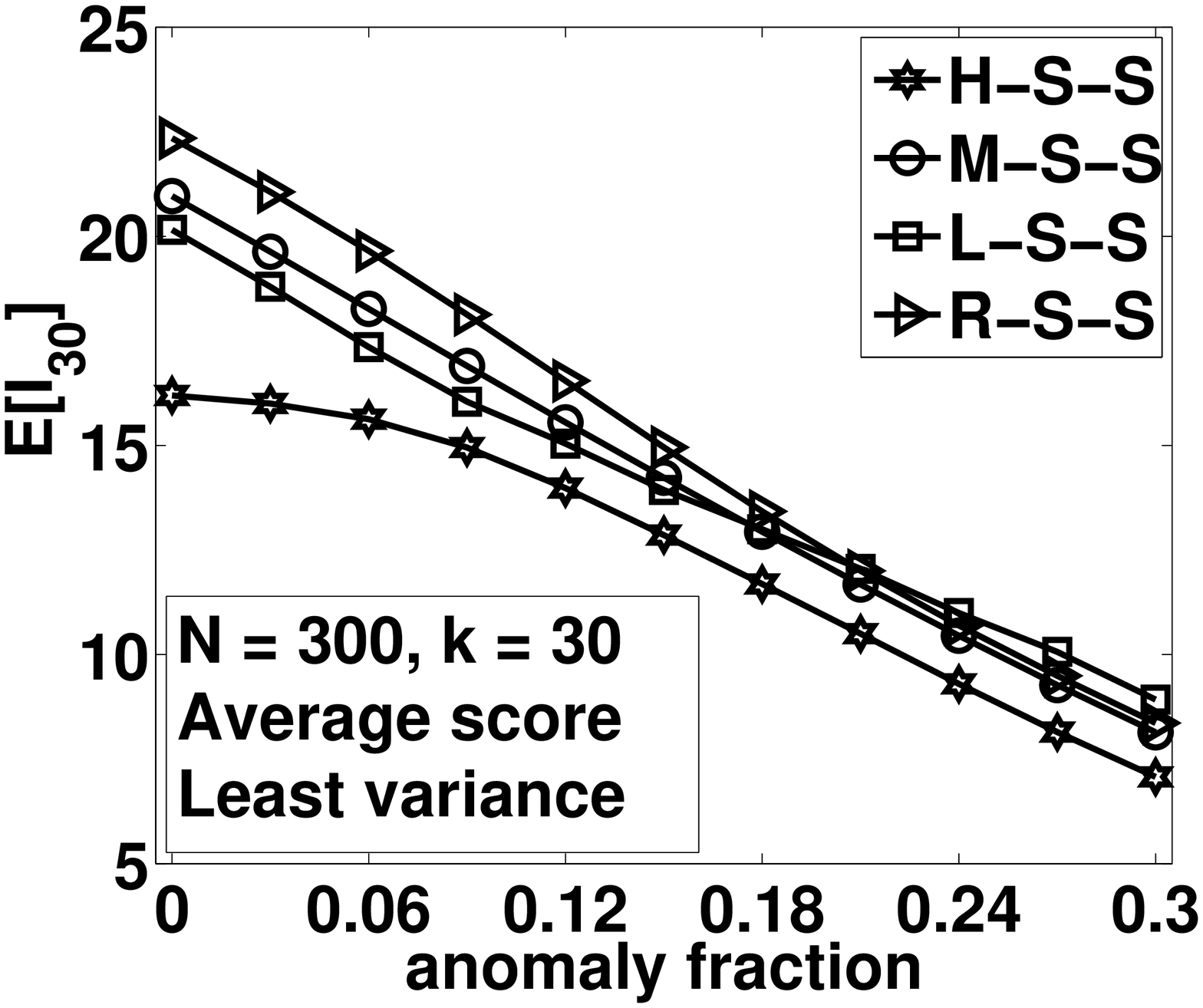}
 \label{fig:sec_anomaly_behav_crazy:exp_top_30}
 }
\caption{Impact of bias scoring anomaly behavior when $n \! = \! 4$.}
\label{fig:sec_anomaly_behav_crazy:exp}
\end{figure}

Let us explore the effect of another potential anomaly behavior on the accuracy
of a conference recommendation system. Specifically, we consider {\em bias
scoring behavior}, under which a misbehaving reviewer gives a high score $m$ to
a paper if the evaluated quality is low, say less than three, otherwise gives a
low score 1 to those papers whose evaluated quality is above three.
%To see the effect of this kind of anomaly behavior
%on the accuracy of conference recommendation system,
We vary the fraction of anomaly behavior from 0 to 0.3 and we use expectation
as our performance measure. We explore the effect of
this type of anomaly behavior on the
conference recommendation system specified in Section
\ref{section:prob_expec_var} with number of reviews per paper $ n \! = \! 4 $.
The numerical results of $E[I_1]$, $E[I_5]$, $E[I_{10}]$, and $E[I_{30}]$ are
shown in Fig. \ref{fig:sec_anomaly_behav_crazy:exp}.

In Fig. \ref{fig:sec_anomaly_behav_crazy:exp}, the horizontal axis represents
the fraction of these misbehaving reviewers. The vertical axis shows the
corresponding expectation. From Fig. \ref{fig:sec_anomaly_behav_crazy:exp}, we
have the following observations. When we increase the fraction of anomaly
reviewers slightly, the expectation decreased remarkably. From Fig.
\ref{fig:sec_anomaly_behav_crazy:exp_top_1}, we see that for low {\em
self-selectivity } papers, the chance of accepting the best paper can withstand
a small fraction of misbehaving reviewers, say around 6\%, but for the other
three {\em self-selectivity} types, even a small fraction of misbehaving
reviewers may lead to high inaccuracy. For example, when 6\% of reviewers are
misbehaving, the best paper only has less than 70\% chance of being accepted
for the {\em highly self-selective} paper submissions and it only has less than
80\% chance of being accepted for the submitted papers which are of medium or
random {\em self-selectivity}. Similar deterioration can be said for the top
five papers. Around 15\% of misbehaving reviewers can drastically disrupt the
accuracy of a conference recommendation system, since less than 15 papers from
top 30 papers will get accepted.

\noindent {\bf Lessons learned:} Bias scoring anomaly behavior can
significantly affect the accuracy of the conference recommendation system. A
small fraction of this kind of misbehaving reviewers can decrease the accuracy
of the conference recommendation system dramatically. This is especially true
for submitted papers which are of high, medium, or random { \em
self-selectivity} (prestigious, medium, or newly started conferences). The
conference recommendation system suffers severely from this kind of anomaly
behavior, say 15\% of misbehaving reviewers will disrupt the accuracy of the
conference recommendation system.

\subsection{Improving Conference Recommendation Systems}

In reality, most conferences use homogeneous review strategy, with which each
paper is reviewed by the {\em same number} of reviewers. One obvious advantage
of this strategy is its fairness for all papers. But its efficiency in using
the workload is low. Here we propose a {\em heterogeneous review strategy} that
that can increase the efficiency of homogeneous review strategy. In other
words, with the same reviewing workload $W$, it has a much higher chance to
finally include the top $k$ papers in the final acceptance recommendation.
Assume that the the total reviewing workload for the homogeneous review
strategy is $W=N*n$.  Our heterogeneous review strategy works in two rounds:

\noindent {\bf \em Round 1}: Eliminate half of the submitted papers using only
half of the workload. Specifically, each paper will receive $\lfloor n/2
\rfloor$ reviews in the first round. After this reviewing round, apply a voting
rule and tie breaking rule to eliminate $N/2$ papers.

\noindent {\bf \em Round 2}: Select $k$ papers from the remaining $N/2$ papers
to accept. Each paper entering this round will receive $2\lceil n/2 \rceil$
reviews. After the reviewing process finished, combine their reviews in round 1
and round 2. Then apply a voting rule and a tie breaking rule to select the top
$k$ papers to accept.

%Before showing our results, let us define:
\begin{definition}
Let $E[I_i \: | \: \mbox{hom}]$ and $E[I_i \: | \: \mbox{hetero}]$ represent
the expectation of $I_i$ under homogeneous or heterogeneous review strategy
applied respectively, where $I_i$ is defined by Definition
\ref{definition:I_i}.
\end{definition}

\begin{definition}
The improvement of heterogeneous review strategy over homogeneous review
strategy is:
%$E[I_i \: | \: \mbox{hom}]$ minus $E[I_i \: | \: \mbox{hetero}]$, or
\[
    \Delta E[I_i] = E[I_i \: | \: \mbox{hetero}] -  E[I_i \: | \: \mbox{hom}],
\]
and the improvement ratio is:
\[
    \Delta E[I_i] / E[I_i \: | \: \mbox{hom}].
\]
\end{definition}

\begin{figure}[htb]
\centering
\subfigure[Improvement]{
\includegraphics[width=0.42\textwidth]{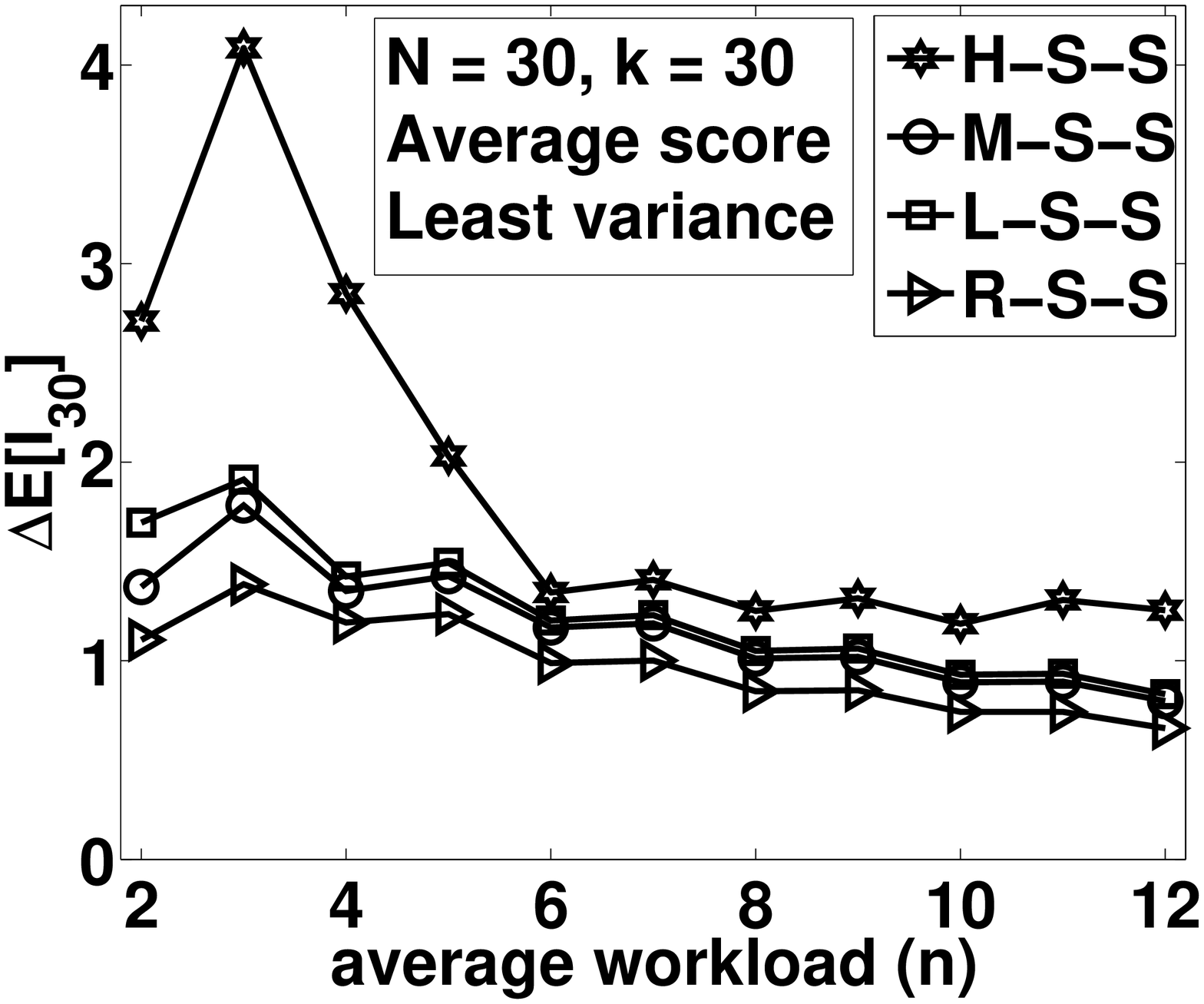}
\label{fig:sec_impro:hetero_improvement_abso}
}
\subfigure[Improvement ratio]{
 \includegraphics[width=0.42\textwidth]{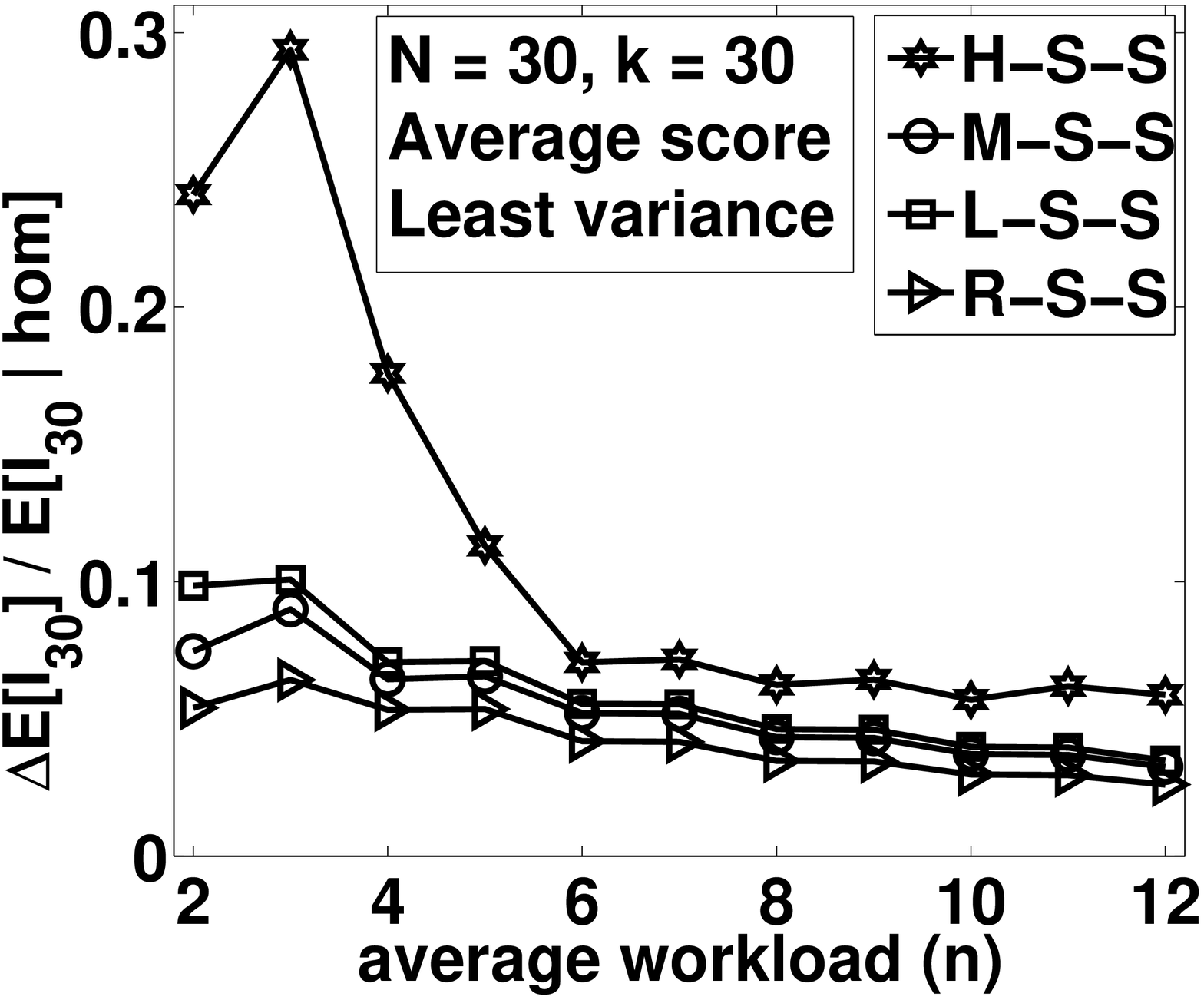}
 \label{fig:sec_impro:hetero_improvement_pct}
 }
\caption{Improvement of heterogeneous review strategy on homogeneous review strategy}
\label{fig:sec_impro:hetero_improvement_comparison}
\end{figure}

We evaluate these two strategies on a conference recommendation system
specified in Section \ref{section:prob_expec_var}. The numerical results of
$\Delta E[I_{30}]$ and $\Delta E[I_{30}] / E[I_{30} \: | \: \mbox{hom}]$ are
shown in Fig. \ref{fig:sec_impro:hetero_improvement_comparison}, where the
horizontal axis represents the average reviewing workload $n$. The vertical
axis shows the corresponding improvement or improvement ratio.  From Fig.
\ref{fig:sec_impro:hetero_improvement_comparison}, we see an improvement of
heterogeneous review strategy over homogeneous review strategy. When the
reviewing workload is small, say $n\!=\!3$, with heterogeneous review strategy
at least one more paper from the top 30 papers will get accepted. For papers
with high {\em self-selectivity}, we have more improvement wherein three or
more papers from the top 30 papers will get accepted. When the average
reviewing workload increased to six, the improvement becomes stabilized. When
the reviewing workload is three, the improvement is the highest, or around four
more papers from the top 30 papers will get accepted for papers submitted with
high {\em self-selectivity}, and the improvement ratio for this case is around
30\%.

\begin{figure}[htb]
\centering
\subfigure[\# of top 1 papers get in]{
\includegraphics[width=0.42\textwidth]{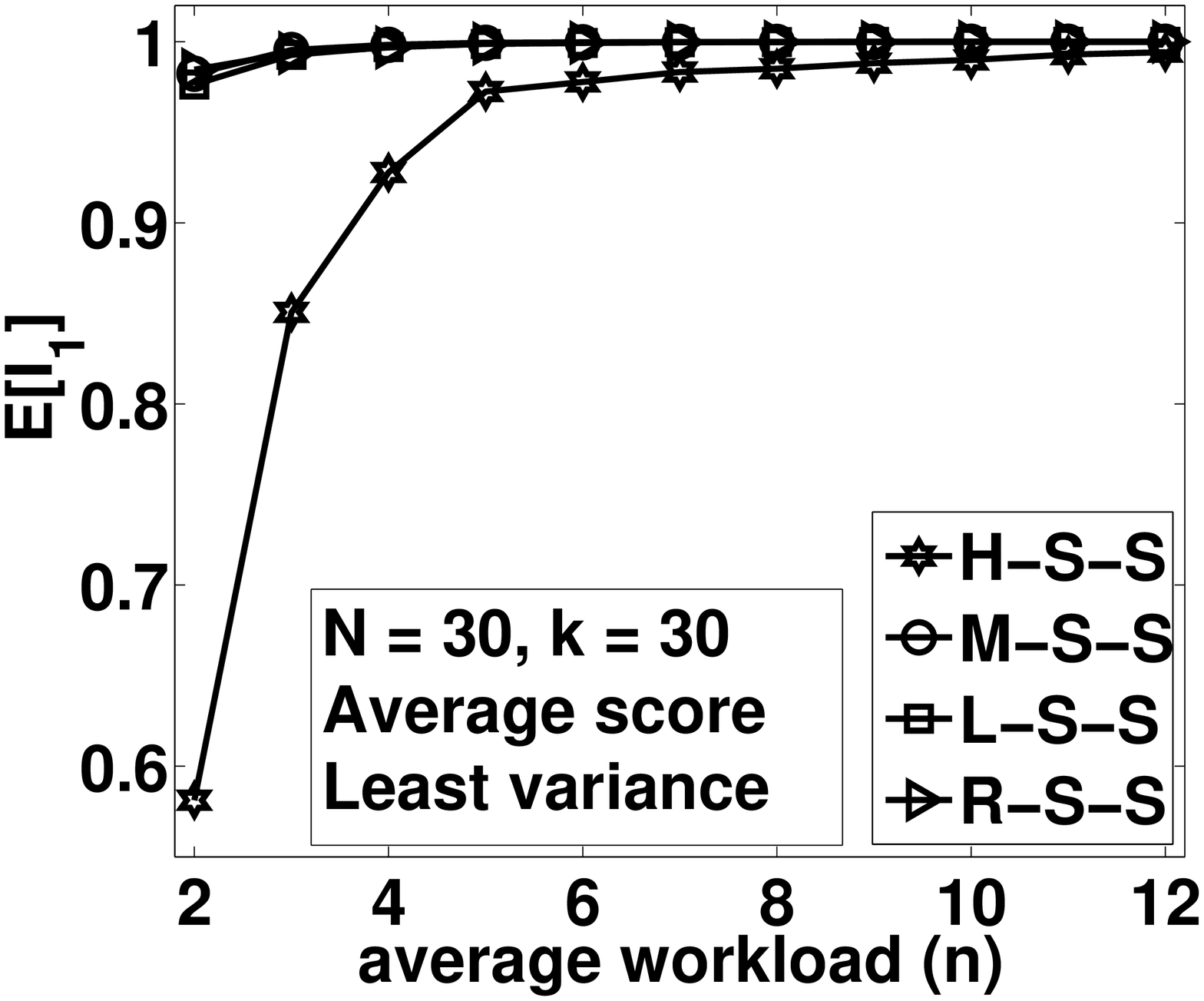}
\label{fig:sec_impro:hetero_review_top_1}
}
\subfigure[\# of top 5 papers get in]{
 \includegraphics[width=0.42\textwidth]{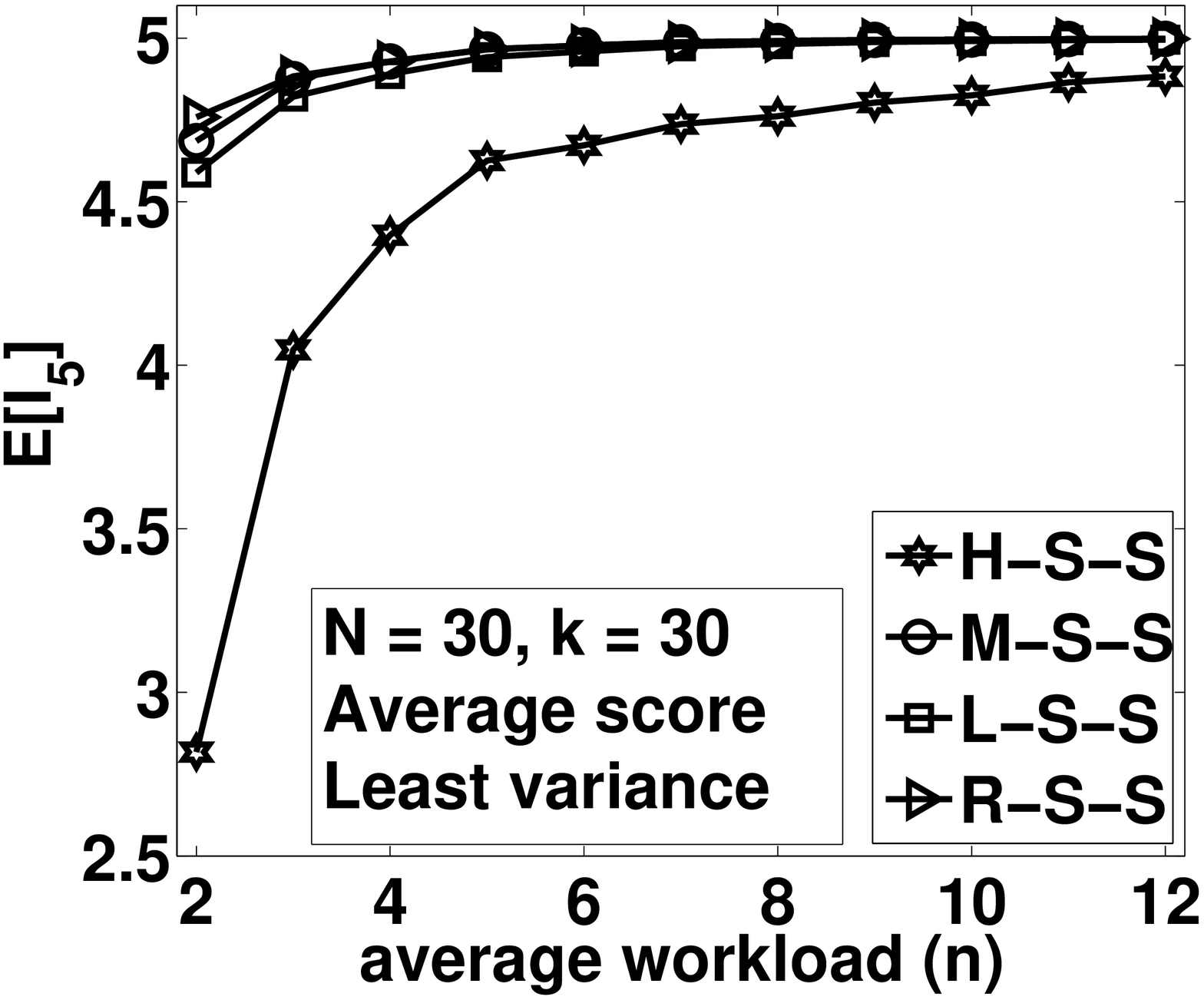}
 \label{fig:sec_impro:hetero_review_top_5}
 }
\\
\subfigure[\# of top 10 papers get in]{
\includegraphics[width=0.42\textwidth]{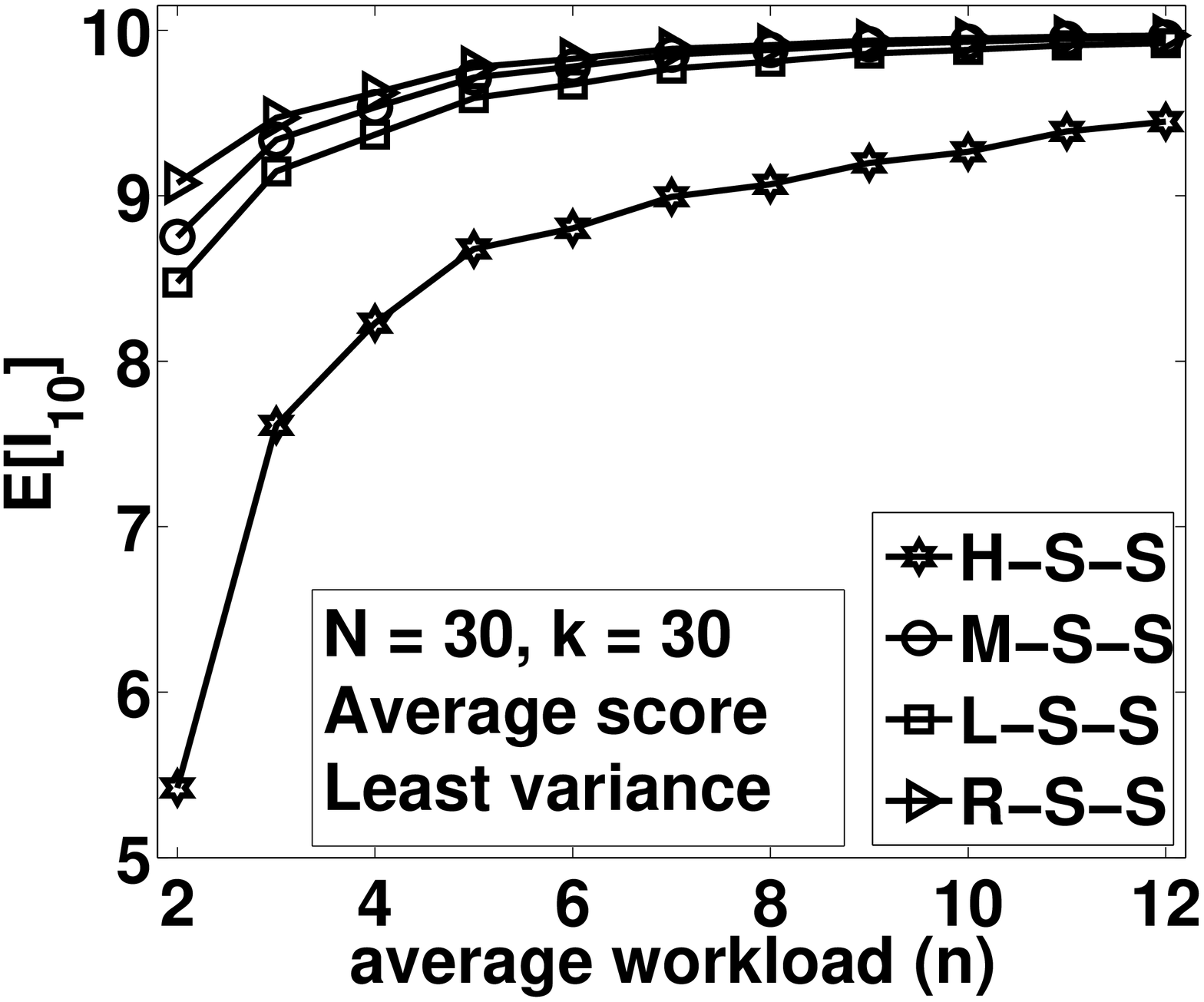}
\label{fig:sec_impro:hetero_review_top_10}
}
\subfigure[\# of top 30 papers get in]{
 \includegraphics[width=0.42\textwidth]{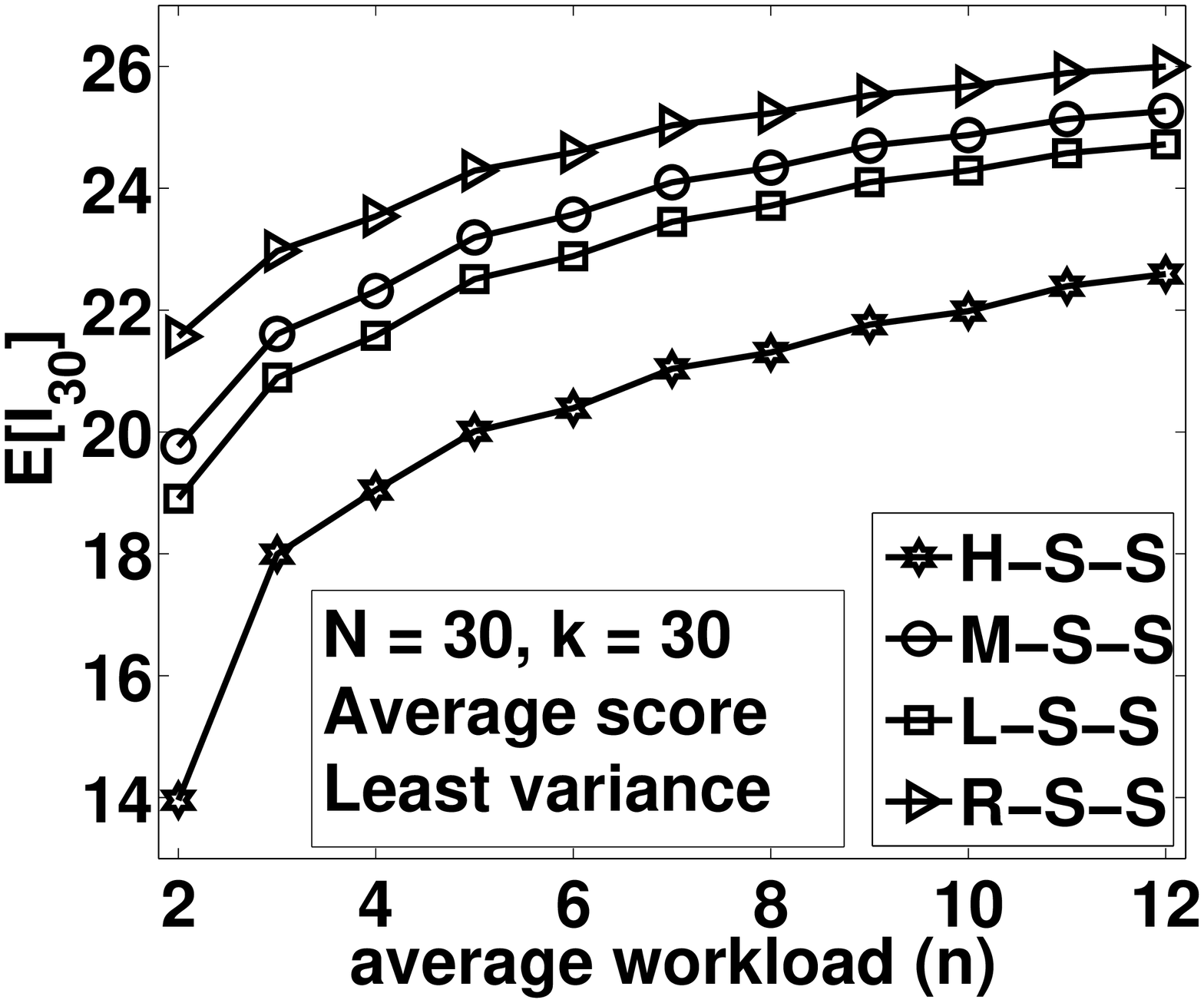}
 \label{fig:sec_impro:hetero_review_top_30}
 }
\caption{Expectation of $I_1$, $I_5$, $I_{10}$ and $I_{30}$ with heterogeneous review strategy.}
\label{fig:sec_impro:hetero_review}
\end{figure}

An interesting question is that with heterogeneous review strategy, how large
the average reviewing workload do we need. We apply our heterogeneous strategy
to the conference recommendation system specified in Section
\ref{section:workload}. We use expectation as performance measure. The
numerical results of $E[I_1]$, $E[I_5]$, $E[I_{10}]$ and $E[I_{30}]$ are shown
in Fig. \ref{fig:sec_impro:hetero_review}, where the horizontal axis represents
the average reviewing workload $n$. The vertical axis shows the corresponding
expectation.  From Fig. \ref{fig:sec_impro:hetero_review}, we see that we need
to increase the workload to at least five such that we have a strong guarantee
that the best paper will get accepted, which {\em reduces} the average review
workload by two as compared with the homogeneous review strategy, in which we
need to increase the average reviewing workload to at least seven as stated in
Section \ref{section:workload}.

\noindent {\bf Summary:} \noindent {\bf Lessons learned:} Our heterogeneous
review strategy uses equal or less resource (e.g., reviewing workload) than the
homogeneous review strategy and at the same time, achieve higher accuracy.

\section{Related Work}
\label{section: related work}

In \cite{auto_assign_manuscript, conf_paper_reviewer_assign_problem,
committee_reviewer_assign,a_algo_for_deter_peer_reviewer}, authors studied peer
review systems. Typically, the main issue is the reviewer assignment problem
which contains three phrases: specifying the assignment constraint, computing
the matching degree between reviewers and submissions, and optimizing the
assignment with constraints. Disciplines like information retrieval
\cite{auto_assign_manuscript}, artificial intelligence
\cite{min_reviewer_for_prosal,a_algo_for_deter_peer_reviewer} and operations
research \cite{opt_allocate_proposal, conf_paper_reviewer_assign_problem,
committee_reviewer_assign}, etc address these assignment problems.

Authors in \cite{rat_scale_meta_analysis, rat_scale_optmal,
rating_how_diff_is_it, rat_scale_comp_contin_discrete} worked on the {\em group
recommendation systems} and address issues on rating scale and
\cite{group_rec_sys_prefer_agg, group_rec_sys_opt_pre,
group_rec_sys_creat_group_ranking} on preference aggregation. Rating is used to
show individuals' preferences, and in \cite{rat_scale_comp_contin_discrete},
authors stated that discrete rating scales (number of rating points) outperform
continuous rating scales. In \cite{rat_scale_meta_analysis}, authors evaluated
the reliability of rating scales and showed evidence that more rating points
will have a more reliable rating. In \cite{rat_scale_optmal}, authors stated
that the best rating scale is around five to ten rating points. Preference
aggregation is the process to merge the preference of multiple people so as to
make recommendations. Basically the aggregation method can be divided into two
classes based on the preference type: cardinal ranking or ordinal ranking. For
cardinal ranking case, {\em weighted average strategy}\cite{group_modeling} is
the most popular strategy, and it is used in {\em PolyLen}. The second class is
the ordinal ranking preference, for which, each individual's preference is
shown by a ranked list of a subset of the candidates. For this case, users'
preferences are treated as a set of constraints and a preference aggregation
approach attempts to find recommendations that satisfy the constraints of all
users\cite{group_rec_sys_prefer_agg, group_rec_sys_creat_group_ranking,
group_rec_sys_opt_pre}. As far as we know, our work is the first that study the
mathematical modeling of competitive group recommendation systems and apply it
to peer review systems.

\section{Conclusions}
\label{section: conclusion}

This is the first paper that provides a mathematical model and analysis on a
competitive group recommendation system. We apply it to a conference peer
review system and show how various factors may influence the overall accuracy
of the final recommendation. We formally analyze the model and through this
analysis, we gain the insight on developing a randomized algorithm, which is
both computationally efficient and can provide performance guarantees on
various performance measures. Number of interesting observations are found,
e.g., for a medium tier conference, three reviews per paper are sufficient to
achieve high accuracy in the final recommendation, but for some prestigious
conferences, we need at least seven reviews per paper. Lastly, we propose a
heterogeneous review strategy that requires equal or less reviewing workload
but can produce a more accurate recommendation than the homogeneous review
strategy. We believe our model and methodology are important building blocks
for researchers to study competitive group recommendation systems.

%{\small
\bibliographystyle{abbrv}
\bibliography{peer_review_modeling}

\begin{thebibliography}{10}

\bibitem{group_rec_sys_sem_eff}
S.~Amer-Yahia, S.~Roy, A.~Chawlat, G.~Das, and C.~Yu.
\newblock Group recommendation: Semantics and efficiency.
\newblock {\em Proc. of VLDB, 2009}.

\bibitem{group_rec_sys_prefer_agg}
J.~Baskin and S.~Krishnamurthi.
\newblock Preference aggregation in group recommender systems for committee
  decision-making.
\newblock In {\em Proc. of ACM RecSys, 2009}.

\bibitem{group_rec_sys_state_of_the_art}
L.~Boratto and S.~Carta.
\newblock State-of-the-art in group recommendation and new approaches for
  automatic identification of groups.
\newblock {\em Information Retrieval and Mining in Distributed Environments},
  pages 1--20, 2011.

\bibitem{rat_scale_meta_analysis}
G.~Churchill~Jr and J.~Peter.
\newblock Research design effects on the reliability of rating scales: a
  meta-analysis.
\newblock {\em Journal of Marketing Research}, pages 360--375, 1984.

\bibitem{opt_allocate_proposal}
W.~Cook, B.~Golany, M.~Kress, M.~Penn, and T.~Raviv.
\newblock Optimal allocation of proposals to reviewers to facilitate effective
  ranking.
\newblock {\em Management Science}, pages 655--661, 2005.

\bibitem{group_rec_sys_creat_group_ranking}
W.~Cook, B.~Golany, M.~Penn, and T.~Raviv.
\newblock Creating a consensus ranking of proposals from reviewers partial
  ordinal rankings.
\newblock {\em Computers \& operations research}, 34(4):954--965, 2007.

\bibitem{rec_sys_item_top_n_algo}
M.~Deshpande and G.~Karypis.
\newblock Item-based top-n recommendation algorithms.
\newblock {\em ACM TOIS}, 22(1):143--177, 2004.

\bibitem{auto_assign_manuscript}
S.~Dumais and J.~Nielsen.
\newblock Automating the assignment of submitted manuscripts to reviewers.
\newblock In {\em Proc. of ACM SIGIR, 1992}.

\bibitem{conf_paper_reviewer_assign_problem}
D.~Hartvigsen, J.~Wei, and R.~Czuchlewski.
\newblock The conference paper-reviewer assignment problem.
\newblock {\em Decision Sciences}, 30(3):865--876, 1999.

\bibitem{rec_sys_eval_coll_fil_algo}
J.~Herlocker, J.~Konstan, L.~Terveen, and J.~Riedl.
\newblock Evaluating collaborative filtering recommender systems.
\newblock {\em ACM TOIS}, 22(1):5--53, 2004.

\bibitem{min_reviewer_for_prosal}
S.~Hettich and M.~Pazzani.
\newblock Mining for proposal reviewers: lessons learned at the national
  science foundation.
\newblock In {\em Proc. of SIGKDD'06}.

\bibitem{group_rec_sys_challenge}
A.~Jameson.
\newblock More than the sum of its members: challenges for group recommender
  systems.
\newblock In {\em Proc. of ACM working conference on Advanced visual
  interfaces, 2004}.

\bibitem{travel_decision_forum}
A.~Jameson, S.~Baldes, and T.~Kleinbauer.
\newblock Two methods for enhancing mutual awareness in a group recommender
  system.
\newblock In {\em Proceedings of ACM working conference on Advanced visual
  interfaces, 2004}.

\bibitem{committee_reviewer_assign}
M.~Karimzadehgan and C.~Zhai.
\newblock Constrained multi-aspect expertise matching for committee review
  assignment.
\newblock In {\em Proc. of CIKM'09}.

\bibitem{group_rec_sys_opt_pre}
F.~Lorenzi, F.~dos Santos, P.~Ferreira, and A.~Bazzan.
\newblock Optimizing preferences within groups: a case study on travel
  recommendation.
\newblock {\em Advances in Artificial Intelligence-SBIA}, pages 103--112, 2008.

\bibitem{group_modeling}
J.~Masthoff.
\newblock Group modeling: Selecting a sequence of television items to suit a
  group of viewers.
\newblock {\em User Modeling and User-Adapted Interaction}, 14(1):37--85, 2004.

\bibitem{probability_and_computing}
M.~Mitzenmacher and E.~Upfal.
\newblock {\em Probability and computing: Randomized algorithms and
  probabilistic analysis}.
\newblock Cambridge Univ Pr, 2005.

\bibitem{PolyLen}
M.~O¡¯connor, D.~Cosley, J.~Konstan, and J.~Riedl.
\newblock Polylens: A recommender system for groups of users.
\newblock In {\em Proc. of ECSCW 2001}.

\bibitem{rat_scale_optmal}
C.~Preston and A.~Colman.
\newblock Optimal number of response categories in rating scales: reliability,
  validity, discriminating power, and respondent preferences.
\newblock {\em Acta psychologica}, 104(1):1--15, 2000.

\bibitem{rec_sys}
P.~Resnick and H.~Varian.
\newblock Recommender systems.
\newblock {\em Communications of the ACM}, 40(3):56--58, 1997.

\bibitem{a_algo_for_deter_peer_reviewer}
M.~Rodriguez and J.~Bollen.
\newblock An algorithm to determine peer-reviewers.
\newblock In {\em Proc. of ACM CIKM, 2008}.

\bibitem{rec_sys_rec_algo}
B.~Sarwar, G.~Karypis, J.~Konstan, and J.~Reidl.
\newblock Item-based collaborative filtering recommendation algorithms.
\newblock In {\em Proc. of WWW'01}.

\bibitem{rec_sys_ec}
J.~Schafer, J.~Konstan, and J.~Riedi.
\newblock Recommender systems in e-commerce.
\newblock In {\em Proc. of ACM EC, 1999}.

\bibitem{rating_how_diff_is_it}
E.~I. Sparling and S.~Sen.
\newblock {Rating: how difficult is it?}
\newblock In {\em Proc. of ACM RecSys, 2011}.

\bibitem{rat_scale_comp_contin_discrete}
E.~Svensson.
\newblock Comparison of the quality of assessments using continuous and
  discrete ordinal rating scales.
\newblock {\em Biometrical Journal}, 42(4):417--434, 2000.

\end{thebibliography}
%}

\clearpage
\section*{Appendix}

\begin{theorem}
\emph{(Chernoff Bound\cite{probability_and_computing})} Let $X_1, \ldots, X_n$
be independent random variables with $X_i = 1$ with probability $p$ and 0
otherwise. Let $X = \sum_{i=1}^n X_i$ and let $\mu = E[X]= np$. Then we have
\[
    \MP[ |X-\mu|\geq \epsilon \mu ] \leq 2e^{-\mu \epsilon^2/3}.
\]
\label{theorem:chernoff_bound}
\end{theorem}

In the following lemma we derive a loose bound on the number of simulation
rounds $K$ but have a good performance guarantee on the pmd of $I(k)$, or
$\MP[I(k) = i], \forall i$ by carefully applying Theorem
\ref{theorem:chernoff_bound}.

\begin{lemma}
\label{lemma:prob_loose_round} When the following condition holds:
\begin{equation}
\label{inequality:loose_round}
    K \geq \max\nolimits_{ \begin{subarray}{c}
                                i \in\{0,\ldots,k\}   \\
                                \text{Pr}[I(k)=i]\neq 0
                            \end{subarray}
                           }
    \frac{3\ln(2( k+1) / \delta )}{ \MP[I(k)=i]\epsilon^2} ,
\end{equation}
then for each $i = 0, 1, \ldots, k$,
\[
  \left| \HP[I(k)=i] - \MP[I(k)=i] \right| \leq \epsilon \MP[I(k)=i],
\]
holds with probability at least $1 - \delta/(k+1)$.
\end{lemma}

\noindent {\bf Proof:} Without lose of any generality, consider the performance
guarantee on the approximation of $\MP[I(k) = i]$, where $i = 0, 1, \ldots, k$.
Our goal is to show
\[
    \left| \HP[I(k)=i] - \MP[I(k)=i] \right| \leq \epsilon \MP[I(k)=i].
\]

Let $\mathbf{I}_{ij}$ be an indicator random variable defined by
\[
\mathbf{I}_{ij}
= \left\{
    \begin{array}{ll}
        1 & \mbox{if in $j$-th round, $|\mathcal{A}^I(k) \cap \mathcal{A}(k)| = i $}\\
        0 & \mbox{otherwise}
    \end{array},
\right.
\]
where $j =1 , \ldots, K$. There are two cases to explore:
\\
{\bf Case 1}: $\MP[I(k) = i] = 0$. The physical meaning implies that the event
$I(k) =i$ never happen, which result in $\mathbf{I}_{ij} = 0 $ for all $j = 0,
1, \ldots, K$. Hence, $\HP[I(k) =i] = \sum_{j=1}^K \mathbf{I}_{ij} / K = 0$.
Then we have
\[
  \left| \HP[I(k)=i] - \MP[I(k)=i] \right| \leq \epsilon \MP[I(k)=i].
\]
\\
{\bf Case 2}: $\MP[I(k) = i] \neq 0$. Since each round runs independently, thus
random variables $\mathbf{I}_{i1}, \ldots, \mathbf{I}_{iK}$ are independent
random variables with $\mathbf{I}_{ij} = 1$ with probability $\MP[I(k) = i]$
and 0 otherwise. Let $\mathbf{I}_i=\sum_{j=1}^K \mathbf{I}_{ij}$. Then $E[
\mathbf{I}_i ] = K \MP[I(k) = i]$. From Algorithm
\ref{algorithm_monte_carlo_algorithm} we could see that $\HP[I(k) = i]$, the
approximate value of $\MP[I(k) = i]$, is given by $ \HP[I(k) = i] =
\mathbf{I}_i / K$. Then by applying Theorem \ref{theorem:chernoff_bound} we
have,
\begin{align*}
& \MP\left[| \HP[I(k)=i]-\MP[I(k) =i] | \geq \epsilon \MP[I(k) =i] \right]  \\
& = \MP \left[ | \mathbf{I}_i - K\MP [I(k) = i]| \geq \epsilon K \MP[I(k)=i] \right]    \\
& \leq 2e^{-K \text{Pr}[I(k) =i] \epsilon^2/3},
\end{align*}
by substituting $K$ with Inequality (\ref{inequality:loose_round}), we have
\begin{align*}
   &\MP \left[ |\HP[I(k) = i]-\MP[I(k) = i]|\geq \epsilon \MP[I(k) = i]\right] \\
   &\leq \delta / (k+1).
\end{align*}
Finally the proof of this lemma can be completed by
\begin{align*}
   &\MP \left[ |\HP[I(k) = i]-\MP[I(k) = i]| \leq \epsilon \MP[I(k) = i]\right]  \\
   &\geq 1 - \MP \left[ |\HP[I(k) = i]-\MP[I(k) = i]|\geq \epsilon \MP[I(k) = i]\right] \\
   &\geq 1 - \delta / (k + 1).
\end{align*}
This lemma is proved. \done

Lemma \ref{lemma:prob_loose_round} shows the performance guarantee on $\MP[I(k)
= i]$ with success probability at least $1 - \delta/(k+1)$ for each specific $i
= 0, 1, \ldots, k$. Then, what is the success probability for all $i = 0, 1,
\ldots, k$? The answer of this question is stated in the following lemma.

\begin{lemma}
\label{lemma:joint_prob_loose_round} When the following condition holds:
\[
   K \geq  \max\nolimits_{ \begin{subarray}{c}
                                            i \in\{0,\ldots,k\}   \\
                                            \text{Pr}[I(k)=i]\neq 0
                                        \end{subarray}
                                     }
                   \frac{3\ln(2( k+1) / \delta )}{ \MP\:[I(k)=i]\epsilon^2} ,
\]
then
\[
    \left|\HP[I(k) = i] - \MP[I(k) = i] \right|  \leq \epsilon \MP[I(k) = i],
\]
holds for all $x = 0, 1, \ldots, k$, with probability at least $1 - \delta$.
\end{lemma}

\noindent {\bf Proof:} Let $E_i$ denote the event that $|\HP[I(k) = i] -
\MP[I(k) = i]|  \leq \epsilon \MP[I(k) = i]$ holds. From Lemma
\ref{lemma:prob_loose_round} we see that for each $i = 0, 1, \ldots, k$, the
condition
\[
    |\HP[I(k) = i] -\MP[I(k) = i]| \leq \epsilon \MP[I(k) = i]
\]
holds with probability at least $1-\delta/(k+1)$, thus
\[
    \MP[E_i] \geq 1-\delta/(k+1).
\]
Our goal is to derive the probability that condition
\[
    |\HP[I(k) = i] -\MP[I(k) = i]| \leq \epsilon \MP[I(k) = i]
\]
holds for all $i = 0, 1, \ldots, k$. Specifically, the probability that events
$E_0, E_1, \ldots, E_k$ all happens, or
\[
    \MP [E_0, \ldots, E_k].
\]

Note that the physical meaning of $E_i$ is that $\HP[I(k) = i]$, the
approximate value of $\MP[I(k) = i]$, is close to the true value of $\MP[I(k) =
i]$. Thus the physical meaning of $\cap_{i \in \SF} E_i$, where $\SF \subseteq
\{0, \ldots, k \}$, is that the approximate value of $\MP[I(k) = i]$, is close
to the true value of $\MP[I(k) = i]$, for all $i \in \SF$. Thus $\cap_{i \in
\SF} E_i$ happens implies that $\sum_{i \in \SF} \HP[I(k) = i]$, the
approximate value of $\sum_{i \in \SF} \MP[I(k) = i]$ is close to the real
value of $\sum_{i \in \SF} \MP[I(k) = i]$. Since
\[
    \sum\nolimits_{i \in \overline{\SF}} \HP[I(k) = i] -
    \sum\nolimits_{i \in \overline{\SF}} \MP[I(k) = i]
    = \sum\nolimits_{i \in \SF } \HP[I(k) = i] -
    \sum\nolimits_{i \in \SF} \MP[I(k) = i],
\]
where $\overline{\SF} = \{0, 1, \ldots, k\} \backslash \SF$ is the complement
of $\SF$, thus we have that $\sum_{i \in \SF} \HP[I(k) = i]$ is close to the
real value of $\sum_{i \in \SF} \MP[I(k) = i]$ if and only if $\sum_{i \in
\overline{\SF}} \HP[I(k) = i]$ is close the real value of $\sum_{i \in
\overline{\SF}} \MP[I(k) = i]$. Thus $\cap_{i \in \SF} E_i$ implies that
$\sum_{i \in \overline{\SF}} \HP[I(k) = i]$ is close the real value of $\sum_{i
\in \overline{\SF}} \MP[I(k) = i]$.

The event $\cap_{i \in \SG }E_i$, where $\SG \subseteq \overline{\SF}$, is more
likely to happen given the prior information that $\sum_{i \in \overline{\SF}}
\HP[I(k) = i]$ is close the real value of $\sum_{i \in \overline{\SF}} \MP[I(k)
= i]$ than given noting at all. Since $\cap_{i \in \SF} E_i$ implies that
$\sum_{i \in \overline{\SF}} \HP[I(k) = i]$ is close the real value of $\sum_{i
\in \overline{\SF}} \MP[I(k) = i]$, thus  $\cap_{i \in \SG }E_i$, where $\SG
\subseteq \overline{\SF}$, is more likely to happen given prior information
that $\cap_{i \in \SF} E_i$  happens than given nothing at all. Or
mathematically,
\[
    \MP[ \cap_{i \in \SG } E_i | \cap_{i \in \SF} E_i] \geq \MP[ \cap_{i \in \SG } E_i ],
\]
where $\SG \subseteq \overline{\SF}$. Based on this fact, we have
\[
    \MP [E_0, \ldots, E_k] = \prod\nolimits_{i=0}^k \MP[E_i | \cap_{j \in \SF_i } E_j] \geq
\prod\nolimits_{i=0}^k \MP [E_i ],
\]
where $\SF_i =\{0, \ldots, k\}/\{0, \dots, i\}$. By substituting $\MP [E_i]$
with $\MP [E_i] \geq 1 - \delta/(k+1)$ we have
\[
    \MP[E_0, \ldots, E_k] \geq [1 - \delta/(k+1)]^{k + 1} \geq 1 - \delta,
\]
which completes the proof. \done

In the following lemma we derive the error bound of expectation, or $|\HE[I(k)]
- E[I(k)]|$, under the condition that $|\HP[I(k) = i]-\MP[I(k) = i]|\leq
\epsilon \MP[I(k) = i]$ holds for all $i = 0, 1, \ldots, k$.

\begin{lemma}
\label{lemma:expectation_loose_round} When the following condition holds:
\[
    \left|\HP[I(k) = i] - \MP[I(k) = i] \right| \leq \epsilon \MP[I(k) = i],
\]
for all $i = 0, 1, \ldots, k$, then
\[
   \left|\HE[I(k)] - E[I(k)] \right| \leq \epsilon E[I(k)]
\]
holds.
\end{lemma}

\noindent {\bf Proof:} The proof is quite straightforward,
\begin{align*}
\left| \HE[I(k)] \!-\! E[I(k)] \right|
& \!=\! \left|\sum\nolimits_{i=0}^k i \left( \HP[I(k) \!=\! i] \!-\! \MP[I(k) \!=\! i] \right)\right| \\
& \!\leq\! \sum\nolimits_{i=0}^k i \left| \HP[I(k) \!=\! i] \!-\! \MP[I(k) \!=\! i] \right| \\
& \!\leq\! \sum\nolimits_{i=0}^k  \epsilon i \MP[I(k) = i] = \epsilon E[I(k)],
\end{align*}
which complets the proof. \done

In the following, we let $\Delta E[I(k)] =  \HE[I(k)] - E[I(k)] $ and $\Delta
p_i = \HP[I(k) = i] - \MP[I(k) = i]$. In the following we first derive the
bound of  $(\Delta E[I(k)])^2$, and  then apply the bound of $(\Delta
E[I(k)])^2$ to derive the bound of variance, or $| \HV[I(k)] - \MV[I(k)]|$. The
bound of $(\Delta E[I(k)])^2$ is stated in the following lemma.

\begin{lemma}
\label{lemma:dleta_expectation} When the following condition holds:
\[
    \left|\HP[I(k) \!=\! i] \!-\! \MP[I(k) \!=\! i] \right| \!\leq\! \epsilon \MP[I(k) \!=\! i]
\]
for all $i = 0, 1, \ldots, k$, then
\begin{equation}
\label{inequality_delta_expectation}
   \left( \Delta E[I(k)] \right)^2 \leq \epsilon^2 \MV[I(k)]
\end{equation}
holds.
\end{lemma}

\noindent {\bf Proof:} First we have $| \Delta p_i | \leq \epsilon \MP[I(k) =
i]$. It is straightforward to show $\sum\nolimits_{i=0}^k \Delta p_i = 0$, or
\begin{align*}
    \sum\nolimits_{i=0}^k \Delta p_i
    &= \sum\nolimits_{i=0}^k \HP[I(k) = i] - \sum\nolimits_{i=0}^k \MP[I(k) = i] \\
    & = 0 .
\end{align*}
Based on this fact, we could have
\begin{align}
\label{equation_delta_expectation}
(\Delta E[I(k)])^2
& = \left( \sum\nolimits_{i=0}^k i \Delta p_i \right)^2  \nonumber \\
& = \left( \sum\nolimits_{i=0}^k i \Delta p_i - E[I(k)] \sum\nolimits_{i=0}^k \Delta p_i \right)^2  \nonumber \\
& \leq \left(\sum\nolimits_{i=0}^k |i -E[I(k)]| \: | \Delta p_i | \right)^2 .
\end{align}
Then by substituting $| \Delta p_i |$ with $|\Delta p_i|  \leq \epsilon
\MP[I(k) = i]$ we have
\[
    (\Delta E[I(k)])^2  \leq  \left( \sum\nolimits_{i=0}^k |i -E[X]| \epsilon \MP[I(k) = i] \right)^2
\]
by applying Cauchy's Inequality we have,
\begin{align*}
(\Delta E[I(k)])^2
& \leq \left( \sum\nolimits_{i=0}^k  \epsilon^2 \MP[I(k) = i] \right)
       \left( \sum\nolimits_{i=0}^k (i - E[X] )^2 \MP[I(k) = i] \right) \\
&  =  \epsilon^2\MV[I(k)],
\end{align*}
which completes the proof. \done

In the following lemma we apply Lemma \ref{lemma:dleta_expectation} to derive
the error bound of variance, or $| \HV[I(k)] - \MV[I(k)]|$.

\begin{lemma}
\label{lemma:variance_loose_round} When the following condition holds:
\[
    \left|\HP[I(k) \!=\! i] \!-\! \MP[I(k) \!=\! i] \right| \!\leq\! \epsilon \MP[I(k) \!=\! i]
\]
for all $i = 0, 1, \ldots, k$, then
\[
   \left| \HV[I(k)] - \MV[I(k)] \right|\leq \epsilon (1+\epsilon)\MV[I(k)]
\]
holds.
\end{lemma}

\noindent {\bf Proof:} First we can write $\HV[I(k)] - \MV[I(k)]$ in the
following form:
\begin{align*}
\HV[I(k)] - \MV[I(k)]
& = \sum\nolimits_{i=0}^k i^2 \Delta p_i -
    2E[I(k)]\Delta E[I(k)] - (\Delta E[I(k)])^2 \\
& = \sum\nolimits_{i=0}^k ( i -E[I(k)] )^2 \Delta p_i -
    (E[I(k)])^2 \sum\nolimits_{i=0}^k \Delta p_i - (\Delta E[I(k)])^2
\end{align*}
Since $\sum_{i=0}^k \Delta p_i = 0$, thus we have
\begin{align}
\label{inequality_delta_variance}
\left|\HV[I(k)] - \MV[I(k)]\right|
& = \left| \sum\nolimits_{i=0}^k ( i - E[I(k)] )^2 \Delta p_i -
    (\Delta E[I(k)])^2 \right|    \nonumber \\
& \leq \sum\nolimits_{i=0}^k (i - E[I(k)])^2 | \Delta p_i | + (\Delta E[I(k)])^2,
\end{align}
by substituting $\Delta p_i$ with $|\Delta p_i| \leq \epsilon \MP[I(k) = i]$,
and substituting $(\Delta E[I(k)])^2$ with Inequality
(\ref{inequality_delta_expectation}) we have
\[
   | \HV[I(k)] - \MV[I(k)] |\leq \epsilon (1+\epsilon)\MV[I(k)],
\]
which completes the proof. \done

In the following lemmas, we show how to tradeoff the number of simulation
rounds $K$ with the performance guarantees. The tradeoff between $K$ and the
performance guarantee on pmf is stated in the following theorem.

\begin{lemma}
\label{lemma:prob_tight_bound} When the following condition holds:
\begin{equation}
\label{inequality:tight_rounds}
    K \geq  3\ln(2( k+1) / \delta ) / \epsilon^2 ,
\end{equation}
then for each $i=0, \ldots, k$,
\[
  \left| \HP[I(k)=i] - \MP[I(k)=i] \right| \leq \epsilon \sqrt{\MP[I(k)=i]},
\]
holds with probability at least $1 - \delta/(k+1)$.
\end{lemma}

\noindent {\bf Proof:} Without lose of any generality, consider the performance
guarantee on the approximation of $\MP[I(k) = i]$, where $i = 0, 1, \ldots, k$.
Our goal is to show
\[
    \left| \HP[I(k)=i] - \MP[I(k)=i] \right| \leq \epsilon \sqrt{\MP[I(k)=i]}.
\]
Following similar method in Lemma \ref{lemma:prob_loose_round} and the same
notations defined in the proof of Lemma \ref{lemma:prob_loose_round}, we could
have the following:
\\
{\bf Case 1}: $\MP[I(k) = i] = 0$. Then $\HP[I(k) = i] = 0$. Hence
\[
    \left| \HP[I(k)=i] - \MP[I(k)=i] \right| \leq \epsilon \sqrt{\MP[I(k)=i]}.
\]
\\
{\bf Case 2}: $\MP[I(k) = i] \neq 0$. Following similar method in Lemma
\ref{lemma:prob_loose_round} we have
\begin{align*}
& \MP \left[ |\HP[I(k)=i] - \MP[I(k)=i]| \leq \epsilon \sqrt{ \MP[I(k)=i] } \right] \\
& \geq 1 - \MP \left[ |\HP[I(k)=i] - \MP[I(k)=i]| \geq \epsilon \sqrt{ \MP[I(k)=i] } \right] \\
& =1 - \MP \left[ | \mathbf{I}_i - K\MP [I(k) = i]| \geq \frac{\epsilon K \MP[I(k)=i] }{\sqrt{\MP[I(k)=i]}} \right] \\
& \geq 1 - 2 e^{ - K \epsilon^2 / 3 },
\end{align*}
by substituting $K$ with Inequality (\ref{inequality:tight_rounds}) we can
prove this lemma. \done

Using the same method in Lemma \ref{lemma:prob_loose_round}, we can prove the
following lemma.
\begin{lemma}
\label{lemma:joint_prob_tight_round} When $K \geq  3\ln(2( k+1) / \delta ) /
\epsilon^2$ , then
\[
    \left|\HP[I(k) = i] - \MP[I(k) = i] \right|  \leq \epsilon \sqrt{\MP[I(k) = i]}
\]
holds for all $i = 0, 1, \ldots, k$ with probability at least $1 - \delta$.
\end{lemma}

In the following lemma we derive the error bound of expectation, or $|\HE[I(k)]
- E[I(k)]|$, under the condition that $|\HP[I(k) = i]-\MP[I(k) = i]|\leq
\epsilon \sqrt{\MP[I(k) = i]}$ holds for all $i = 0, 1, \ldots, k$.

\begin{lemma}
\label{lemma:expectation_tight_round} When the following condition holds:
\[
    \left|\HP[I(k) = i] - \MP[I(k) = i] \right| \leq \epsilon \sqrt{ \MP[I(k) = i]}
\]
holds for all $i = 0, 1, \ldots, k$, then
\begin{displaymath}
   \left| \HE[I(k)] - E[I(k)] \right| \leq \epsilon \sqrt{(k+1)k E[I(k)] /2}
\end{displaymath}
holds.
\end{lemma}

\noindent {\bf Proof:} First we have $\Delta p_i \leq \epsilon \sqrt{ \MP[I(k)
= i]}$. Then we have the following
\begin{align*}
\left| \HE[I(k)] - E[I(k)] \right|
& \leq \sum\nolimits_{i=0}^k i | \Delta p_i | \\
& \leq \sum\nolimits_{i=0}^k  \epsilon i \sqrt{ \MP[I(k) = i]},
\end{align*}
by applying Cauchy's Inequality we have,
\begin{align*}
\left| \HE[I(k)] - E[I(k)] \right|
& \leq \epsilon \sqrt{ \!\! \left(\sum\nolimits_{i=0}^k i \right) \!\! \left( \sum\nolimits_{i=0}^k i \MP[I(k) = i] \right)} \\
& = \epsilon \sqrt{(k+1)k E[I(k)] /2},
\end{align*}
which completes the proof. \done

In the following lemma we derive the bound of $(\Delta E[I(k)])^2$ under the
condition that $|\HP[I(k) = i]-\MP[I(k) = i]|\leq \epsilon \sqrt{\MP[I(k) =
i]}$ holds for all $i = 0, 1, \ldots, k$.

\begin{lemma}
\label{lemma:dleta_expectation_sqrt} When the following condition holds:
\[
    \left|\HP[I(k) = i] - \MP[I(k) = i] \right| \leq \epsilon \sqrt{ \MP[I(k) = i]}
\]
holds for all $i = 0, 1, \ldots, k$, then
\begin{equation}
\label{inequality_delta_expectation_sqrt}
   (\Delta E[I(k)])^2 \leq \epsilon^2 ( k + 1 ) \MV[I(k)]
\end{equation}
holds.
\end{lemma}

\noindent {\bf Proof:} First we have $\Delta p_i \leq \epsilon \sqrt{ \MP[I(k)
= i]}$. From Inequality (\ref{equation_delta_expectation}) we have
\[
    (\Delta E[I(k)])^2 \leq \left(\sum\nolimits_{i=0}^k |i -E[I(k)]| \: | \Delta p_i | \right)^2,
\]
by substituting $| \Delta p_i |$ with $\Delta p_i \leq \epsilon \sqrt{ \MP[I(k)
= i]}$ we have
\[
    (\Delta E[I(k)])^2  \leq  \left(\sum\nolimits_{i=0}^k |i -E[I(k)]| \: \epsilon \sqrt{\MP[I(k) = i]} \right)^2,
\]
by applying Cauchy's Inequality we have,
\begin{align*}
(\Delta E[I(k)])^2
&\leq \left( \sum\nolimits_{i=0}^k  \epsilon^2 \right)
      \left( \sum\nolimits_{i=0}^k (i - E[I(k)] )^2 \MP[I(k) = i] \right) \\
&=\epsilon^2 (k+1) \MV[I(k)],
\end{align*}
which completes the proof. \done

In the following lemma we apply Lemma \ref{lemma:dleta_expectation_sqrt} to
derive the error bound of variance, or $| \HV[I(k)] - \MV[I(k)]|$, under the
condition that $|\HP[I(k) = i]-\MP[I(k) = i]|\leq \epsilon \sqrt{\MP[I(k) =
i]}$ holds for all $i = 0, 1, \ldots, k$.

\begin{lemma}
\label{lemma:variance_tight_round} When the following condition holds:
\[
    \left|\HP[I(k) = i] - \MP[I(k) = i] \right| \leq \epsilon \sqrt{ \MP[I(k) = i]}
\]
holds for all $i = 0, 1, \ldots, k$, then
\begin{align*}
\left|\HV[I(k)] - \MV[I(k)] \right|
&\leq   \epsilon ( k +1 ) \left( \sqrt{ (2k + 1) \MV\:[I(k)]/6} + \epsilon \MV\:[I(k)] \right)
\end{align*}
holds.
\end{lemma}

\noindent {\bf Proof:} First we have $\Delta p_i \leq \epsilon \sqrt{ \MP[I(k)
= i]}$. From Inequality (\ref{inequality_delta_variance}) we have
\begin{align*}
\left|\HV[I(k)] - \MV[I(k)] \right|
&\leq   \sum\nolimits_{i=0}^k ( i - E[I(k)] )^2 | \Delta p_i | \!+  (\Delta E[I(k)])^2 ,
\end{align*}
by substituting $\Delta p_i$ with $|\Delta p_i|  \leq \epsilon \sqrt{ \MP[I(k)
= i]}$ we have
\begin{align*}
 | \HV[I(k)] - \MV[I(k)] |
&\leq \sum\nolimits_{i=0}^k ( i - E[I(k)] )^2 \epsilon \sqrt{\MP[I(k) = i]}+(\Delta E[I(k)])^2,
\end{align*}
by applying Cauchy's Inequality we have,
\begin{align*}
| \HV[I(k)] - \MV[I(k)] |
&\leq \epsilon \left( \sum\nolimits_{i=0}^k (i - E[I(k)])^2 \MP[I(k) = i] \right)^{1/2}
    \left( \sum\nolimits_{i=0}^k ( i - E[I(k)] )^2  \right)^{1/2} +(\Delta E[I(k)])^2  \\
& \leq  \epsilon \left((k + 1)\sqrt{ (2k + 1) \! \MV[I(k)]/6} \right)+(\Delta E[I(k)])^2 ,
\end{align*}
the proof can be completed by substituting $(\Delta E[X])^2$ with Inequality
(\ref{inequality_delta_expectation_sqrt}). \done

\end{document}